\documentclass[12pt,a4paper]{article}
\usepackage{geometry}
\usepackage[round]{natbib}
\usepackage{mathtools}
\usepackage{amsfonts}
\usepackage{amssymb}
\usepackage{graphicx}
\usepackage{placeins}
\usepackage[labelfont=bf]{caption}
\usepackage[skip=-0.25\textwidth]{subcaption}
\usepackage{enumitem}
\usepackage{titlesec}
\titlelabel{\thetitle.\quad} 
\newcommand{\subsubsubsection}[1]{\paragraph{#1}\mbox{}\\ \vspace{-2ex}}
\usepackage{color}
\usepackage[titletoc,title]{appendix}
\usepackage{xpatch}
\xpretocmd{\eqref}{equation\,}{}{}
\usepackage{authblk}
\title{\textbf{Electrohydrodynamic migration of a  surfactant-coated deformable drop in Poiseuielle flow}}
\author[1]{Antarip Poddar}
\author[1,2]{Shubhadeep Mandal}
\author[1]{Aditya Bandopadhyay\thanks{Email: aditya@mech.iitkgp.ernet.in}}
\author[1]{Suman Chakraborty\thanks{Email: suman@mech.iitkgp.ernet.in}}

\affil[1]{Department of Mechanical Engineering, Indian Institute of Technology Kharagpur, Kharagpur, West Bengal - 721302, India}
\affil[2]{Max Planck Institute for Dynamics and Self-Organization, Am Fassberg 17, D-37077 G\"{o}ttingen, Germany}
\date{}                     
\begin{document}
			\maketitle
	\begin{abstract}
\noindent
A coating of surface active agents on the drop surface is known to severely affect the motion and deformation characteristics of a drop in a flow field with a uniform shear stress gradient. Also the drop motion can be altered by exploiting the advantages of applying an external electric field. In this study we attempt to explore the consequences of surfactant coating on the  electrohydrodynamic manipulation of a drop motion in a plane Poiseuielle flow. We employ a `leaky dielectric model' in solving the electric potential field. In addition we consider bulk insoluble surfactants and a linear dependency of the surface tension on the surfactant concentration.  Subsequently a double asymptotic perturbation method is used in terms of small electric Reynolds number and capillary number in the limit of a diffusion-dominated surfactant transport mechanism. Also going beyond the widely employed axisymmetric framework, the coupled system of governing differential equations in three dimensions are then solved by adopting the `generalized Lamb solution technique'. The resulting important expressions are in agreement with various previous works in the literature in different limiting conditions. The expressions of key variables suggest that the flow curvature of the external flow, the electric field effects and the surfactant effects are coupled in a non-trivial manner, well beyond a linear superposition. A careful investigation shows that surfactant-induced Marangoni stresses interacts with the electrohydrodynamic stresses in a highly coupled fashion. Owing to this, under different combinations of electrical conductivity and permittivity  ratios, the Mason number  and the applied electric field direction, the  surfactants affect differently on the longitudinal as well as cross-stream  migration velocity of the drop. The effect of drop deformation on the drop speed is also found to be altered with increased sensitivity of the surface tension on the surfactant concentration, depending on the competing effects of the electrohydrodynamic flow modification and the `tip stretching' phenomena. Further with a suitable choice of electrical property ratios, the Marangoni effects can be used to either direct the drop in reaching a final transverse position towards or away from the channel centerline. The present results may be of utmost importance in providing a deep insight to the underlying complex physical mechanisms. Most importantly the ability of surfactants in selectively controlling the drop motion in different directions, makes them suitable for achieving a new degree of freedom in the electrical actuation of droplets in the microfluidic devices. 
	\end{abstract}
\section{Introduction}

Electrical actuation of drops and bubbles provides a useful means of manipulating the motion and shape deformation of such objects in a controllable fashion inside microfluidic environments. As a consequence it has found importance in various applications of practical relevance which include but are not limited to electrohydrodynamic atomization \citep{Wu2008}, enhanced mixing of reagents \citep{Hoburg1977,Tsouris2003}, enhanced coalescence and separation \citep{Ptasinski1992}, ink-jet printing \citep{Basaran2002}, electrowetting \citep{Banerjee2012} etc. Adding to this, various nonlinear phenomenon such as the shape deformation \citep{Haber1971,Taylor1966}, fluid inertia \citep{Mortazavi2000,Magnaudet2003}, viscoelasticity \citep{Mukherjee2013} and presence of surface active agents \citep{Anna2016,Baret2011} typically interacts with the drop dynamics, either being intentionally introduced or present in the system owing to unavoidable physical circumstances.

It is quite common in droplet-based microfluidic platforms to transport a drop through a channel using pressure driven flow. In these cases the externally imposed Poiseuielle flow greatly influences the drop movement. Since the  fluid flow in these occurrences mostly falls within a creeping flow limit, in the absence of any nonlinear effect, a symmetric flow reversal results. Thus the drop can only move in the applied flow direction only. However the nonlinear effects being coupled with the flow gradients, can break this symmetry and a cross-stream velocity component of the drop is observed \citep{Stan2011,Griggs2007,Haber1971,Mortazavi2000,Chen2014,Chan1979}.

Owing to their ability in modifying the interfacial properties of the drops, the surfactants have been proven to be immensely useful in applications such as emulsion stabilization, droplet generation and so on \citep{Tucker2002,Lequeux1998,Fischer2007,Baroud2010}. Thus a fundamental understanding of the physical consequences of surfactant coating drop surface under an externally imposed  Poiseuielle flow, is necessary to optimize the various aspects involved in the practical situations.   \citet{Hanna2010} found that a non-deformable drop with a coating of non-diffusing surfactant molecules on the interface, can migrate perpendicular to the background flow direction in a Poiseuielle flow and finally settles along the centerline of the channel. The study of \citet{Pak2014} revealed the existence of a similar cross-stream motion when the surfactant transport is predominantly diffusion dominated and convection is weak. In both the cases only an asymmetric distribution of surfactants about the axial plane is the triggering mechanism for such motion. Later \citet{Das2017a} investigated the coupled effect of drop deformation and non-uniformity in surfactant distribution on the  lateral movement of a drop. The surfactants are generally amphiphilic molecules which impart rigidity to the surface thus decreasing the mobility in those locations. When the drop is exposed to a dynamic condition, these molecules assumes a non-uniform distribution which further generates a surface tension gradient over the surface. This triggers a flow acting opposite to the flow direction which would have existed otherwise. Such effect is known as the so called `Marangoni convection' \citep{Subramanian2001}.

\citet{Taylor1966} was the first to provide a profound theoretical understanding of the deformation behaviour of a leaky diectric drop in another leaky dielectric fluid under a static electric field. The theoretical model introduced by Taylor was subsequently termed as the `leaky dielectric model' \citep{Saville1997} and has been employed in studying the buoyant ans well as gravity driven drops \citep{Ajayi1978,Tsukada1993,Xu2006}. According to this model the bulk is considered to be charge free due to small electrical conductivity of the fluids while the electric field effects comes through the accumulation of charges at the two fluid interface depending on the contrast in the electrical properties. These interfacial charges then gives rise to a net interfacial electrostatic stress known as the Maxwell stress.  In later studies it was also identified that charge convection  at the interface plays an important role in modifying the electrohydrodynamic flow pattern and the shape deformation \citep{Spertell1974,Feng1999,Lanauze2015,Yariv2016,Sengupta2017}. Only recently \citet{Bandopadhyay2016} observed the capability of the charge convection mechanism to break the symmetric flow pattern and cause a lateral movement of a sedimenting drop even without any shape distortion being taken into account. In a later study \citep{Mandal2016} they  brought out the pivotal contribution of charge convection as  well as the shape deformation in causing a cross-stream motion of a drop in a background Poiseuelle flow. They found that the drop can reach a steady state final position above or below the channel centerline, in contrary to the case of a drop final position along the centerline in the absence of an electric field \citep{Chan1979}.

The impact of the Marangoni effects in modifying the electrohydrodynamic flow and subsequent drop deformation and breakup was explored through experiments by \citet{Ha1998} and \citet{Zhang2015}. \citet{Ervik2018} showed that surfactants can also have significant  effect on the drop behaviour when an AC electric field is applied. Parallel theoretical models were also developed to gain a sound understanding of the underpinning physical mechanisms \citep{Ha1995,Teigen2010,Nganguia2013,Mandal2016b,Poddar2017}. {However in all these studies the authors have mostly concentrated on the case of a neutrally buoyant or a  sedimenting drop. The simplified configurations of these problems allowed them to consider only an axisymmetric variation of the variables.}
   
To the best of our knowledge no such study exists in the literature that takes into account of the intricate interplay  among the electrical Maxwell stress, a shear stress gradient typical to a Poiseuielle flow and the surfactant-induced Marangoni stress, on the migration characteristics of a deformable drop. In the present work we consider the motion of a leaky dielectric drop through another leaky dielectric medium through a plane Poiseuielle flow and under the influence of a applied uniform electric field which can also be adjusted to a tilted configuration. The drop surface is considered to be covered with bulk insoluble surfactant molecules. {To capture the ability of the considered physical mechanisms in breaking the symmetry about the axial plane, we employ a complete three dimensional mathematical framework going beyond the widely employed axisymmetric assumption in the earlier studies.}  Towards making an analytical treatment we solve the fluid flow equations under the Stokes flow regime and  adopted a leaky dielectric framework to obtain the  electric field potential inside and outside the drop. We then employ a double asymptotic perturbation with the capillary number $ (Ca) $ and electric Reynolds number $ (Re_E) $ as small perturbation parameters and subsequently consider that the surface P\'eclet number $ (Pe) $ is low. Finally, the drop trajectory is  obtained under the `quasi steady state' condition. 

\section{Problem Formulation}
\label{sec:mathform}
\subsection{Description of the physical system}
\label{ssec:system}

In the present study we consider a viscous drop of radius $ a $  migrating through another viscous medium which obeys a plane Poiseuielle flow denoted by $\widetilde{\boldsymbol{V}}_\infty=\widetilde{V}_0\left (c_0+c_1\left ( \dfrac{\widetilde{x}}{a} \right ) +c_2\left ( \dfrac{\widetilde{x}}{a} \right )^2  \right )\boldsymbol{i}_z$, where $ c_0=\dfrac{4\,\widetilde{x}_d}{\widetilde{H}}\left ( 1-\dfrac{\widetilde{x}_d}{\widetilde{H}} \right ) $, $ c_1=\dfrac{4\,a}{\widetilde{H}}\left ( 1-\dfrac{2\widetilde{x}_d}{\widetilde{H}} \right ) $, $ c_2=-\dfrac{4\,a^2}{\widetilde{H}^2} $ and $ \widetilde{V}_0 $ is the fluid velocity at the channel centerline. The imposed velocity field is obtained by considering a pressure driven flow in a parallel plate channel which is infinitely extended in $ y $ and $ z $ directions. The channel  height $ \widetilde{H} $ is considered to be much larger than the drop diameter.  The drop is simultaneously acted upon by an imposed DC electric field $  \widetilde{\boldsymbol{E}}_\infty = \widetilde{E}_0(E_x \boldsymbol{i}_x+E_z \boldsymbol{i}_z)$ in the $ x-z $ plane, where $ \widetilde{E}_0 $ is the chosen characteristic electric field strength and the cartesian components are related as $ \sqrt{E_x^2+E_z^2}=1 $. {The physical configuration of the present problem prohibits us from adopting a simplified axisymmetric co-ordinate system. Hence we employ  a complete three-dimensional co-ordinate  system capable of capturing the possible variations of the variables in all three spatial dimensions. }  The drop surface is considered to be contaminated with non-ionic surfactant molecules with a concentration distribution $ \widetilde{\Gamma}(\theta, \phi) $. This non-uniform distribution of surfactant molecules modifies the interfacial tension of the drop given by $ \widetilde{\gamma}(\theta, \phi) $. Since the drop surface can deform, it assumes a shape with varying radius $ r_S(\theta,\phi) $ while migrating in the cross-stream as well as far  downstream along the channel. 
\begin{figure}
	\centering
	\includegraphics[width=1.0\textwidth]{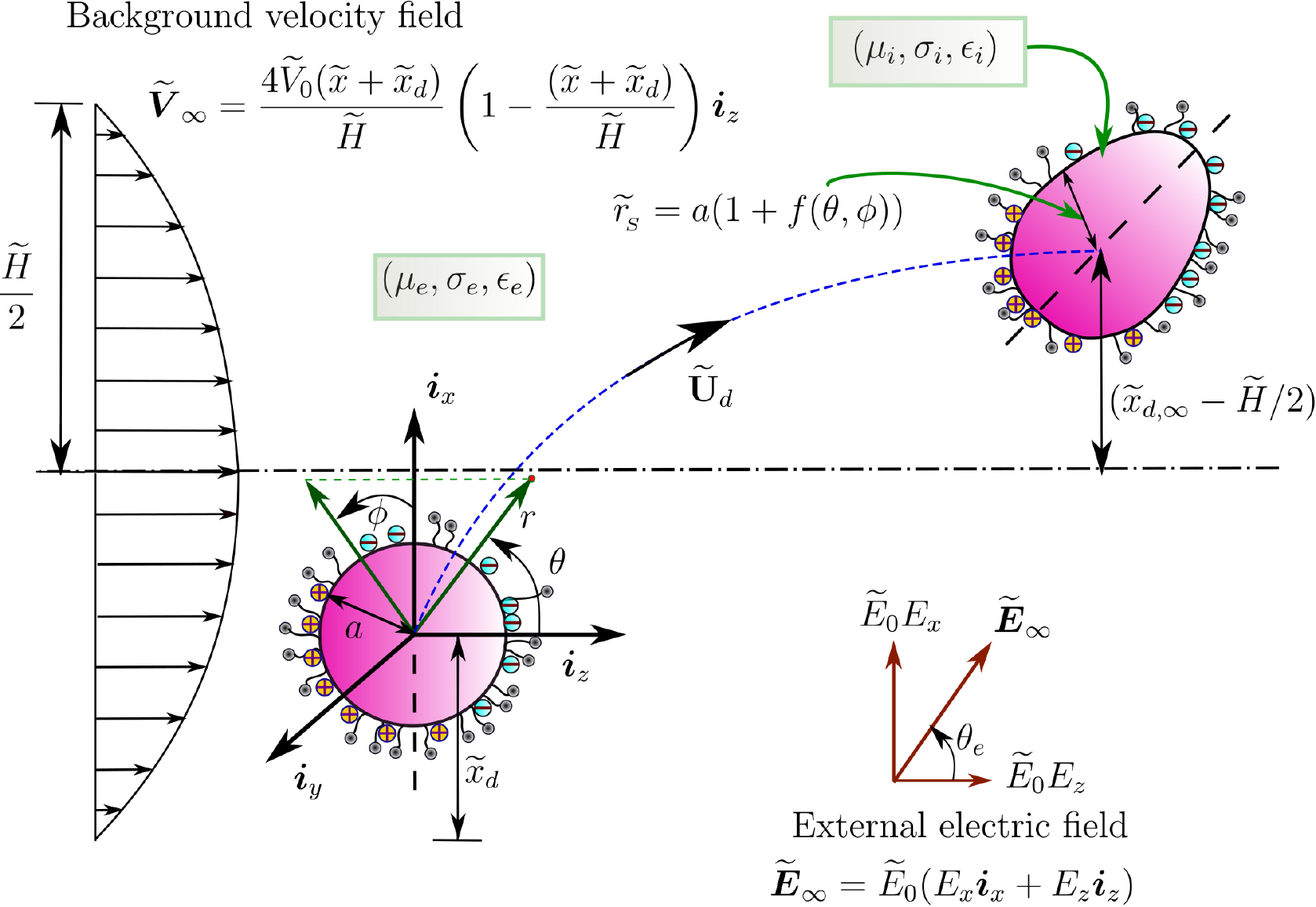}
	\caption{ Schematic diagram of a viscous, deformable and surfactant-covered drop which is simultaneously acted upon by background plane Poiseuielle flow $ (\widetilde{\mathbf{V}}_\infty) $ and a tilted electric field $\widetilde{\mathbf{E}}_\infty$ (with a tilt angle $ \theta_e $).  A non-uniform  accumulation of charge on the drop interface is also shown. A three dimensional  co-ordinate system ($\widetilde r,\theta,\phi$) is fixed at the centroid of the moving drop. The drop can translate in the plane of the imposed fields with a constant velocity, $ \widetilde{\mathbf{U}}_d=\widetilde{V}_0(U_{d,x}\mathbf{i}_x+U_{d,y}\mathbf{i}_y+U_{d,z}\mathbf{i}_z) $. The initial transverse position of the drop from the channel bottom plane is denoted by $ \widetilde{x}_d $ while its final steady state position is given by $ \widetilde{x}_{d,\infty} $. }
	\label{fig:schematic}
\end{figure}
\subsection{Governing equations and boundary conditions}
\label{sec:math_gov_eq}
\subsubsection{Surfactant transport}
\label{sss:surf trans}
Under the assumption of bulk insolubility of surfactant, the local concentration distribution of surfactant molecules on the drop surface is dictated  by a competition between diffusion and advection processes. The governing equation for surfactant transport takes a form \citep{Leal2007,Li1997,Stone1990a}
\begin{equation}
\widetilde{\nabla}_{\!_S}\cdot(\widetilde{\mathbf{u}}_{\!_S}\widetilde{\Gamma})=D_{\!_S}\widetilde{\nabla}^2_{\!_S}\widetilde{\Gamma}.
\label{eq:dim_surf_transport}
\end{equation}
Here $\widetilde{\mathbf{u}}_{\!_{\,S}}$ and $D_{\!_S}$ denote the fluid velocity at the drop interface ($ \widetilde{\mathbf{u}}_i|_{\widetilde{r}=\widetilde{r}_{\!_S}} $) and the surface-diffusion coefficient, respectively.  This surfactant concentration distribution $ \widetilde{\Gamma}(\theta, \phi) $, in turn, trigger a variation of surface tension $ \widetilde{\gamma}(\theta, \phi) $. Here we assume a dilute concentration of surfactant and a small deviation from the equilibrium throughout the system. This allows us to consider a linearized version of the equation  of state given as \citep{Stone1990a,Li1997}:
\begin{equation}
\label{eq:linear}
\widetilde{\gamma}(\widetilde{\Gamma})=\widetilde{\gamma}_c-R_gT\,\widetilde{\Gamma},
\end{equation}
where $ R_g $ is the ideal gas constant, $ T $ is the absolute temperature and $\widetilde{\gamma}_c$ is the surface tension of a surfactant-free drop which are related mutually by $\widetilde{\gamma}_{eq}=\widetilde{\gamma}_c-R_gT\,\widetilde{\Gamma}_{eq}$. Under these considerations \eqref{eq:linear} becomes
\begin{equation}
\label{eq:mod_linear}
\widetilde{\gamma}(\widetilde{\Gamma})=\widetilde{\gamma}_{eq}+R_gT(\widetilde{\Gamma}_{eq}-\widetilde{\Gamma}).
\end{equation}  

To identify the various important dimensionless numbers and property ratios we choose the drop radius $ a $ as a reference length scale, the centerline velocity of the plane Poiseuielle flow $ \widetilde{V}_0 $ as reference velocity scale and equilibrium surfactant concentration $ \widetilde{\Gamma}_{eq} $ as surfactant concentration scale. In addition to that we choose the physical quantities of the surrounding fluid $ (\mu_e ,\epsilon_e\, \text{and}\, \sigma_e) $ as references for property values. As a consequence of the above non-dimensionalization scheme, the relative importance between  electric stress ($\sim$ $\epsilon_eE_0^2$) and hydrodynamic stress ($\sim$ $\mu_e \widetilde{V}_0/{a}$) is depicted by Mason number, $ M=\epsilon_eaE_0^2/\mu_e \widetilde{V}_0 $.  The deformable nature of drop shape is quantified by the capillary number $Ca=\mu_e \widetilde{V}_0/\widetilde \gamma_{eq}$ which signifies the relative strength of viscous and capillary stresses. The ratio between two important time scales associated with the electric charge, namely the charge relaxation time scale $ t_{cr}=\epsilon_e/\sigma_e $ and the charge convection time scale $ t_{cc}= a/\widetilde{V}_0 $, is denoted by electric Reynolds number $Re_E=\epsilon_e \widetilde{V}_0/a\sigma_e$. Along similar lines the surface P\'eclet number $ Pe=a\,\widetilde V_0/D_{\!_S} $ stands for the relative strength of surfactant convection and diffusion at the drop surface. The property ratios are defined as: electrical
permittivity ratio, $S=\epsilon_i/\epsilon_e$; electrical conductivity ratio, $R=\sigma_i/\sigma_e$; and viscosity ratio, $\lambda=\mu_i/\mu_e$. Henceforth all the $\,\,\widetilde{}\,\,$  symbols will be dropped to denote the non-dimensional forms of the field variables and property ratios.Thus the dimensionless form of \eqref{eq:dim_surf_transport} takes the below form:
\begin{equation}
Pe\nabla_S\cdot({\mathbf{u}_{\!_{\,S}}}\Gamma)=\nabla^2_{\!_S}\Gamma
\label{eq:surf_transport}
\end{equation}
The surfactant concentration also adhere to a mass conservation of at the drop surface denoted as
\begin{equation}
\label{mass-constraint}
\int_{\phi=0}^{2\pi}\int_{\theta=0}^{\pi} \Gamma(\theta, \phi)\, r^2_{\!_S}(\theta,\phi)\sin{\theta} \,d\theta\,d\phi=4 \pi 
\end{equation}

\subsubsection{Electrostatic problem}
\label{ssec:E_prob}
Following the Taylor-Melcher leaky dielectric model 
\citep{Taylor1966,Melcher1969,Saville1997}, in the present study we have considered absence of volumetric charges and only surface accumulation of charge is taken into account. Thus electric potential in and around the drop ($ \varphi_i, \varphi_e $) are governed by the Laplace equations, given as  
\begin{equation}
\begin{aligned}
\nabla^2\varphi_i&=0  \\
\nabla^2\varphi_e&=0  
\end{aligned}
\label{eq:laplace}
\end{equation}

The electric potential takes finite values inside the drop, i.e. $\varphi_i$ is bounded at $r\le r_S$. Also the far stream potential-gradient is determined by the imposed electric field by the following relation:
as $r\to\infty,$\,\,$\nabla\varphi_e=-\mathbf{E}_\infty $. At the drop interface potential continuity is satisfied. The interface balance of charges  is governed by a competition between Ohmic conduction and surface convection of accumulation charges. In addition we consider the approximation of quasi-steady-state which assumes instantaneous adjustment of surface charges to the background flow-field. Thus in mathematical notation the charge balance at the interface becomes
\begin{equation}
\textrm{\\at} \,\, r=r_{\!_S}(\theta, \phi)\quad \mathbf{n}\cdot (R\nabla \varphi_i-\nabla\varphi_e)=-Re_E\nabla_{\!_S}\cdot(q\,_{\!_S}\mathbf{u}_{{\,\!_S}}), 
\label{eq:cc}
\end{equation}
In the above equation $\mathbf{n}$ denotes the unit normal vector at the drop surface and $\nabla_{\!_S}$ is the surface divergence operator. The interfacial charge density  ($q_{\!_{\,S}}$) appearing in \eqref{eq:cc} is triggered by a mismatch between the electrical properties of the two fluids and  is determined by the relation:
\begin{equation}
\label{eq:charge-formula}
q_{\!_{\,S}}(\theta,\phi)=\mathbf{n}\cdot (S\nabla \varphi_i-\nabla\varphi_e)\big|_{r=r_{\!_S}}^{}.
\end{equation}
The influence of fluid flow on the electric potential distribution is embedded in \eqref{eq:cc}, hence suggesting  a two-way coupled nature of the electrostatics and  hydrodynamics in the present problem.

\subsubsection{Hydrodynamic problem}
\label{ssec:H_prob}
Here we assume the creeping flow limit $(Re \ll 1)$ \citep{Happel1983} for flow inside and outside of the drop and thus neglect the inertia term in the Navier-Stokes equation.  Thus the hydrodynamics can be fully described by the Stokes-flow equations and the incompressibility conditions given as 
\begin{equation}
\begin{aligned}
-\nabla{p}_i&+\lambda\nabla^2\mathbf{u}_i=0, &&\nabla\cdot{\mathbf{u}_i}=0\\
-\nabla{p}_e&+\,\,\,\,\nabla^2\mathbf{u}_e=0,        &&\nabla\cdot{\mathbf{u}_e}=0
\end{aligned}
\label{eq:momentum}
\end{equation}

The velocity and pressure inside the drop ($\mathbf{u}_{\!\,_i},p_{\!\,_i}$) satisfy boundedness at the origin of the reference frame which is fitted at the drop centroid and translates with the drop $ (\mathbf{U}_d) $. In the far-field,  flow velocity approaches  the imposed plane Poiseuielle flow condition, i.e. at
$r\to\infty,$\quad$\mathbf{u}_e=\mathbf{V}_\infty-\mathbf{U}_d$.
At the drop interface fluid velocities are continuous $(\mathbf{u}_i=\mathbf{u}_e)$  and follow a quasi-steady-state no-penetration condition $(\mathbf{u}_i\cdot{\mathbf{n}}=\mathbf{u}_e\cdot{\mathbf{n}}=0)$. 

For a  clean drop the hydrodynamic and electric stresses balance each other at the interface. However when the drop surface is coated with a non-uniform distribution of surfactants, an additional interface stress component known as the Marangoni stress, contributes to this balance.  Thus the resultant stress balance takes the following form:   
\begin{equation}
\\\textrm{at\,} \,r=r_{\!_S}(\theta,\phi),\;
(\boldsymbol{\tau}^H_e+M\,\boldsymbol{\tau}^E_e)\cdot{\mathbf{n}} -(\boldsymbol{\tau}^H_i+M\,\boldsymbol{\tau}^E_i)\cdot{\mathbf{n}}=\frac{1}{Ca}(\nabla\cdot{\mathbf{n}})\mathbf{n}+\underbrace{Ma\,[(1-\Gamma)(\nabla\cdot{\mathbf{n}})\mathbf{n}+\nabla_{\!_S}\Gamma]}_\text{Marangoni stress},
\label{eq:stress_bal}
\end{equation}
where  $ Ma $ denotes the Marangoni number defined as $ Ma=R_gT\,\widetilde{\Gamma}_{eq}/\mu_e \widetilde V_0 $  signifying the ratio between the characteristic Marangoni stress and viscous stress. The viscous stress tensors $ 	(\boldsymbol{\tau}^H_i,\boldsymbol{\tau}^H_e) $ and electric stress tensors $ (\boldsymbol{\tau}^E_i,\boldsymbol{\tau}^E_e) $ are defined as:  
\begin{equation}
\left.\begin{matrix}\,\boldsymbol{\tau}^H_i=-p_i\mathbf{I}+{\lambda}[\nabla\mathbf{u}_i+(\nabla\mathbf{u}_i)^T] 
\\ \boldsymbol{\tau}^H_e=-p_e\mathbf{I}+[\nabla\mathbf{u}_e+(\nabla\mathbf{u}_e)^T]
\end{matrix}\right\}   \text{ and }  
\left.\begin{matrix}\, \,\,\boldsymbol{\tau}^E_i=S\left[\mathbf{E}_i\mathbf{E}^T_i-\frac{1}{2}\left|{\mathbf{E}_i}\right|^2\mathbf{I}\right] 
\\ \boldsymbol{\tau}^E_e=\left[\mathbf{E}_e\mathbf{E}^T_e-\frac{1}{2}\left|{\mathbf{E}_e}\right|^2\mathbf{I}\right]
\end{matrix}\right\} 
\end{equation}

In order to quantify the impact of non-uniformity in surfactant concentration on the surface tension, we introduce a phiscochemical parameter known as the elasticity number $ \beta (0 \le \beta \le 1) $ \citep{Stone1990a,Li1997,Mandal2016b}. This is defined as $\beta=-\dfrac{d(\widetilde{\gamma}/\widetilde\gamma_c)}{d\,\widetilde\Gamma}=\dfrac{R_gT\,\widetilde\Gamma_{eq}}{\widetilde\gamma_c}$. Here the surface elasticity number $ (\beta) $ is varied between a range of 0 and 0.8. 
To elucidate the effect variability in of surfactant concentration distribution on the stress balance condition, we represent the tangential and normal stress balances as follows
\begin{equation}
\\\textrm{at } \,r=r_{\!_S}(\theta,\phi),\;
\left [(\boldsymbol{\tau}^H_e+M\,\boldsymbol{\tau}^E_e)\cdot{\mathbf{n}} -(\boldsymbol{\tau}^H_i+M\,\boldsymbol{\tau}^E_i)\cdot{\mathbf{n}}  \right ]\cdot (\mathbf{I}-\mathbf{nn})=\underbrace{\frac{\beta}{Ca(1-\beta)}\,(\nabla_{\!_S}\Gamma)\cdot (\mathbf{I}-\mathbf{nn})}_{\textrm{ Marangoni effect}}
\label{eq:t_stress_bal}
\end{equation}
and
\begin{equation}
\\\textrm{at } \,r=r_{\!_S}(\theta,\phi),\;
\left [(\boldsymbol{\tau}^H_e+M\,\boldsymbol{\tau}^E_e)\cdot{\mathbf{n}} -(\boldsymbol{\tau}^H_i+M\,\boldsymbol{\tau}^E_i)\cdot{\mathbf{n}}\right ]\cdot \mathbf{n} =\frac{1}{Ca}(\nabla\cdot{\mathbf{n}})+\underbrace{\frac{\beta}{Ca(1-\beta)}\,[(1-\Gamma)(\nabla\cdot{\mathbf{n}})]}_\text{Marangoni effect}
,
\label{eq:n_stress_bal} 
\end{equation}
respectively. Here $ (\mathbf{I-nn}) $ is the surface projection operator.

\subsection{Solution using asymptotic perturbation} \label{ss:perturbation}
The governing equations of electrical potential, hydrodynamic quantities  and surfactant concentration along with their boundary conditions, are highly nonlinear and are coupled to each other in such a manner that it prohibits a closed form analytical solution for arbitrary range of relevant dimensionless numbers namely $ Ca, Pe, Re_E \text{\,and\,} M $. Also drop surface deviates from the initially assumed spherical shape upon application of flow field. This renders the application-points of the boundary conditions unknown beforehand. Thus we proceed by considering a feasible range of the dimensionless parameters involved. In most of the practical conditions these parameters are found to be in the range: $ Ca \ll 1, Re_E \ll 1 \text{ and } M \sim 1$ \citep{Mhatre2013,Xu2006a,Bandopadhyay2016}. Hence we choose $ Ca $ and $ Re_E $ as perturbation parameters and proceed to obtain a solution using a double asymptotic perturbation. 

It is important to note that a perturbation scheme based on the surface P\'eclet number $ (Pe) $ has been previously employed in several earlier studies \citep{Pak2014,Das2017b,Subramanian2005} related to spherical drop surfaces. However, with the present consideration of a deformable drop surface it is not possible to adopt $ Ca $ and $ Pe $ as independent dimensionless numbers  \citep{Stone1990a,Mandal2016c}.  Thus we take a different route and define a dimensionless physicochemical parameter composed of different property constants  and the length scale as  \citep{Stone1990a,Mandal2016c,Das2017a} 
\begin{equation}\label{eq:k-def}
k=\frac{Pe}{Ca}=\frac{\widetilde\Gamma_{eq}\,a}{\mu_e\,D_{\!_S}},
\end{equation}
We can now express P\'eclet number in terms of $ k  $ and $ Ca $ as $ (Pe=k\,Ca) $.  In the present study we focus on the condition where the surface P\'eclet number is low ($Pe\ll1$) \citep{Pak2014a,Das2017a,Mandal2016c}. Such a condition physically interprets the situation where the diffusion of surfactant molecules is much stronger than the convection effect. Thus  the parameter $ k $ turns out to be an $ O(1) $ constant. 

Based on the above perturbation scheme, any field variable $\zeta$ can now be expressed in a general form as
\begin{equation}
\zeta=\zeta^{(0)}+Ca\,\zeta^{(Ca)}+Re_E\,\zeta^{(Re_E)}+O(Ca^2,CaRe_E,Re_E^2).
\label{eq:general_perturb}
\end{equation}
However, the stress and pressure variables inside the drop region should be expanded in a slightly different manner as follows \citep{Bandopadhyay2016a,Mandal2016a,Chan1979}:
\begin{equation}
\label{eq:cap-press}
\left.
\begin{split}
p_i &=\frac{1}{Ca}p^{(1/Ca)}_i+p^{(0)}_i+Ca\,p^{(Ca)}_i+Re_E\,p^{(Re_E)}_i+O(Ca^2,CaRe_E,Re_E^2)\\
\boldsymbol{\tau_i} &=\frac{1}{Ca}\boldsymbol{\tau_i}^{(1/Ca)}+\boldsymbol{\tau_i}^{(0)}+Ca\,\boldsymbol{\tau_i}^{(Ca)}+Re_E\,\boldsymbol{\tau_i}^{(Re_E)}+O(Ca^2,CaRe_E,Re_E^2)
\end{split}
\right \}
\end{equation} In a similar fashion the dimensionless drop radius can be written as 
\begin{equation}
r_{\!_S}(\theta, \phi)=1+Ca\,f^{(Ca)}+CaRe_E\,f^{(CaRe_E)}+Ca^2\,f^{(Ca^2)}+\ldots
\label{eq:shape_perturb}
\end{equation} 
In the above equation $ f^{(w)}(\text{where } w=Ca,CaRe_E \text{ and } Ca^2) $ are functions of $ \theta $, $ \phi $  and denote the shape function in various order of perturbation. The surfactant concentration $ \Gamma(\theta , \phi) $ can be similarly represented as \citep{Ha1995,Stone1990a,Mandal2016b}, given as 
\begin{equation}
\Gamma=\Gamma^{(0)}+Ca\,\Gamma^{(Ca)}+CaRe_E\,\Gamma^{(CaRe_E)}+Ca^2\Gamma^{(Ca^2)}+\ldots
\label{eq:surf_perturb}
\end{equation}

When the drop shape deviates from sphericity owing to imposed conditions, the boundary conditions for electric and hydrodynamic problems are to be determined at the modified drop-bulk fluid interface locations. On the other hand, shape distortion is dependent on the flow behavior. However, under the assumption of small deviation of drop shape from sphericity, we can apply the domain perturbation technique and represent any generic variable at the deformed surface as 
\begin{equation}\label{eq:taylor-expansion}
\left[\zeta\big|_{ \\at\,r=r_{\!_S}(\theta, \phi)=1+Ca\,f^{(Ca)}(\theta, \phi) }^{}\right]^{(Ca)}={\zeta}^{(Ca)}\big|_{ \\at\,r=1 }^{}+f^{(Ca)}(\theta, \phi)\frac{\partial\zeta^{(0)}}{\partial r}\bigg|_{r=1}.
\end{equation} 
This allows us to apply the boundary conditions at $ r=1 $ with appropriate account of the correction terms presented above.

Based on the above formulation the surfactant transport equation for various orders of expansion takes the following forms:
\begin{equation}
\text{\textbf{Leading order:}  } \textrm{\qquad at}  \,\,r=1,\quad 
k\nabla_{\!_S}\cdot({\mathbf{u}^{(0)}_{\!_S}})=\nabla^2_{\!_S}\Gamma^{(Ca)}
\end{equation}
\begin{equation}
\;\text{\textbf{Order }}Re_E: 
\textrm{\qquad at} \,\,r=1,\quad 
k\nabla_{\!_S}\cdot({\mathbf{u}^{(Re_E)}_{\!_S}})=\nabla^2_{\!_S}\Gamma^{(CaRe_E)}
\end{equation}
\begin{equation}
\;\text{\textbf{Order }}Ca:\quad
\quad\textrm{at} \,\,r=1+Ca f^{(Ca)},\quad 
k\nabla_{\!_S}\cdot({\mathbf{u}^{(0)}_{\!_S}}\Gamma^{(Ca)}+{\mathbf{u}^{(Ca)}_{\!_S}})=\nabla^2_{\!_S}\Gamma^{(Ca^2)}.
\label{eq:Ca-order-surf}
\end{equation}

\subsubsection{Description of field variables}
\label{sss:spherical-harmonics}
Since the Laplace equation (\eqref{eq:laplace}) is satisfied by the electric potentials both inside and outside of the drop, the electric potentials can be conveniently expressed in the form
\begin{equation}
\label{eq:potential-general}
\left.
\begin{split}
\varphi_i=\sum_{n=0}^{\infty}r^n\sum_{m=0}^{n}\left [ a_{n,m}\cos(m \phi)+\hat{a}_{n,m}\sin(m \phi) \right ]P_{n,m}(\eta)\qquad\qquad\\
\varphi_e=\varphi_\infty+\sum_{n=0}^{\infty}r^{-n-1}\sum_{m=0}^{n}\left [ b_{-n-1,m}\cos(m \phi)+\hat{b}_{-n-1,m}\sin(m \phi) \right ]P_{n,m}(\eta).
\end{split}
\right \}
\end{equation}
Here $ P_{n,m}(\eta) $ denotes the associated Legendre polynomial of degree $ n $ and order $ m $ with an argument $ \eta=\cos(\theta) $. Also $ \varphi_\infty $ stands for the unperturbed electric potential in the far-stream and represents the externally applied electric potential by means of a tilted electric field $ \mathbf{E}_\infty $. Thus the far-stream potential exists only in the leading order and takes the form $ \varphi_\infty^{(0)} = -r(E_z P_{1,0}+E_x P_{1,1}\cos(\phi))$. In the subsequent sections $ a_{n,m},\hat{a}_{n,m},b_{-n-1,m} $ and $ \hat{b}_{-n-1,m} $ will be determined for various perturbation orders ($ w=0,Ca,Re_E $) using appropriate set of boundary conditions and thereafter applying orthogonality of $ P_{n,m} (\eta)$.

From \eqref{eq:momentum} it is evident that the flow fields both inside and outside the drop are Stokesian and satisfy the incompressibility condition as well. Hence we adopt the Lamb's general solution method \citep{Lamb1975,Happel1983,Hetsroni1970} to obtain the velocity and pressure at different orders of expansion. According to this method velocity and pressure inside the drop can be expressed as a summation of growing solid spherical harmonics $(p_n,\Phi_n,\chi_n)$ in the form:
\begin{equation}
\mathbf{u}_i=\sum_{n=1}^{\infty} \left [\nabla\times (\mathbf{r}\chi_n)+\nabla \Phi_n+\frac{n+3}{2(n+1)(2n+3)\lambda}r^2\nabla p_n-\frac{n}{(n+1)(2n+3)\lambda}\mathbf{r}p_n  \right ]
\label{eq:in-vel-lamb}
\end{equation}
\begin{equation}
\text{and }\quad p_i=\sum_{n=1}^{\infty}p_n
\label{eq:in-press-lamb}
\end{equation}
while the outside velocity and pressure fields outside the drop can be expressed in terms 
of the decaying solid spherical harmonics $(p_{-n-1},\Phi_{-n-1},\chi_{-n-1})$ and far-stream conditions:	 
\begin{equation}
\mathbf{u}_e=\mathbf{V}_\infty - \mathbf{U}_d +\sum_{n=1}^{\infty} \left [\nabla\times (\mathbf{r}\chi_{-n-1})+\nabla \Phi_{-n-1}-\frac{n-2}{2n(2n-1)}r^2\nabla p_{-n-1}+\frac{n+1}{n(2n-1)}\mathbf{r}p_{-n-1}  \right ]
\label{eq:out-vel-lamb}
\end{equation}
\begin{equation}
\text{and }\quad p_e=\sum_{n=1}^{\infty}p_{-n-1}
\label{eq:out-press-lamb}
\end{equation}
The detailed expressions of the growing and decaying harmonics in terms of the associate Legendre polynomials $ P_{n,m} $ can be found in earlier studies  \citep{Hetsroni1970,Happel1983,Bandopadhyay2016a}, and thus we do not repeat them here for sake of brevity.

Now following \citet{Haber1972}, we can express the surfactant concentration as a summation of spherical harmonics given in the form:
\begin{equation}
\Gamma=\sum_{n=0}^{\infty}\sum_{m=0}^{n}\left [ \Gamma_{n,m}\cos(m \phi)+\hat{\Gamma}_{n,m}\sin(m \phi) \right ]P_{n,m}(\eta).
\end{equation} Similarly the shape functions associated with drop deformation are represented as
\begin{equation}
f=\sum_{n=0}^{\infty}\sum_{m=0}^{n}\left [ L_{n,m}\cos(m \phi)+\hat{L}_{n,m}\sin(m \phi) \right ]P_{n,m}(\eta).
\end{equation}
In order to determine the constants involved $( A_{n,m},B_{n,m},C_{n,m},A_{-n-1,m},B_{-n-1,m},C_{-n-1,m},\\ \Gamma_{n,m} \text{ and }\Gamma_{-n-1,m} )$ in the expressions of spherical harmonics, the boundary conditions are  transformed in consistent vectorial forms similar to the method adopted by several authors previously \citep{Hetsroni1971,Haber1972,Bandopadhyay2016a,Mandal2016,Das2017a}. 

Next we simultaneously solve the surfactant transport equation, all the boundary conditions for electric potential and hydrodynamics except the normal stress boundary condition (\eqref{eq:n_stress_bal}). Consequently, the force-free condition on the drop:
\begin{equation}\label{eq:force-free}
\boldsymbol{F}^H+M\boldsymbol{F}^E=\boldsymbol{0},
\end{equation}
is utilized at different perturbation orders. Here electrical and hydrodynamic forces on the drop can be obtained as 
\begin{equation}
\label{eq:force-integrations}
\left.
\begin{split}
\mathbf{F}^E=\int_{\phi=0}^{2\pi} \int_{\theta=0}^{\pi} \left(\boldsymbol{\tau}^E_e\cdot\mathbf{n} \right)r^2_{\!_S} \sin{\theta} d\theta  d\phi
\\
\mathbf{F}^H=\int_{\phi=0}^{2\pi} \int_{\theta=0}^{\pi} \left(\boldsymbol{\tau}^H_e\cdot\mathbf{n} \right)r^2_{\!_S} \sin{\theta} d\theta  d\phi 
\end{split}
\right \},
\end{equation}
respectively. Thus we obtain the arbitrary constants associated with various spherical expansions and a complete description of the field variables in terms of the physicochemical and geometric parameters, can be provided. Finally the normal stress boundary condition (\eqref{eq:n_stress_bal})  is used to obtain the harmonics $ (L_{n,m},\hat{L}_{n,m})
$ associated with the shape deformation of the drop.  

\subsubsection{Combination of surfactant, electric field and imposed flow}
\label{sss:combination-surf-E}
In this section we elaborate  the coupling between the non-uniform surfactant distribution and the imposed electric field  along with the flow effects presented due to an imposed Poiseuielle flow.
\subsubsubsection{Leading order description}
\label{ssssec:leading-order}\\
Here we only consider a spherical drop in the absence of surface charge convection. In this case the electrical problem is uncoupled from either of the flow field or the surfactant concentration distribution and thus can be represented as a function of electrical properties only. 
       In this order the surfactant concentration distribution takes the form:
\begin{equation}
\label{eq:leading-surf-distr}
\Gamma^{(Ca)}=\sum_{n=1}^{3}\sum_{m=0}^{n} \Gamma_{n,m}^{(Ca)}\cos(m \phi)P_{n,m}(\eta).
\end{equation}
Various surface harmonics appearing in the above equation are provided in Appendix-\ref{sec:GCa_nm}. Following  \eqref{eq:leading-surf-harmonics}, it can be inferred that the term $ c_2 $ in the applied flow field, denoting a gradient in the shear stress, has significant consequences in governing the surfactant distribution. Moreover $ \Gamma^{(Ca)} $ is not only affected by the imposed Poiseuielle flow parameters but also electrical effects have an important role to play in it. While in some of the terms  both effects are coupled (e.g. $  \Gamma^{(Ca)}_{2,1} $), some other terms only exist when an electric field is present (e.g. $  \Gamma^{(Ca)}_{2,0}  $ and $  \Gamma^{(Ca)}_{2,2}  $).

The pressure and velocity fields in both interior and exterior to the drop can be completely expressed with non-zero harmonics upto $ n=3 $. The detailed description of these variables are provided in the supplementary material. Solving the force-free condition (\eqref{eq:force-free}) it is found that only the longitudinal component of drop velocity exists ($ U^{(0)}_{d,z} \ne 0 $) in the leading order and all the other components become unconditionally zero ($ U^{(0)}_{d,x},U^{(0)}_{d,y} = 0 $). The longitudinal drop velocity component is obtained as

\begin{equation}
\label{eq:Ud-leading-low}
U^{(0)}_{d,z}=\left (c_{{0}}+{\frac {\lambda c_{{2}}}{2+3\lambda}}  \right )+\underbrace{\frac{2}{3}{\frac { k \beta c_{{2}}}{ (2+3\lambda)^2 
			\left[  (1-\beta)+\dfrac{k\beta}{2+3\lambda}
			\right] }}}_{\mathcal{F}^{(0)}_z=\text{ Surfactant contribution}}
\end{equation}   
From the above expression of $ U_{d,z}^{(0)} $ two important observations can be made. Firstly, the electrical parameters do not have any influence in deciding the longitudinal flow velocity in the leading order, and $ U^{(0)}_{d,z} $ comes solely from hydrodynamic origins. Secondly, in the diffusion dominated regime $ (Pe \ll 1) $ the presence of surfactant plays a role in modifying the leading order longitudinal drop velocity only when there exists a flow curvature (i.e. $ c_2 \ne 0 $), thus confirming the coupling between the surfactant effects with the imposed flow field conditions in the leading order itself. Moreover, the surfactant contribution to $ U_{d,z}^{(0)} $ turns out to be a linear superposition to the case of a clean drop.

With a complete knowledge of the hydrodynamic variables of the leading order we now invoke the normal stress balance equation (\eqref{eq:n_stress_bal}) at the interface to get the $ O(Ca) $ shape function as 
\begin{equation}\label{eq:leading-shape}
f^{(Ca)}=\sum_{n=2}^{3}\sum_{m=0}^{n} L^{(Ca)}_{n,m}\cos(m \phi)P_{n,m}(\eta).
\end{equation}
The expressions of non-zero spherical harmonics $( L^{(Ca)}_{n,m}) $ in the shape function are presented in the Appendix-\ref{sec:LCa_nm}. A closer look into the correction terms in \eqref{eq:sol-leading-shape} reveals that they still contain electrical terms apart from the imposed flow and surfactant effects. This demonstrate the fact that beyond a linear superposition of the surfactant and electrical effects, a complex interplay of these two effects decide the shape deformation characteristics of the drop. 
\subsubsubsection{$ O(Re_E):$ Effect of finite surface charge convection}
\label{ssssec:order-ReE}\\
 In this order of perturbation we only consider a finite amount of surface charge convection $ (Re_E \ne 0) $ which from mathematical viewpoint only modifies the charge balance condition (\eqref{eq:cc}). In addition, the drop deformability is also not considered. As a consequence, similar to the leading order case,  stress balance conditions are evaluated at the spherical surface itself and they also take similar mathematical forms as that of the leading order case. The main difference with the leading order case lies in the charge continuity equation which in this case takes a form: 
\begin{equation}
\textrm{\\at} \,\, r=1, \quad \mathbf{i}_r\cdot (R\nabla \varphi^{(Re_E)}_i-\nabla\varphi^{(Re_E)}_e)=-\nabla_{\!_S}\cdot(q^{(0)}_{\,\!_S}\mathbf{u}^{(0)}_{{\,\!_S}}). 
\label{eq:cc-O(ReE)}
\end{equation}
The above equation shows that the $ O(Re_E) $ electric potential distribution carries the information about the charge distribution and fluid velocity at the interface. It has been pointed out earlier that the leading order charge distribution $(q^{(0)}_{\,\!_S})$ is not influenced by the surfactant effects. However, the leading order surface velocity $ (\mathbf{u}^{(0)}_{{\,\!_S}} )$ does contain the effects of surfactant parameters. Thus unlike the leading order, in the present case we find the electrical potential distribution getting affected by the inhomogeneity in surfactant  distribution. 

In this case the surfactant distribution is can be expressed in the form
\begin{equation}
\label{eq:O(ReE)-surf-distr}
\Gamma^{(CaRe_E)}=\sum_{n=1}^{5}\sum_{m=0}^{n} \Gamma_{n,m}^{(CaRe_E)}\cos(m \phi)P_{n,m}(\eta).
\end{equation}
The detailed expression of the $ O(Re_E) $ charge distribution $ (q^{(Re_E)}_{\,\!_S}) $ and the surfactant harmonics $ (\Gamma_{n,m}^{(CaRe_E)}) $  can be found in the supplementary material. 
  The non-zero spherical harmonics, in the $ O(Re_E) $ pressure and velocity field descriptions, exists upto $ n=5 $. The expressions of the constants, which make up these spherical harmonics, convey both the charge convection and surfactant distribution effects.
  
Contrary to the leading order case, in this order, the solution of the force free condition
gives both the longitudinal and transverse components of drop velocity. The transverse component is given as 
\begin{equation}\label{eq:Udx-ReE-low}
U_{d,x}^{(Re_E)}=U_{d,x}^{(Re_E)}\big{|}_\text{Clean}\times \mathcal{F}^{(Re_E)}_x,
\end{equation}
where	\begin{equation}
\label{eq:Udx-clean-ReE}
U_{d,x}^{(Re_E)}\big{|}_\text{Clean} =-\small {\frac {3 \left( R-S \right) M E_{{x}} E_{{z}} \left( R+1-S/3 \right) c_{{
				2}}}{ 14 \left( \lambda+4 \right)  \left( R+2 \right) ^{2} \left( 
		2+3\,\lambda \right) ^{2} \left( R+3/2 \right)  \left( 1+1\,\lambda
		\right) } \left( -54-{\frac {12\,{\lambda}^{2}}{5}}-{\frac {378\,
			\lambda}{5}} \right) }
\end{equation}
\begin{equation}
\label{eq:Udz-SC-ReE}
\begin{aligned}
& \text{and \quad} \mathcal{F}^{(Re_E)}_x = 
\frac {105\, \left( 2/3+\lambda \right) ^{2} \left( -1+\beta
	\right)  \left( 1+\lambda \right) }{4\, \left(  \left( {\it k}-3\,
	\lambda-2 \right) \beta+3\,\lambda+2 \right) ^{2} \left(  \left( {
		\it k}-7\,\lambda-7 \right) \beta+7\,\lambda+7 \right)} \times \\
& \qquad \qquad \qquad \qquad \qquad \left( 
\left( -{\frac {12\,{\lambda}^{2}}{5}}+ \left( -3\,{\it k}-{\frac {
		378}{5}} \right) \lambda+{{\it k}}^{2}+19\,{\it k}-54 \right) {\beta}^{2} \right.+ 
\\ 
&\left.
\left( {\frac {24\,{\lambda}^{2}}{5}}+ \left( 3\,{\it k}+{
	\frac {756}{5}} \right) \lambda-19\,{\it k}+108 \right) \beta-{
	\frac {12\,{\lambda}^{2}}{5}}-{\frac {378\,\lambda}{5}}-54 \right) 
\left( {\frac {45}{2}}+{\lambda}^{2}+{\frac {63\,\lambda}{2}}
\right) ^{-1}. 
\end{aligned}
\end{equation}
It is evident from the above expressions that $ U_{d,x}^{(Re_E)}\big{|}_\text{Clean} $ is a function of $ M, E_x, E_z, R, S, \lambda \text{ and } c_2 $ while the surfactant correction term $ \mathcal{F}^{(Re_E)}_x $ contains only $ \beta , k  $ and $ \lambda $. 
Similarly the axial migration velocity can be represented as 
\begin{equation}\label{eq:Udz-ReE-low}
U_{d,z}^{(Re_E)}=U_{d,z}^{(Re_E)}\big{|}_\text{Clean}\times \mathcal{F}^{(Re_E)}_z.
\end{equation}
In contrast to the cross-stream velocity, for the longitudinal component $ (U_{d,z}^{(Re_E)}) $, the surfactant correction term $ \mathcal{F}^{(Re_E)}_z $ is a function of the externally applied electric field components $ E_x \text{ and } E_z $ (in addition to $ \lambda, \beta  $ and $ k $). This again shows a complex interplay between electrical and surfactant parameters. The expression of $ U^{(Re_E)}_{d,z} $ being too lengthy to present here, has been provided in the supplementary information. It is to be noted that unlike the leading order, the Marangoni effects comes through a multiplicative factors in the  $O(Re_E)$ drop velocity components. This suggests that the Maxwell and Marangoni stresses at the interface interact in a non-trivial manner beyond just an additive superposition.

Next we employ the normal stress balance equation, solving which gives us the $ O(CaRe_E) $ shape function can in the form
\begin{equation}\label{eq:ReE-shape}
f^{(CaRe_E)}=\sum_{n=2}^{5}\sum_{m=0}^{n} L^{(CaRe_E)}_{n,m}\cos(m \phi)P_{n,m}(\eta).
\end{equation} Again all the harmonics are provided in the supplementary material. 
\subsubsubsection{$ O(Ca): $ Effect of drop-surface deformability}
\label{ssssec:order-Ca}
\\
In this step all the boundary conditions are applied on the deformed surface ($ r=1+Ca f^{(Ca)}$). Hence the information about spherical harmonics $ (L_{n,m}^{(Ca}) $ consisting the $ O(Ca) $ shape functions is carried along in all the important quantities in $ O(Ca) $.
In this order the electric potential inside the drop $ \varphi^{(Ca)} _i$ consists of growing spherical harmonics upto $ n=2 $ while for the outside medium the potential $  \varphi^{(Ca)}_e $ contains decaying harmonics upto $ n=4 $.

Using the force-free condition (\eqref{eq:force-free}) leaves us with the drop velocity components in both cross-stream and longitudinal directions $ U_{d,x}^{(Ca)} $ and $ U_{d,z}^{(Ca)} $. Although the excessive lengths of the expressions $ U_{d,x}^{(Ca)} $ and $ U_{d,z}^{(Ca)} $ prohibit us from presenting those here, the expressions again suggest (please refer to the supplementary material for detail expressions) that the electrical and surfactant effects cannot be decoupled in evident forms of linear superposition similar to the leading order case. 

It is to be noted that the drop velocity components in any order of perturbation can be obtained by determining only the decaying spherical harmonic $ p_{-2}^{(w)} $ which further necessitates evaluating the expressions of $ A_{-2,0}^{(w)}, A_{-2,1}^{(w)} $ and $ \hat A_{-2,1}^{(w)}$ (since $p_{-2}^{(w)}= A_{-2,0}^{(w)}P_{1,0}+(A_{-2,1}^{(w)} \cos(\phi)+\hat A_{-2,1}^{(w)} \sin(\phi)) P_{1,1}$ \citep{Lamb1975,Happel1983,Hetsroni1970}). Due to highly cumbersome algebraic expressions in $ O(Ca) $, it is quite difficult to obtain all other constant terms which are required for a complete description of the flow field and the subsequent higher order shape deformation $ (O(Ca^2)) $ through normal stress balance. {However, since the primary focus of the present study is to calculate the drop velocity and the subsequent drop motion, it is sufficient to calculate only the constants $ A_{-2,0}^{(Ca)}, A_{-2,1}^{(Ca)} $ and $ \hat A_{-2,1}^{(Ca)}$. A similar  strategy has been adopted earlier by \citet{Xu2006a} for unidirectional sedimentation of a clean drop under electric field effects.}  

\subsubsubsection{Depiction of drop trajectory}
\label{ssssec:trajectory-calc}\\
Since the above calculation shows the drop having velocity components in both parallel and perpendicular to the applied flow direction, it will be interesting to visualize a complete picture of the temporal variation of the drop motion. Towards this we assume a quasi-steady-state migration of the drop. Similar treatment has been adopted by many investigators previously \citep{Hanna2010,Pak2014,Schwalbe2011,Das2017a,Mandal2016}.   To investigate the relevance of such consideration, we look into the various time scales governing the flow physics. As the drop traverses in a continuously varying flow field, the flow behavior in both interior and exterior to the drop, distribution of accumulated surface charges and surfactant concentration as well as the deformation of the drop shape, are expected to face changes. Now under the realm of `quasi-steady state' we assume that these temporal processes are occuring at much smaller time scales as compared to the drop migration time scales in either of the directions $ (t_{d,z}, t_{d,x}) $. It can be shown (please refer to Appendix \ref{sec:quasi}) that the present considerations of a drop motion in the Stokes flow regime $ (Re \ll 1)  $, small surface charge convection $ (Re_E \ll 1) $, small shape deformation $ (Ca \ll 1) $ and a diffusion dominated transport of surfactant molecules $ (Pe \ll 1) $, make the `quasi-steady state' assumption applicable in the present context.       
Under such a  condition the resultant drop velocity takes the form:
\begin{equation}\label{eq:total-drop-vel}
\mathbf{U}_d=(Re_E U_{d,x}^{(Re_E)}+Ca U_{d,x}^{(Ca)})\mathbf{i}_x+(U^{(0)}_{d,z}+Re_E U_{d,z}^{(Re_E)}+Ca U_{d,z}^{(Ca)})\mathbf{i}_z
\end{equation}
Now the various orders of drop velocity contain the external flow terms $ c_0, c_1  $ and $ c_2 $, which in turn are functions of drop transverse position  at that instant $ x_d(t) $. Thus we can obtain a temporal depiction of the position of the drop $ (x_d(t), z_d(t)) $ by solving the following two ordinary differential equations:
\begin{subequations}
	\label{eq:trajectory-diff-eq}
	\begin{equation}\label{eq:xd-dt-eq}
	\frac{dx_d(t)}{dt}=Re_E U_{d,x}^{(Re_E)}+Ca U_{d,x}^{(Ca)}
	\end{equation}
	\begin{equation}\label{eq:zd-dt-eq}
	\frac{dz_d(t)}{dt}=U^{(0)}_{d,z}+Re_E U_{d,z}^{(Re_E)}+Ca U_{d,z}^{(Ca)}.
	\end{equation}
\end{subequations}
The above two equations suggest that a functional form of the cross-stream position can be found by integrating \eqref{eq:xd-dt-eq} alone and is given as
\begin{equation}\label{eq:xd_t_form}
x_d(t)=x_{d,\infty}+\{x_{d,0}-x_{d,\infty}\}\exp(-t/t_S),
\end{equation} 
where $ t_S $ is a characteristic time constant to reach the drop its steady state position $ x_{d,\infty} $ from the initial position $ x_{d,0} $. However a complete description of drop position can be obtained only by a simultaneous determination of $ x_d $ and the longitudinal distance traveled by the drop  $ (z_d)$.  It has been previously found \citep{Mandal2016} that for a clean drop the steady state time scale $ t_S $ is only a function of $ H, \lambda $ and $ Ca $. However in the present case, the surfactant parameters $ \beta  $ and $ k $ are also responsible for variations in $ t_S $. The other important quantity $ x_{d,\infty} $ is decided by a  functional dependence on $ H, Ca $; electrical variables $ M, E_x, E_z, R, S, Re_E $ as well as surfactant effects (quantified by $ \beta , k, \lambda $). The detail expressions of $ t_S$ and $x_{d,\infty} $ are provided in the supplementary information.

Before proceeding with a detail discussion of the results obtained we compare them against various previously published works in different limiting conditions (please refer to Appendix-\ref{sec:Validation}) and thus obtain a reasonable confidence on the accuracy of the present analytical calculations.
\section{Results and Discussions}

In this section we consider a leaky dielectric drop suspending through an another leaky dielectric medium and explore the consequences of the surfactant coating on the drop surface.  The imposed Poiseuille flow affects the surfactant distribution and generates Marangoni stress. This Marangoni stress not only affects the flow field but the resulting drop shape is also affected. As a consequence of these two effects  the pattern of accumulation of interfacial charge is also altered. Again the electrical Maxwell stress triggers a n electrohydrodynamic flow, which modifies the surfactant distribution. Thus all the three important aspects (Poiseuille flow, surfactant distribution and electric charge distribution) interacts with each other. In the following sub-sections we describe how such a coupled phenomena can modulate the drop motion.
We first consider a spherical drop without the deformations taken into account. In this case only the charge convection effect is highlighted. In the next step the deformed surface is considered separately. Finally the impact of the non-uniformity in surfactant distribution on the drop motion, with a simultaneous consideration of charge convection and shape deformation, is discussed.
\subsection{Effects on a spherical drop}
\label{ss:spherical drop}
\subsubsection{Mechanisms of surfactant effects}
\label{sssec:beta-k-physiscs} 
\begin{figure}
	\centering
	\begin{subfigure}[!htb]{0.45\textwidth}
		\centering
		\includegraphics[width=1.15\textwidth]{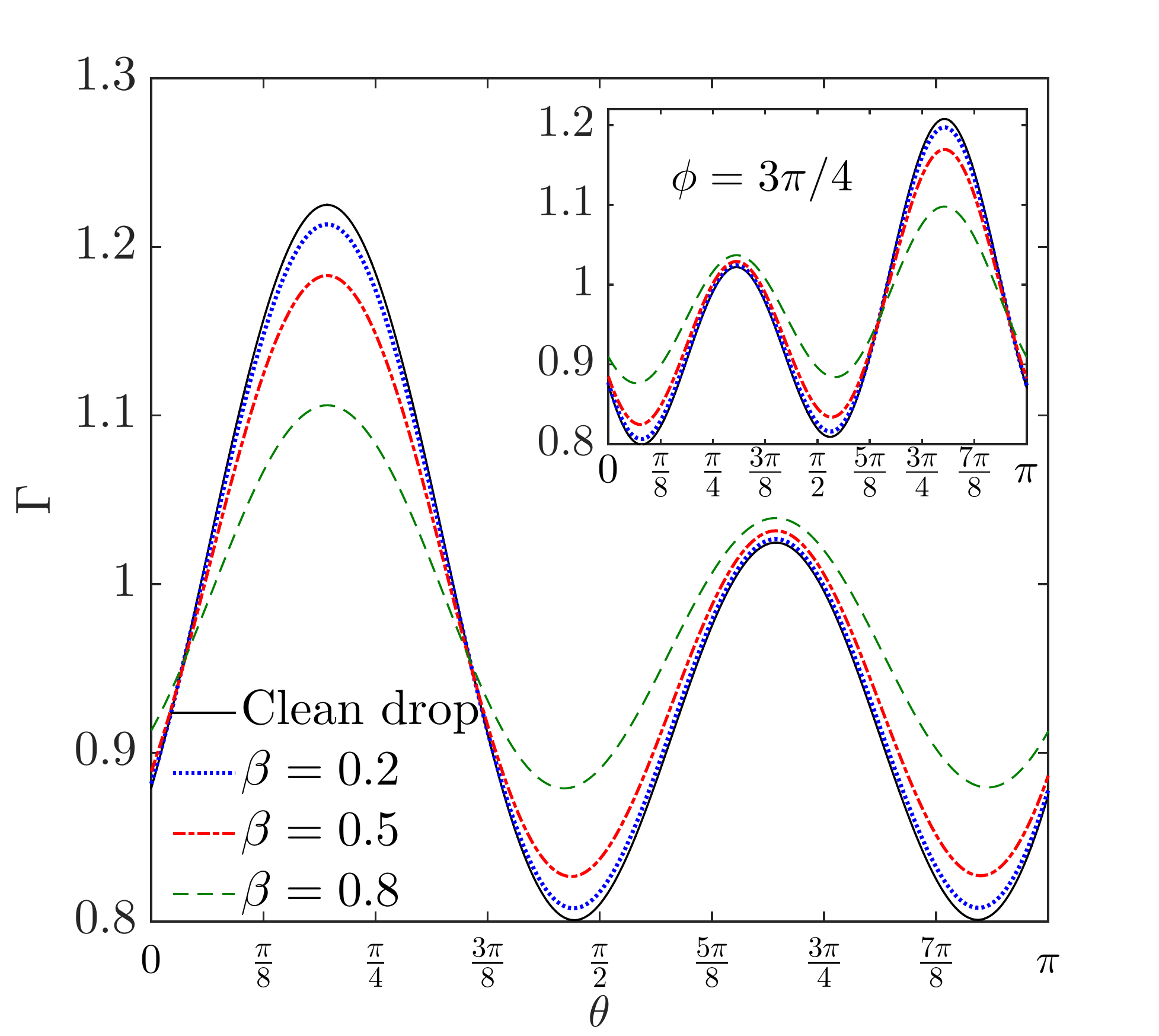}		
		\vspace{8ex}
		\caption{}
		\label{fig:SC_vs_theta_vary_beta}
	\end{subfigure}
	\quad 
	\begin{subfigure}[!htb]{0.45\textwidth}
		\centering
		\includegraphics[width=1.15\textwidth]{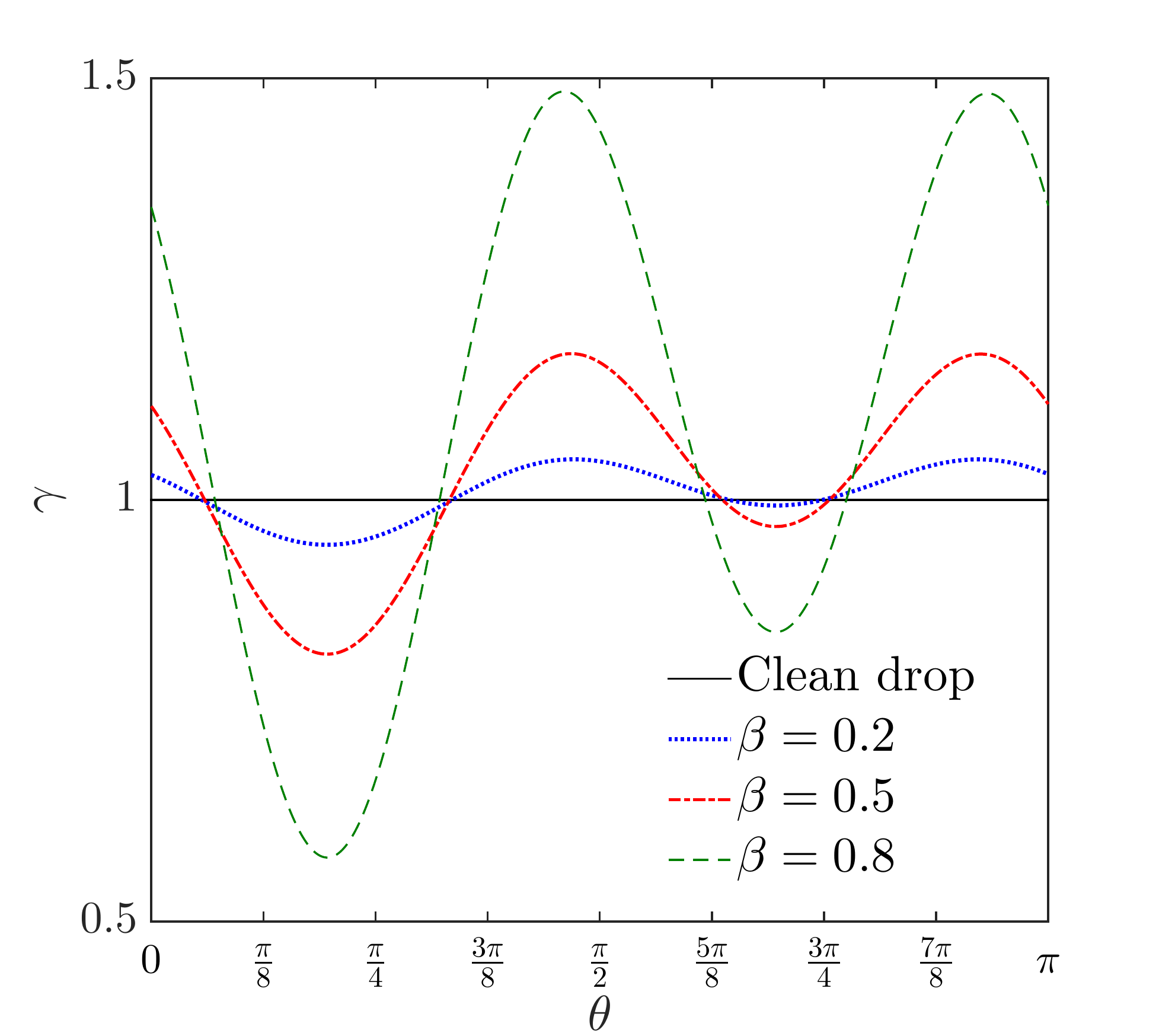}
		\vspace{8ex}
		\caption{}
		\label{fig:ST_vs_theta_vary_beta}
	\end{subfigure}
	\begin{subfigure}[!htb]{0.45\textwidth}
		\centering
		\includegraphics[width=1.15\textwidth]{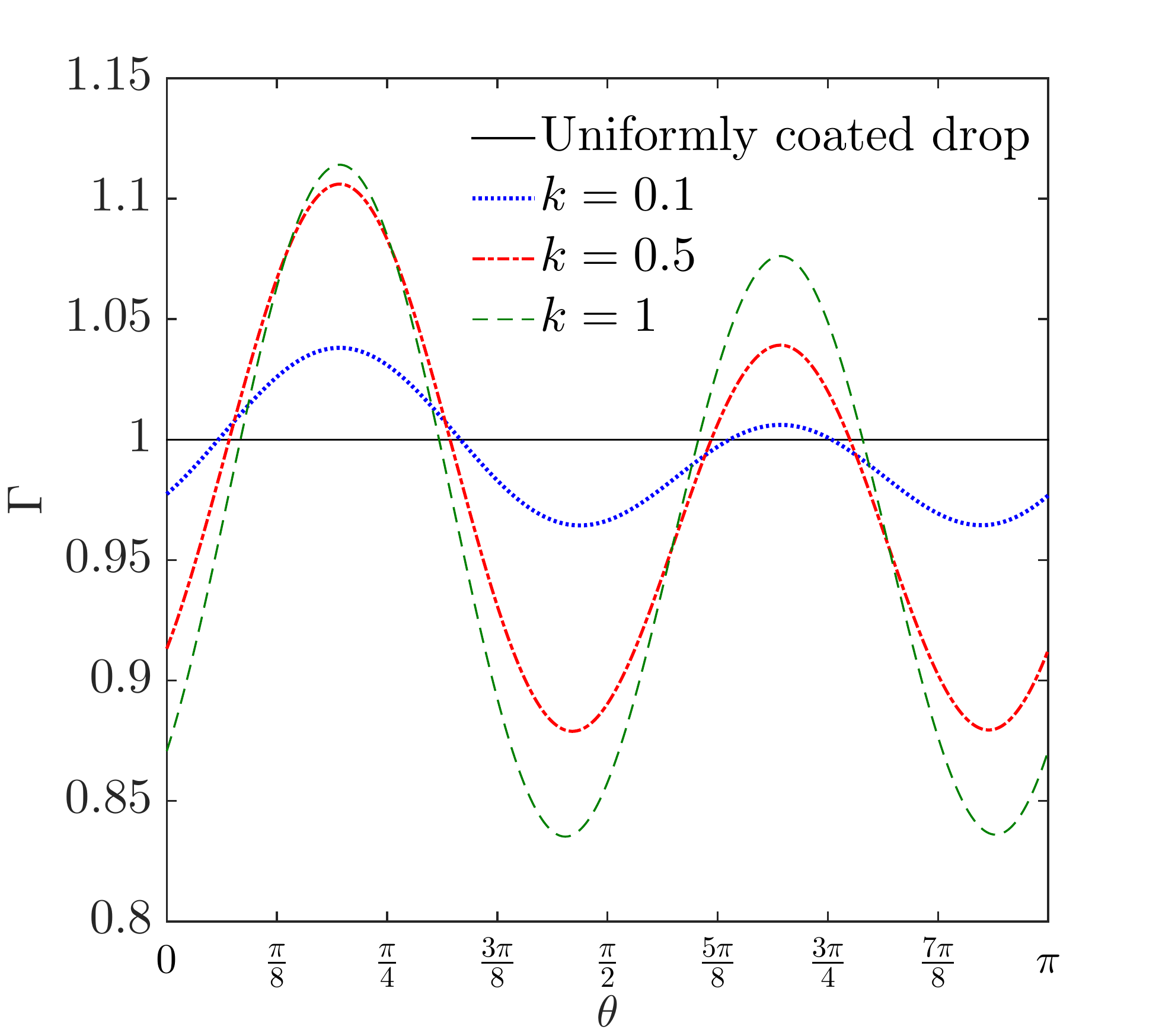}		
		\vspace{8ex}
		\caption{}
		\label{fig:SC_vs_theta_vary_k}
	\end{subfigure}
	\quad 
	\begin{subfigure}[!htb]{0.45\textwidth}
		\centering
		\vspace{1.8ex}
		\includegraphics[width=1.15\textwidth]{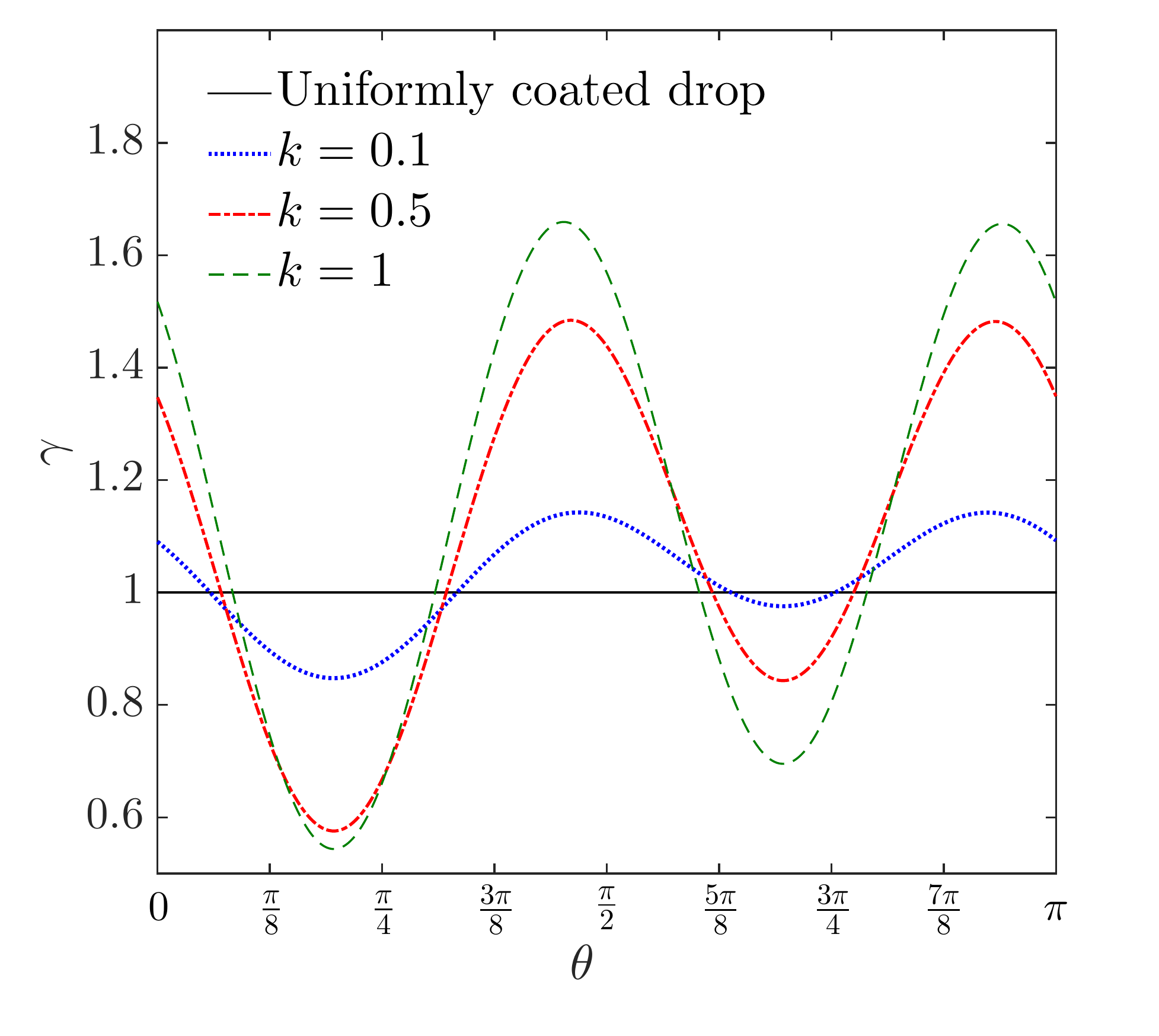}
		\vspace{6ex}
		\caption{}
		\label{fig:ST_vs_theta_vary_k}
	\end{subfigure}
	\caption{Polar distribution of surfactant concentration $ \Gamma $ defined as $\left(\Gamma=\Gamma^{(0)}+Ca\,\Gamma^{(Ca)}+Ca\,Re_E\,\Gamma^{(CaRe_E)} \right)$ and surface tension ($ \gamma $) for various values of elasticity parameter ($ \beta $) and the physicochemical constant $ (k) $. For all the subplots $ \phi=\pi/4 $, $ (\theta_e=\pi/4) $, $ Re_E=0.2, M=1, Ca=0.2 ,R=0.1, S=10, x_d=4$ and $H=10$. For subplots (a) and (b), $ k=0.5 $ while for subplots (c) and (d), $ \beta=0.8 $ is taken.} 
	\label{fig:SC_ST_var_beta_k}
\end{figure}

To get an insight of the physical mechanisms responsible for various effects caused by the surfactant parameters, namely the eleasticity constant $ (\beta) $ and the physicichemical parameter $ (k), $ it is necessary to understand how the surfactant concentration $ (\Gamma)$ and interfacial tension $(\gamma)$ behave with those parameters. The parameter $ \beta $ has a direct consequence of affecting the tangential  Marangoni stress \eqref{eq:t_stress_bal} by altering the gradients in surface tension. On the other hand the parameter $ k $  symbolizes an enhancement in the convection of surfactant molecules. Comparing the  figures \ref{fig:ST_vs_theta_vary_beta} and \ref{fig:SC_vs_theta_vary_beta}, it can be observed that an increase in $ \beta $ causes a sharp increase in surface tension gradients $ (|\gamma_\text{max}-\gamma_\text{min}|) $ while the surfactant concentration shows a trend of reaching towards the equilibrium condition, $  \Gamma=1 $. Similar phenomena was also previously observed by \citet{Li1997} for the case of a surfactant-laden drop in linear shear flow. A deeper reflection reveals that, the stronger the surface tension gradient $ (|\gamma_\text{max}-\gamma_\text{min}|) $ becomes the more it opposes the convective transport of surfactant molecules. As a result the gradient of surfactant concentration $ (|\Gamma_\text{max}-\Gamma_\text{min}|) $ is reduced and in effect $ \Gamma $ gets redistributed to a form towards equilibrium.  

The limiting condition of $ k=0 $ or $ Pe=0 $, indicates that surface diffusion of surfactant molecules is so high that the convection effects become negligible and gradients in surface tension becomes diminishingly small. This corresponds to a physical situation when the drop surface is uniformly coated with surfactants \citep{Stone1990a}. An increase in the parameter $ k $, on the other hand, stands for an increase in the surface P\'eclet number $ (Pe) $ for a constant $ Ca $ (refer to \eqref{eq:k-def}). Thus it leads to an increased convective transport of surfactant molecules in a manner that the opposing effect initiated by an increase in surface tension gradient is surpassed. A coherent observation can be made by studying the figures \ref{fig:SC_vs_theta_vary_k} and \ref{fig:ST_vs_theta_vary_k}. 

\begin{figure}[!htb]	
	\centering
	\includegraphics[width=1\textwidth]{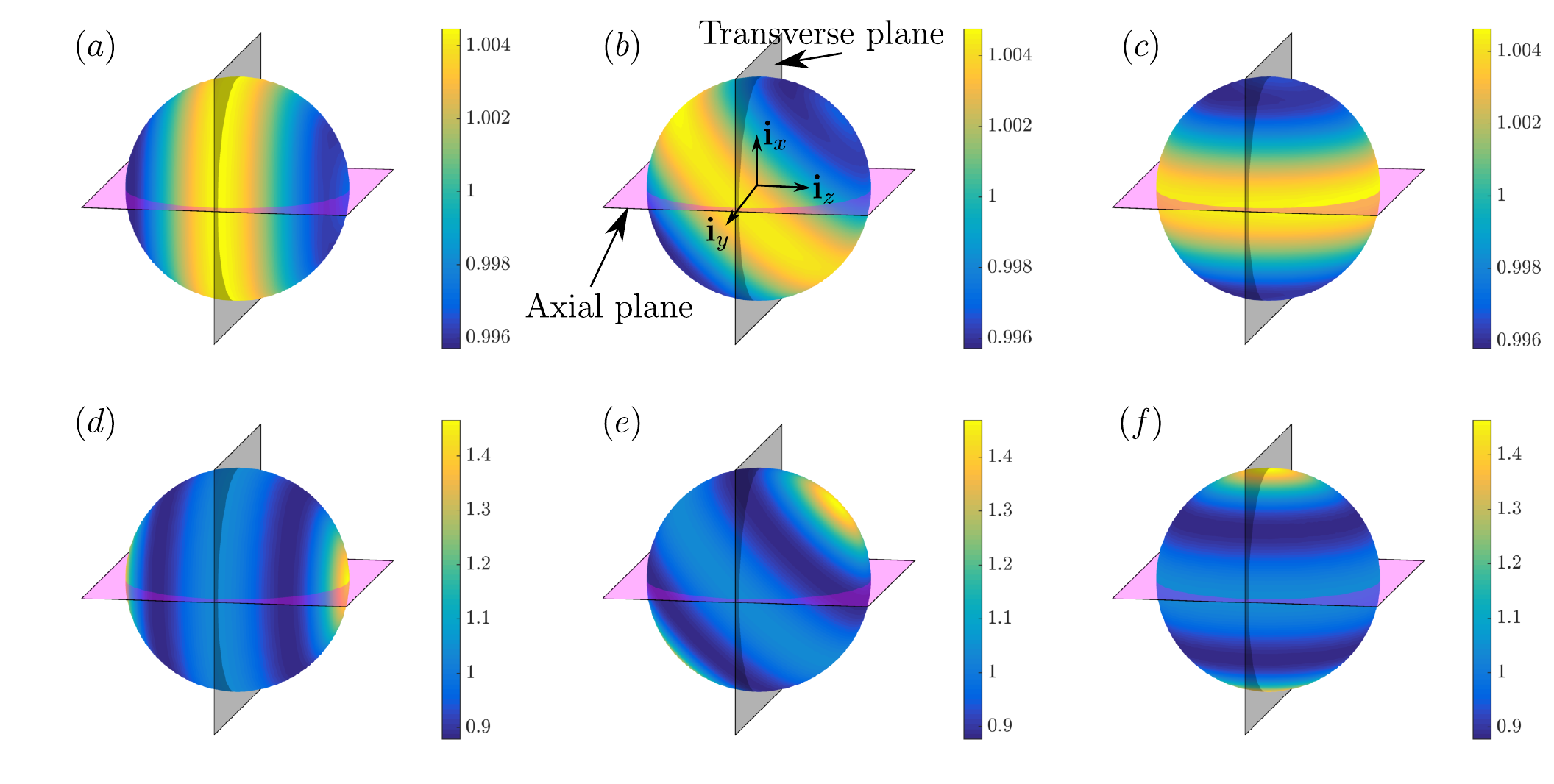}
	\vspace{-35pt} 
	\caption{Surfactant concentration  $\left(\Gamma=\Gamma^{(0)}+Ca\,\Gamma^{(Ca)}+Ca\,Re_E\,\Gamma^{(CaRe_E)} \right)  $ variation on the drop surface for different and tilt angle of the applied electric field $ (\theta_e) $ and Mason number $ M $. Subplots $(a),(b)$ and $ (c) $ are with $ M=0.1 $ while for $(d),(e)$ and $ (f) $ we considered $ M=1 $. Similarly for subplots $(a),(d)$ electric field is axial $ (\theta_e=0) $; for subplots $(b),(e)$ electric field makes an angle $ \theta_e=\pi/4 $ and is for $ Re_E=0 $, $(b)$ is for $ Re_E=0.05 $ and $(c)$ is for $ Re_E=0.1 $. Other parameters are $ Ca=0.2, M=1, R=0.1, S=10, \beta =0.8, k=0.5,  x_d=5, H=10 $ and $ \theta_e=\pi/4$.}
	\label{fig:sph_surf_conc_M_tilt}
\end{figure}

\begin{figure}[!htb]	
	\centering
	\includegraphics[width=1\textwidth]{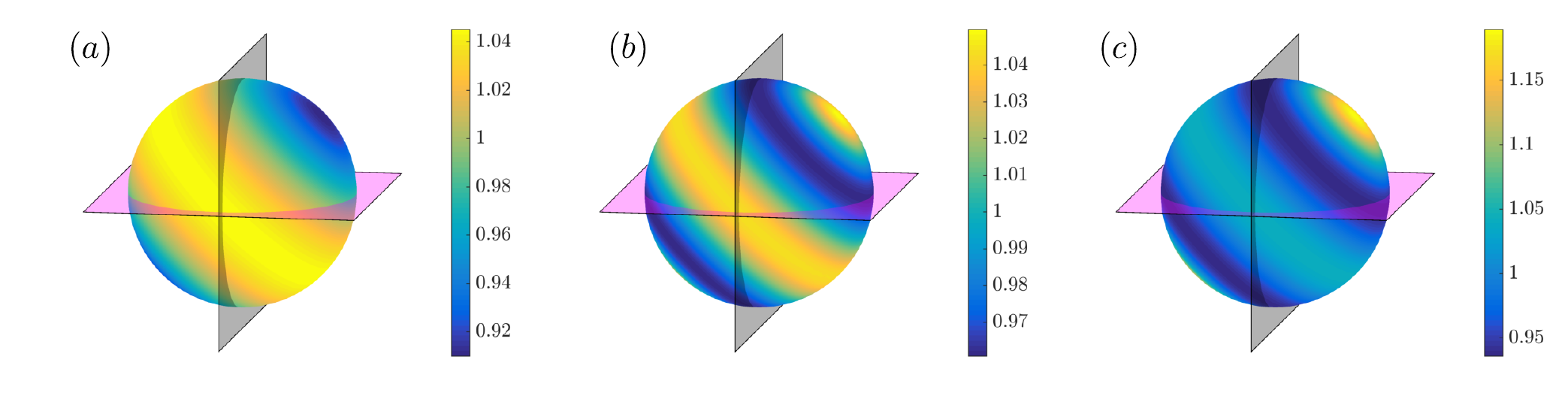}
	\vspace{-35pt} 
	\caption{Surfactant concentration variation on the drop surface for different $ Re_E $. Subplot $(a)$ is for $ Re_E=0 $, $(b)$ is for $ Re_E=0.05 $ and $(c)$ is for $ Re_E=0.1 $. Other parameters are $ Ca=0.2, M=1, R=0.1, S=10, \beta =0.8, k=0.5,  x_d=5, H=10 $ and $ \theta_e=\pi/4$.}
	\label{fig:sph_surf_conc_ReE}
\end{figure}

\begin{figure} 
	\centering
	\begin{subfigure}[!htb]{0.45\textwidth}
		\centering
		\includegraphics[width=1.1\textwidth]{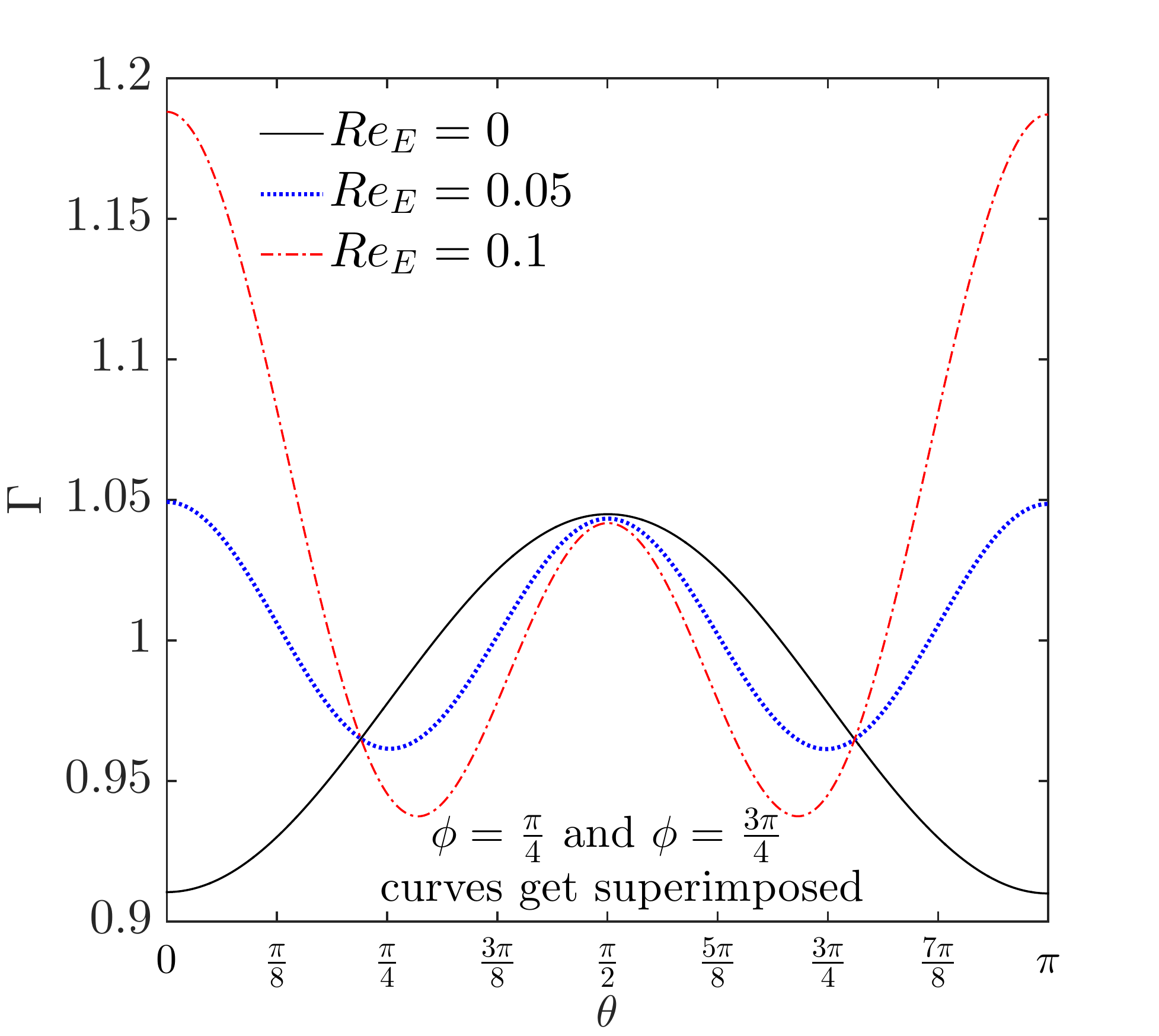}		
		\vspace{8ex}
		\caption{}
		\label{fig:surf_vs_theta_vary_ReE}
	\end{subfigure}
	\quad 
	\begin{subfigure}[!htb]{0.45\textwidth}
		\centering
		\includegraphics[width=1.1\textwidth]{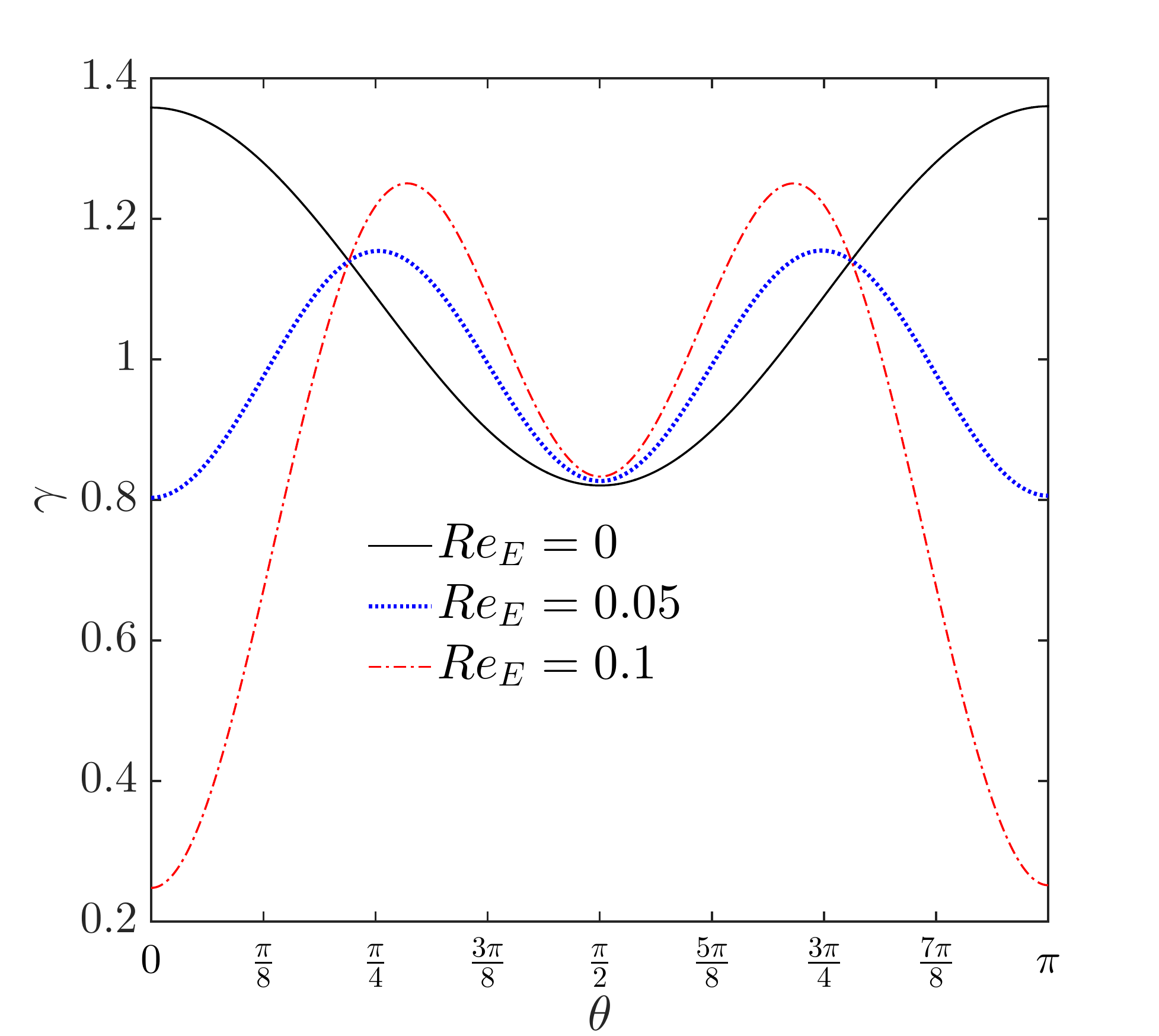}
		\vspace{8ex}
		\caption{}
		\label{fig:ST_vs_theta_vary_ReE}
	\end{subfigure}
	\begin{subfigure}[!htb]{0.49\textwidth}
		\centering
		\includegraphics[width=1.1\textwidth]{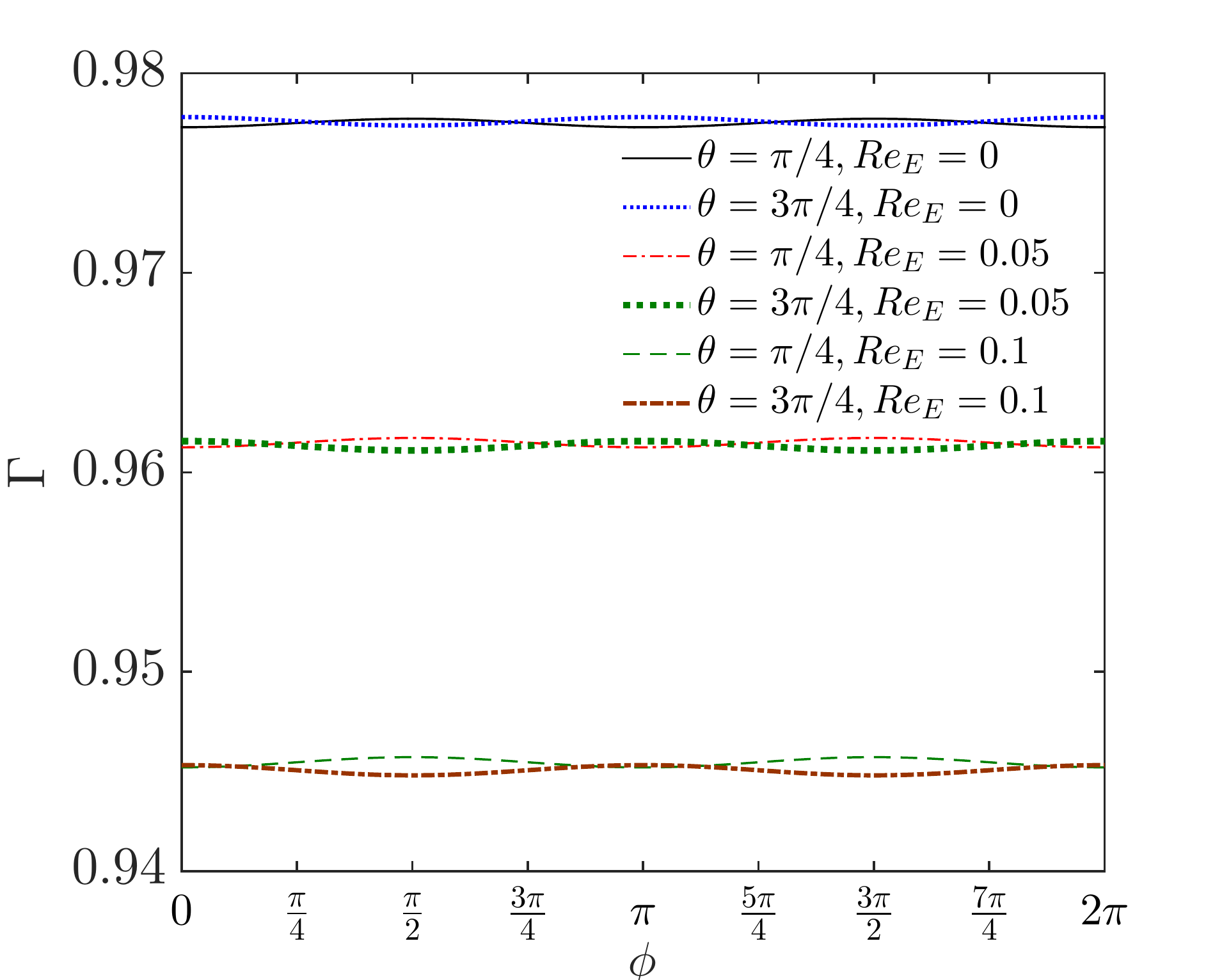}
		\vspace{8ex}
		\caption{}
		\label{fig:surf_vs_PHI_vary_THETA_ReE}
	\end{subfigure}
	\quad
	\begin{subfigure}[!htb]{0.45\textwidth}
		\centering
		\includegraphics[width=1.08\textwidth]{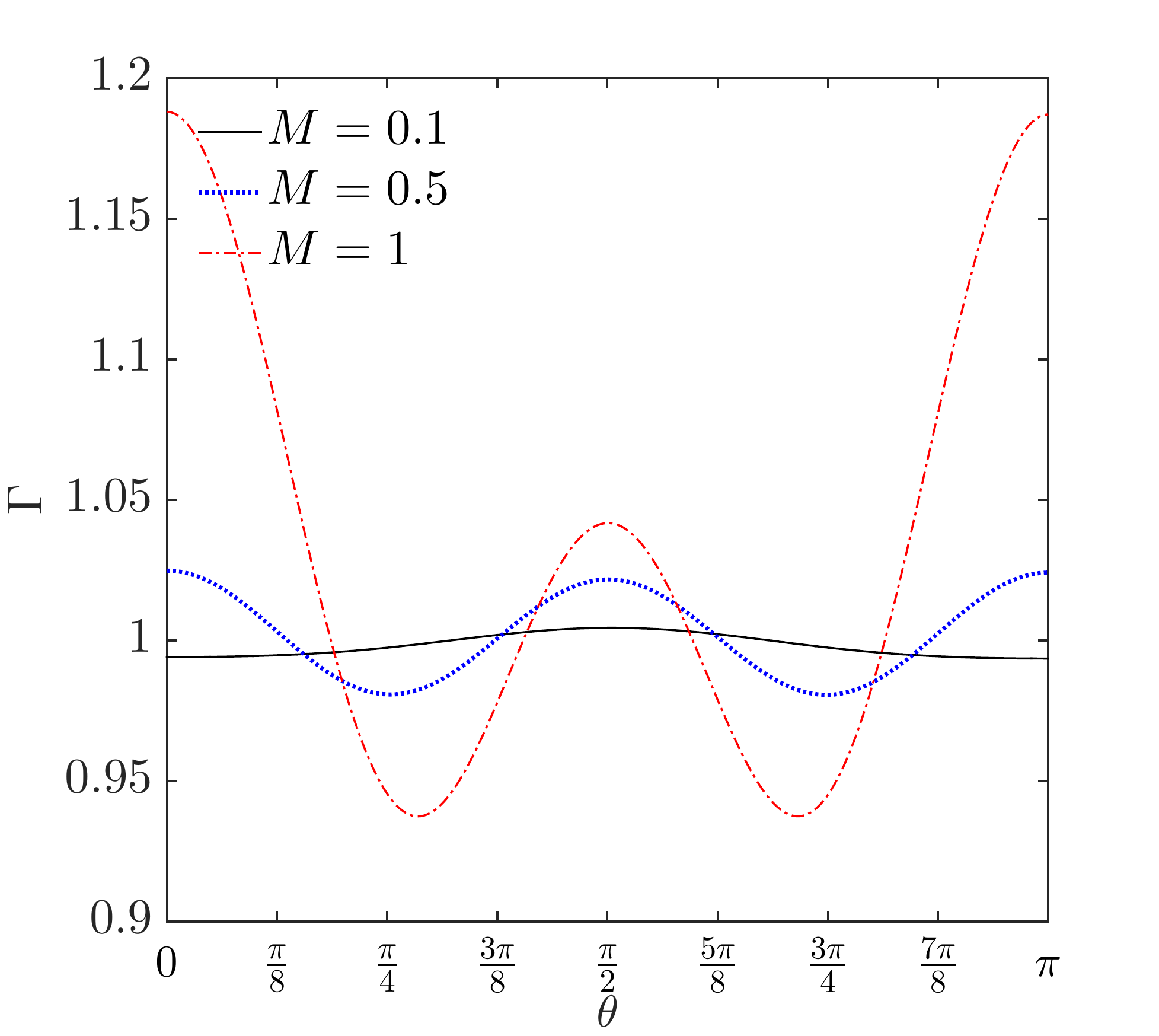}
		\vspace{7ex}
		\caption{}
		\label{fig:surf_vs_theta_vary_Mason}	
	\end{subfigure}
	\caption{Variation of surfactant concentration and surface tension for different values of $  Re_E, M$ and angular positions $ (\theta, \phi) $ for the case when the electric field is axial i.e. $ \theta_e=0 $. Subplot (a): variation in $ \Gamma $ with polar angle $ (\theta) $ for  various values of $ Re_E $ and $ M=1 $; subplot (b): variations in surface tension $ \gamma $ for similar conditions. Subplot (c):$ \Gamma $ vs. azimuthal angle $ \phi $ for different $ \theta $ and $ Re_E $. Subplot (d): $ \Gamma $ vs. $ \theta $ for three different $ M $ at $ \phi =\pi/4 $ and $ Re_E=0.1 $. Other parameters considered are are $ Ca=0.2, R=0.1, S=10, \beta =0.8, k=0.5,  x_d=5, H=10 $ .} 
	\label{fig:surf-line-plot-axial-E}
\end{figure}

\begin{figure}[!htb]
	\centering
	\begin{subfigure}[!htb]{0.45\textwidth}
		\centering
		\includegraphics[width=1.1\textwidth]{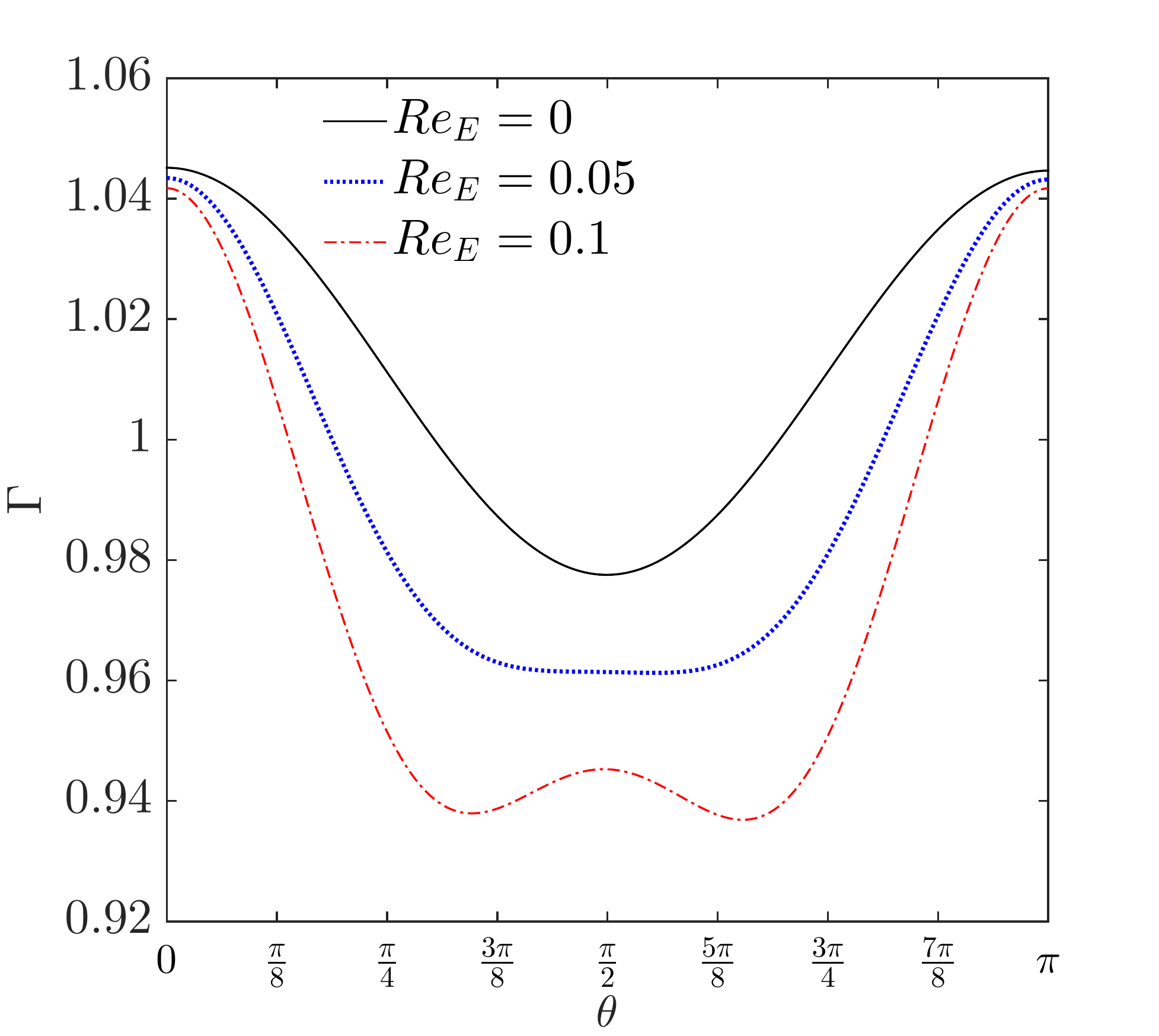}		
		\vspace{8ex}
		\caption{}
		\label{fig:surf_vs_theta_vary_ReE_transverse_E}
	\end{subfigure}
	\quad 
	\begin{subfigure}[!htb]{0.45\textwidth}
		\centering
		\includegraphics[width=1.1\textwidth]{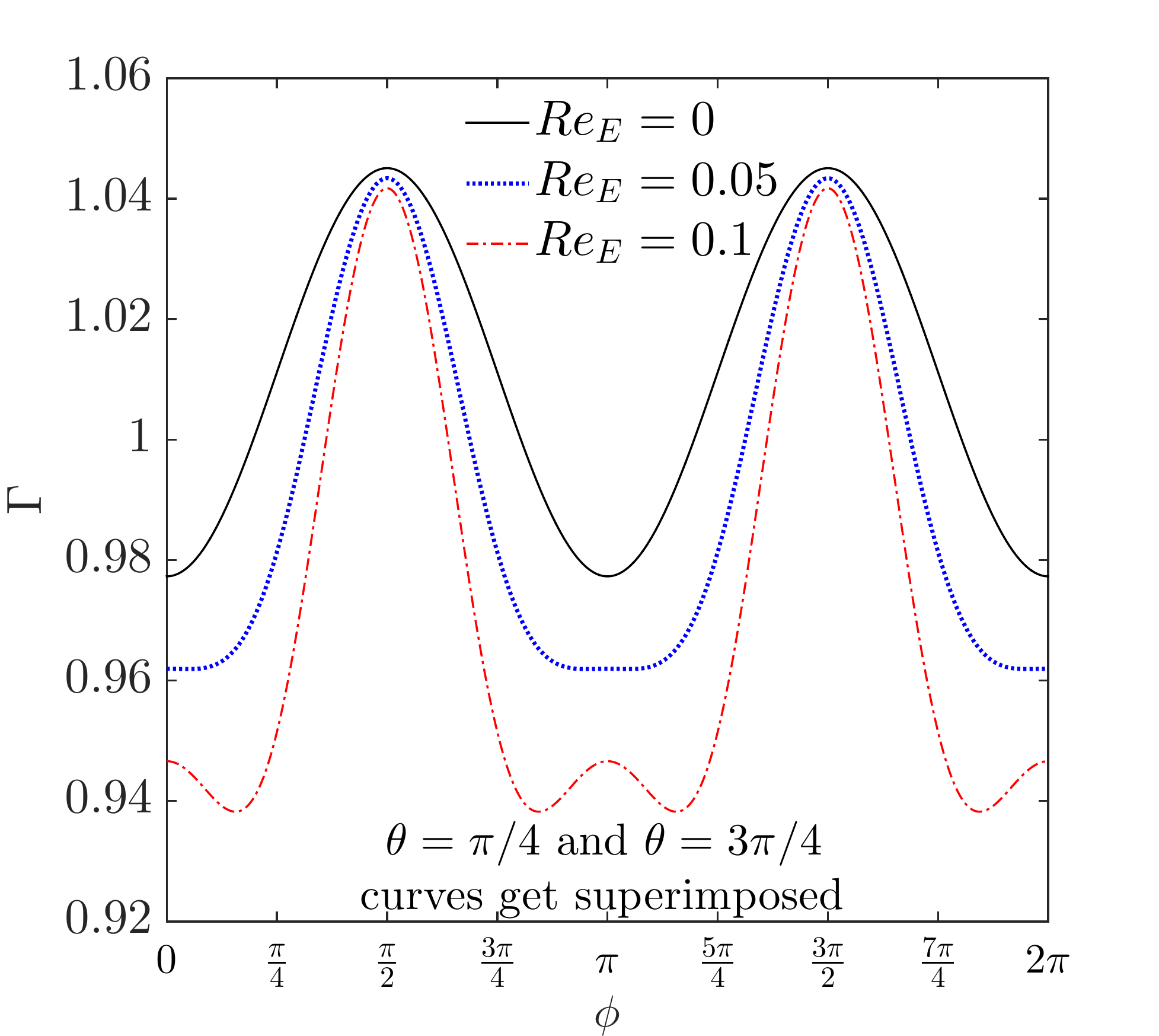}
		\vspace{8ex}
		\caption{}
		\label{fig:surf_vs_phi_vary_ReE_transverse_E}
	\end{subfigure}
	\caption{Surfactant variation for transverse electric field $( \theta_e=\pi/2 )$. Subplot(a): Surfactant concentration variation in the polar direction $ (\theta) $ at $ \phi=\pi/4  $ for different electric Reynolds numbers $( Re_E) $.  Subplot(b): Surfactant concentration variation in the azimuthal direction $ (\phi) $ at $ \theta=\pi/4  $ $ \theta=3\pi/4 $ for different electric Reynolds numbers $( Re_E) $.  Other parameters considered are $ Ca=0.2, R=0.1, S=10, \beta =0.8, k=0.5,  x_d=5 \text{ and } H=10 $. }
	\label{fig:surf-line-plots-transverse-E}
\end{figure}

The electrical parameters have also their part to play in influencing the surfactant concentration distribution $ \Gamma(\theta,\phi) $ on the drop surface. In figure \ref{fig:sph_surf_conc_M_tilt} the effects of both the angle of tilt of the applied electrical field $ (\theta_e) $ and the  relative importance of the electrical stresses compared to hydrodynamic ones (quantified by Mason number $ (M) $), are highlighted. Figures \ref{fig:sph_surf_conc_M_tilt} (a),(b) and (c) reveals that the surfactant distribution follows a virtual symmetry axis similar to the applied electric field direction while the maximum and minimum values of the surfactant concentrations on the drop surface remain unaltered. When the Mason number increases not only the values of $ \Gamma_{\text{max}} $ and $ \Gamma_{\text{min}} $ changes, but also their locations on the surface is altered. In figure \ref{fig:sph_surf_conc_ReE} the effect of intensity of charge convection (quantified by $ Re_E $) is shown. As the charge convection becomes more and more prominent, similar to the effect due to Mason number, the distribution of surfactant molecules is affected in both qualitative and quantitative manners.   For obtaining a better insight, we plotted the surfactant concentration  at various angular locations on the surface in figures \ref{fig:surf-line-plot-axial-E} and \ref{fig:surf-line-plots-transverse-E}. Thus the distribution behaviour of the surfactant molecules are also tremendously sensitive to the electrohydrodynamic conditions. The implications of such effects on the drop migration will be explained in course of the following discussions.  


\subsubsection{Interfacial charge distribution}
\label{ssc:charge-distribution}
\begin{figure}
	\centering
	\begin{subfigure}[!htb]{0.45\textwidth}
		\centering
		\includegraphics[width=1.1\textwidth]{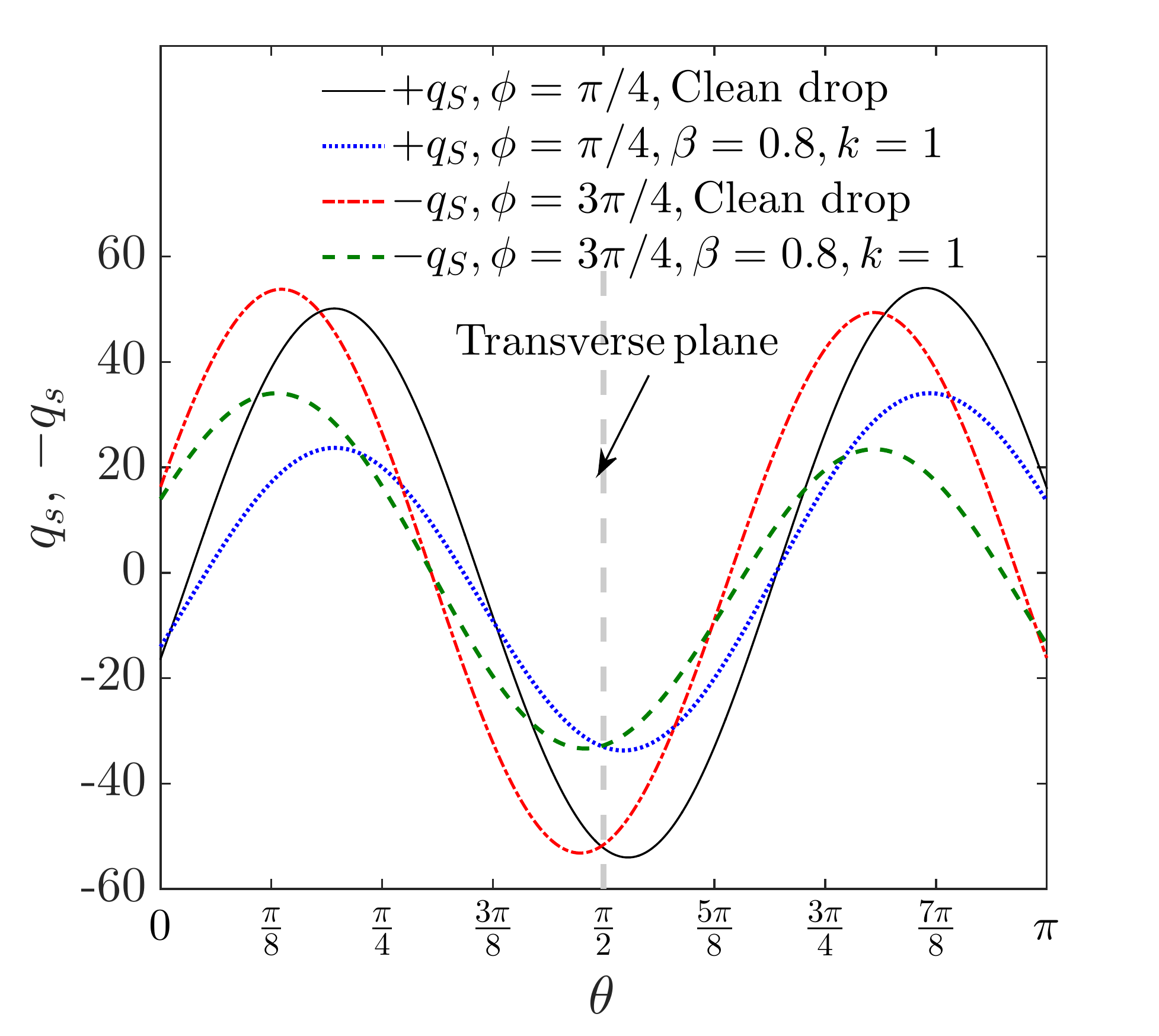}	\vspace{7ex}
		\caption{}
		\label{fig:qs_vs_theta_phi_vary_PHI_beta}
	\end{subfigure}
	\quad 
	\begin{subfigure}[!htb]{0.45\textwidth}
		\centering
		\vspace{-1.5ex}
		\includegraphics[width=1.1\textwidth]{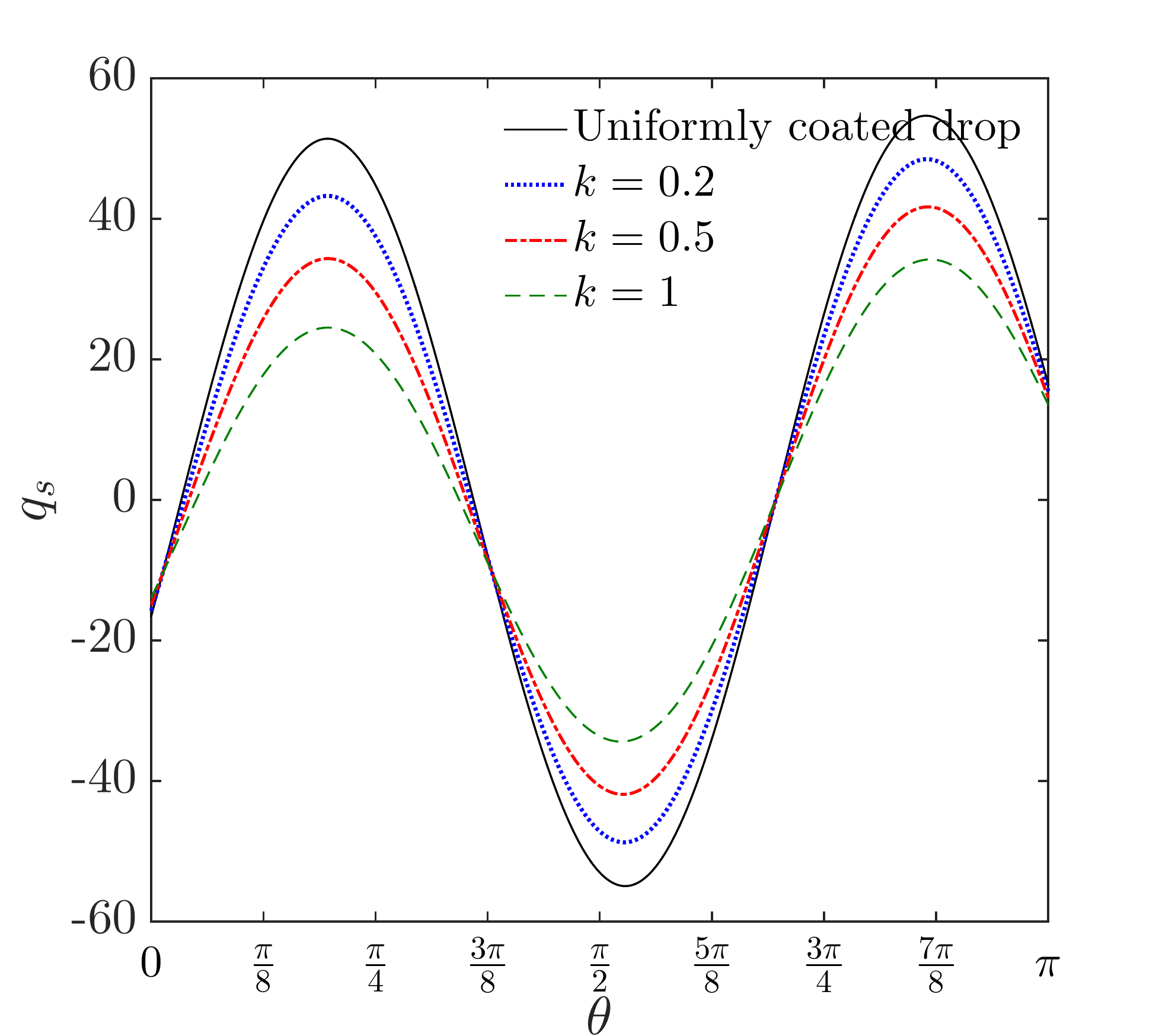}
		\vspace{8ex}
		\caption{}
		\label{fig:qs_vs_theta_phi_pi_by_four_vary_k}
	\end{subfigure}
	\begin{subfigure}[!htb]{0.45\textwidth}
		\centering
		\includegraphics[width=1.1\textwidth]{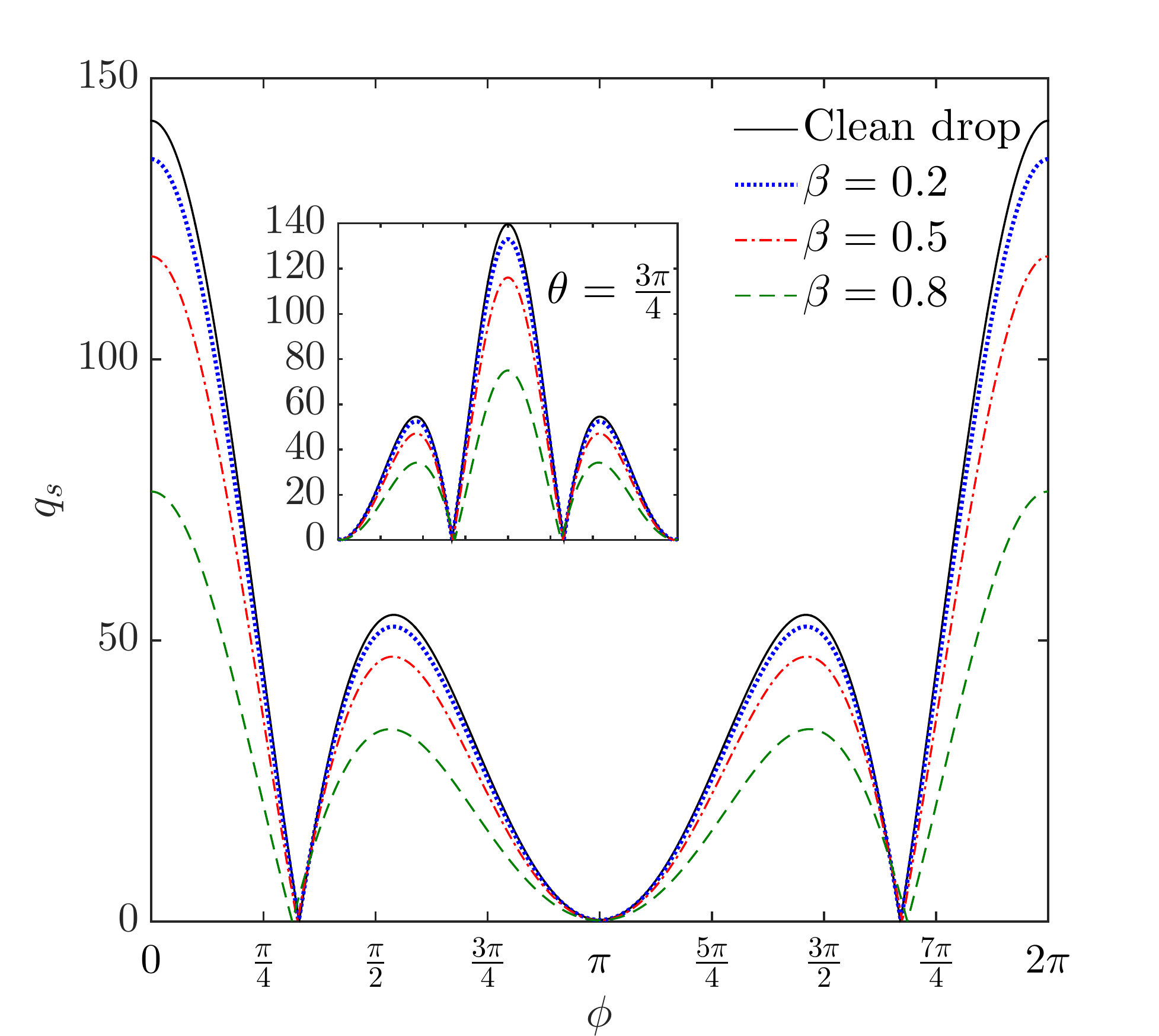}
		\vspace{8ex}
		\caption{}
		\label{fig:qs_vs_phi_theta_pi_by_four_vary_beta}
	\end{subfigure}
	\quad
	\begin{subfigure}[!htb]{0.45\textwidth}
		\centering
		\includegraphics[width=1.1\textwidth]{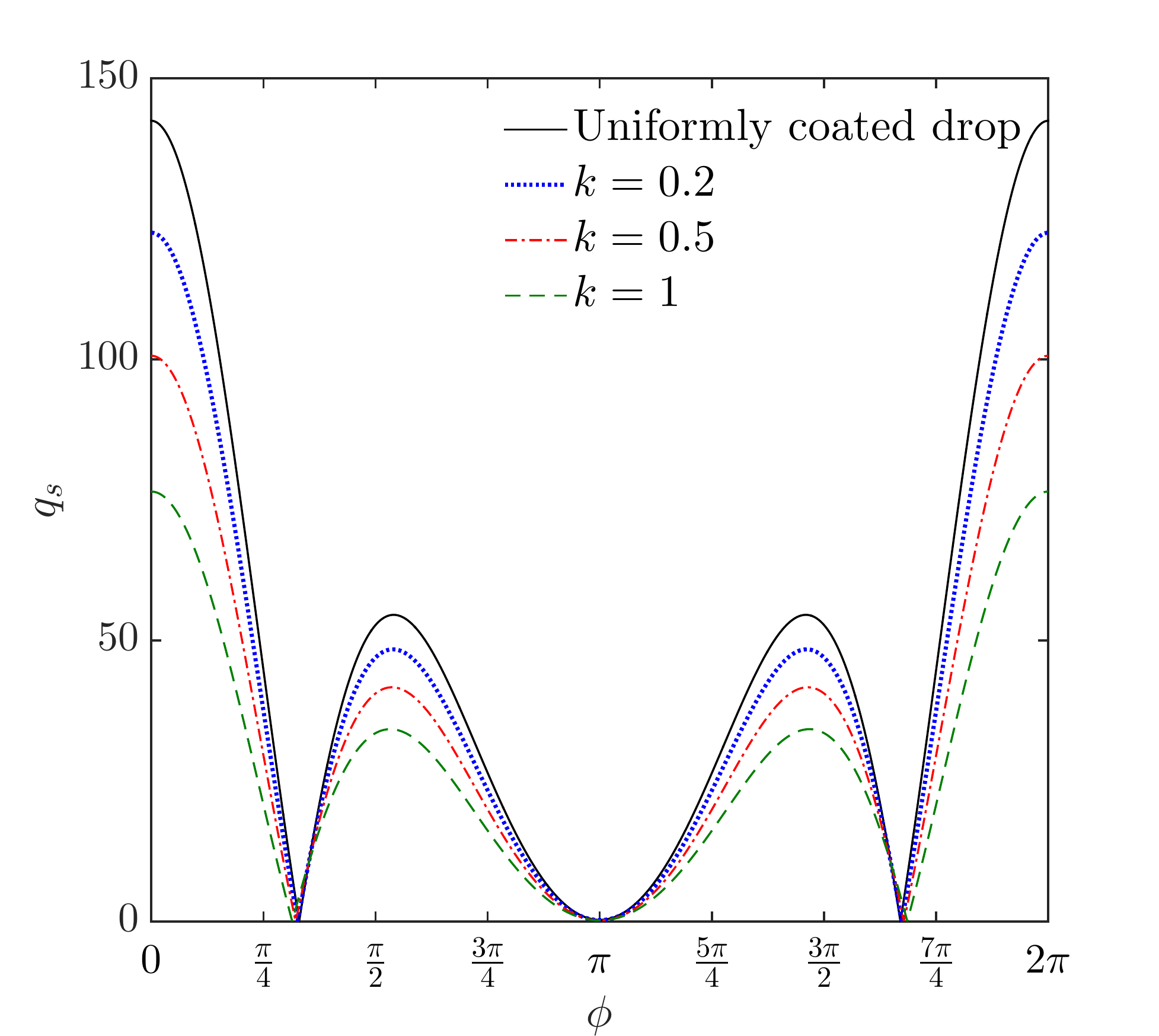}
		\vspace{8ex}
		\caption{}
		\label{fig:qs_vs_phi_theta_pi_by_four_vary_k}	
	\end{subfigure}
	\caption{Polar and azimuthal distribution of accumulated electrical charge $( q_S=q_S^{(0)}+Re_E\, q_S^{(Re_E)}) $ with varying surfactant parameters $ \beta, k $ in the presence of a tilted electric field $ \theta_e=\pi/4 $. The curves for $ \phi=3\pi/4 $ are plotted for $ -q_S $ to compare with the $ \phi=\pi/4 $ cases. The extra negative sign is used to check for any antisymmetry about the axial plane. Subplot (a): $ q_S $ vs. $ \theta $ at $ \phi=\pi/4 $ and $ 3 \pi/4 $ for various values of elasticity parameter $ \beta $ and $ k=1 .$ Subplot (b): $ q_S $ vs. $ \theta $ at $ \phi=\pi/4 $ for various values of physicochemicaL parameter $ k$ and $ \beta=0.8 $.  and $ k=1 .$ Subplot (c): $ q_S $ vs. $ \phi $ at $ \theta=\pi/4 $ and $ 3\pi/4 $ for various values of elasticity parameter $\beta$ and $k=1 $. Subplot (d): $ q_S $ vs. $ \phi $ at $ \theta=\pi/4 $ for various values of physicochemical parameters $ k$ and $ \beta=0.8 $.  The other parameters were taken as  $ Re_E=0.2, M=1, R=0.1, S=10, x_d=4$ and $H=10$.} 
	\label{fig:qs_line_plots}
\end{figure}

Here we show the surfactant induced alterations in the distribution of accumulated charges on the undeformed  drop surface in the presence of a tilted electric field ($ \theta_e=\pi/4 $). In  figure \ref{fig:qs_vs_theta_phi_vary_PHI_beta} we plotted the polar variation of the surface charge at two symmetrically opposite locations about the axial plane $ (\phi=\pi/4 \text{ and } 3\pi/4) $. For the surfactant free case itself the surface charge follows an asymmetric distribution about the transverse plane. When the surface is coated with surfactants, $ q_S $ still displays such symmetry-breaking behaviour for both $ \phi=\pi/4 $ and $ 3\pi/4 $, but its distribution pattern is remarkably different. In regions close to $ \theta=0 $ and $ \pi/2 $ the positive charges get depleted while the negative charges are depleted in the regions close to $ \theta=\pi/2 $. As shown in the preceding section (figure  \ref{fig:ST_vs_theta_vary_beta}) an enhancement in surface tension gradients result due an increase in $ \beta $. This triggers an enhanced Marangoni flow which further affects the convection of surface charges and a redistribution in $ q_S(\theta) $ results. Since the parameter $ k $ was also found to be augmenting the Marangoni flow (refer to figure \ref{fig:ST_vs_theta_vary_k}), a similar nature in the redistribution of $ q_S(\theta) $ is displayed in figure \ref{fig:qs_vs_theta_phi_pi_by_four_vary_k}. In the figures \ref{fig:qs_vs_phi_theta_pi_by_four_vary_beta} and \ref{fig:qs_vs_phi_theta_pi_by_four_vary_k} we simulate the explicit  variation in $ q_S $ in the azimuthal direction. The strong asymmetry in $ q_S $ about the transverse  plane for all values of $ \phi $ can be observed from these figures.  Moreover the redistribution in charges takes place in such a way that the charge density is reduced throughout except some specific points where the accumulation of charges is zero. These locations are indicative of the points where the surface tension is not at all affected by the presence of contaminants. Also in the intermediate points of high charge accumulation, the effects of Marangoni convection is also found to be more prominent.  

\subsubsection{Drop velocity in the cross-stream direction }
\label{ssc:drop-vel-x}
In this section we sought to investigate how the drop migration characteristics is altered by a combined influence of electric field and non-uniform surfactant coating. Even when the charge convection effect is not taken into account, an asymmetry in the surfactant distribution results about the axial plane in the presence of a tilted electric field (shown in figure \ref{fig:sph_surf_conc_ReE}(a)). However this asymmetry is not so severe that it alone cannot cause a cross-stream component of the drop velocity and we found $ U_{d,x}^{(0)}=0 $ (refer to the discussion preceding 	\eqref{eq:Ud-leading-low}).  

\begin{figure}[!htb]	
	\centering
	\includegraphics[width=\textwidth]{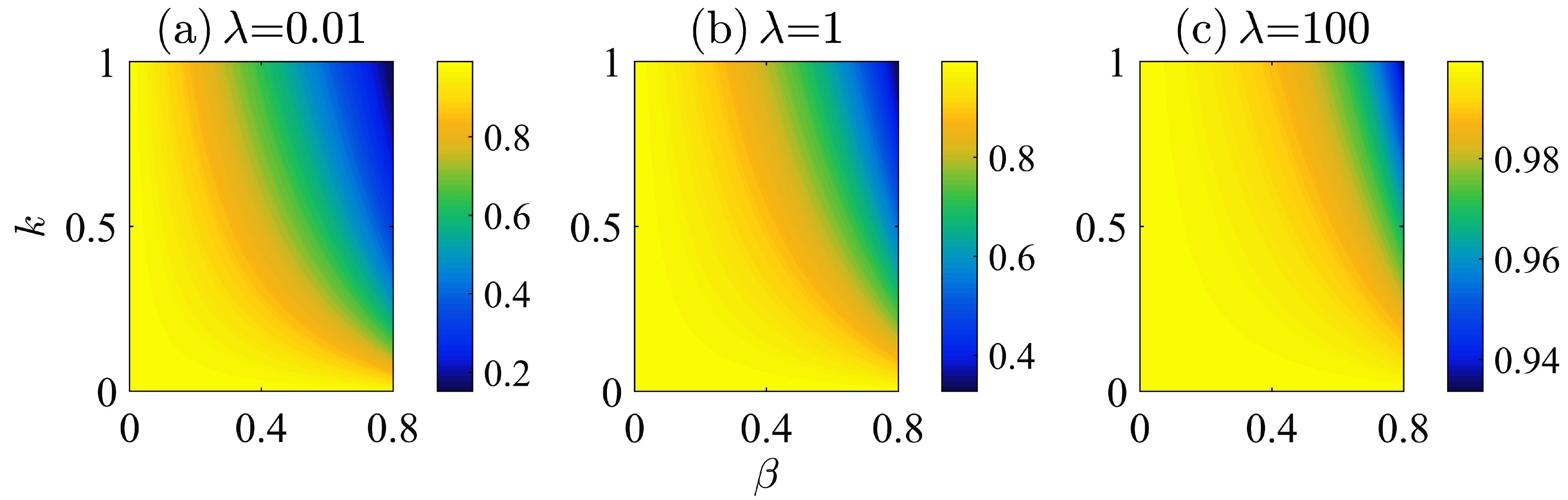}
	\caption{Variation of the correction parameter $ \mathcal{F}^{(Re_E)}_x $ on $ \beta - k $ plane for three different viscosity ratio ($ \lambda $).}
	\label{fig:correction-fact-x}
\end{figure}

For a clean drop it is observed previously \citep{Mandal2016} that even for a non deformable drop-surface the charge convection effect can create asymmetry in charge distribution about the axial plane. This leads to a transverse component of drop velocity $ \left(U_{d,x}^{(Re_E)}\big{|}_\text{Clean}\right) $ which can either be positive or negative depending on the electrical conductivity and permittivity combinations  of the drop-surrounding fluid pair.  However in the present case, \eqref{eq:Udx-ReE-low} suggests that the correction factor $ \mathcal{F}_x^{(Re_E)} $ which quantifies the effect of surfactant on the $ O(Re_E) $ correction to cross-stream velocity, is not influenced by any electrical property of the system. In figures \ref{fig:correction-fact-x}(a),(b) and(c) we have shown the variations of  $ \mathcal{F}^{(Re_E)}_x $ with $ \beta, k \text{ and } \lambda $. 
It can be inferred from the above figures that magnitude of  $ U_{d,x}^{(Re_E)} $  always gets reduced by the presence of non-uniform surfactant distribution (i.e. $ \mathcal{F}^{(Re_E)}_x \le 1  $) while the sign of $ U_{d,x}^{(Re_E)} $ remains unaltered  (i.e. $ \mathcal{F}^{(Re_E)}_x \ge 0 $). To analyze the physical mechanism we appeal to figures \ref{fig:qs_vs_theta_phi_vary_PHI_beta} and \ref{fig:qs_vs_theta_phi_pi_by_four_vary_k} where it is observed that the surfactant parameters $ \beta  $ and $ k $ cause remarkable modulations in the asymmetric pattern of the charge distribution about the axial plane. It was also observed that the surfactant concentration itself assumes an asymmetric pattern of varying extent about the axial plane (shown in figures \ref{fig:SC_vs_theta_vary_beta} and \ref{fig:SC_vs_theta_vary_k}) in the presence of a tilted electric field.  Thus the mechanism of convection of charges due to surface flow, is now counteracted by an decrease in the convective flow strength owing to the gradients in the surface tension. As a direct consequence the  electrohydrodynamic flow around the drop pattern is affected severely. This can be visualized by looking into the modulations in leading order surface velocity $ u_S^{(0)} $ and electrical traction force component $ ([[T^{E}_{\theta}]]) $ in figures \ref{fig:us0_components_vs_theta_vary_beta_phi_pi_by_four} and \ref{fig:FEt_vs_theta_vary_beta}, respectively.

Comparing the three cases presented in figures \ref{fig:correction-fact-x}(a),(b) and(c) it is found that surfactant retardation effect is maximum for $ \lambda = 0.01 $ while the opposite is true for $ \lambda =10 $. When drop fluid shows  very less viscosity in comparison to the suspending media, (equivalently the condition $ \lambda \ll 1 $), the drop resembles the behavior of a bubble and the correction factor $ \mathcal{F}_x^{(Re_E)} $ reaches 
\begin{equation}\label{eq:bubble_Fx_ReE}
\lim_{\lambda \to 0} \mathcal{F}^{(Re_{_E})}_x ={\frac { \left( -756+14\left( {{k}}^{2}+19{k}-54
		\right) {\beta}^{2}+14\left( -19{k}+108 \right) \beta
		\right)  \left( -1+\beta \right) }{27\left( 2+ \left( {k}-2
		\right) \beta \right) ^{2} \left( 7+ \left( {k}-7 \right) \beta\right) }}.
\end{equation}
On the other hand as the drop internal viscosity increases, the internal circulation strength becomes more and more weak and the Marangoni stress has very little to offer in affecting the circulation behaviour.  With $ \lambda \gg 1 $, the drop surface become almost immobile and it behaves as rigid sphere. In that case the surface convection velocity vanishes. As a consequence both the charge convection and surfactant induced Marangoni effects have diminishingly smaller contribution in controlling the drop migration velocity i.e. $  \lim_{\lambda \to \infty} U_{d,x}^{(Re_E)}\big{|}_\text{Clean}=0 \text{ and } \lim_{\lambda \to \infty} \mathcal{F}^{(Re_E)}_x=1 $. Similar trend of $ \mathcal{F}^{(Re_E)}_x $ is observed in figure \ref{fig:correction-fact-x}(c).

\subsubsection{Drop velocity in the longitudinal direction }
\label{ssc:drop-vel-z}
Drop velocity component in the direction parallel to the imposed flow ($ z $-direction) for a non-deformable drop can be expressed as $ U_{d,z}=U_{d,z}^{(0)}+Re_E \, U_{d,z}^{(Re_E)} $.  From mathematical arguments it can be shown that for the realistic range of the controlling dimensionless parameters $ (0 \le k \le 1, 0 \le \beta \le 0.8, \lambda \ge 0, c_2=-\dfrac{4}{H^2} <0) $, the term $ \mathcal{F}^{(0)}_z $ in \eqref{eq:Ud-leading-low}, is always a negative quantity. Unlike $ \mathcal{F}_x^{(Re_E)} $ as discussed before,  $ \mathcal{F}^{(Re_E)}_z $ is dependent on the direction of the externally applied electric field (please refer to \eqref{eq:Udz-ReE-low}). To obtain an overall picture of the surfactant effect, we define a correction factor as 
\begin{equation}\label{eq:corr_upto_ReE_z}
\mathcal{F}^{CC}_{z}=\frac{U_{d,z}^{(0)}+Re_E \, U_{d,z}^{(Re_E)}}{U_{d,z}^{(0)}\big{|}_\text{Clean}+Re_E \, U_{d,z}^{(Re_E)}\big{|}_\text{Clean}},
\end{equation}
which quantifies the extent to which the longitudinal drop velocity is altered in the presence of surfactant coating.

\begin{figure}
	\centering
	\begin{subfigure}[!htb]{0.46\textwidth}
		\centering
		\includegraphics[width=1.075\textwidth]{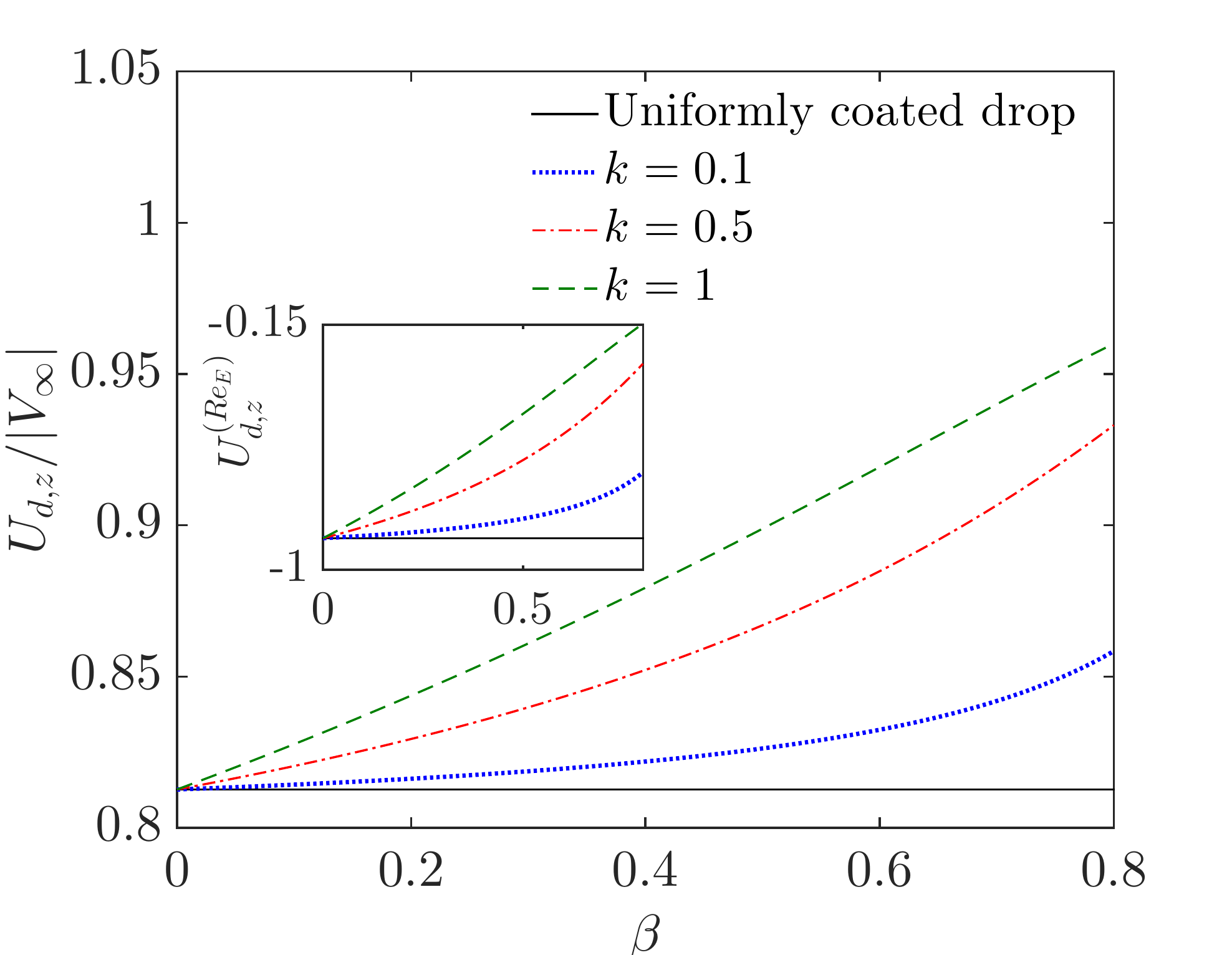}		
		\vspace{12ex}
		\caption{}
		\label{fig:Udz_vs_beta_vary_k}
	\end{subfigure}
	\begin{subfigure}[!htb]{0.48\textwidth}
		\centering
		\includegraphics[width=1.034\textwidth]{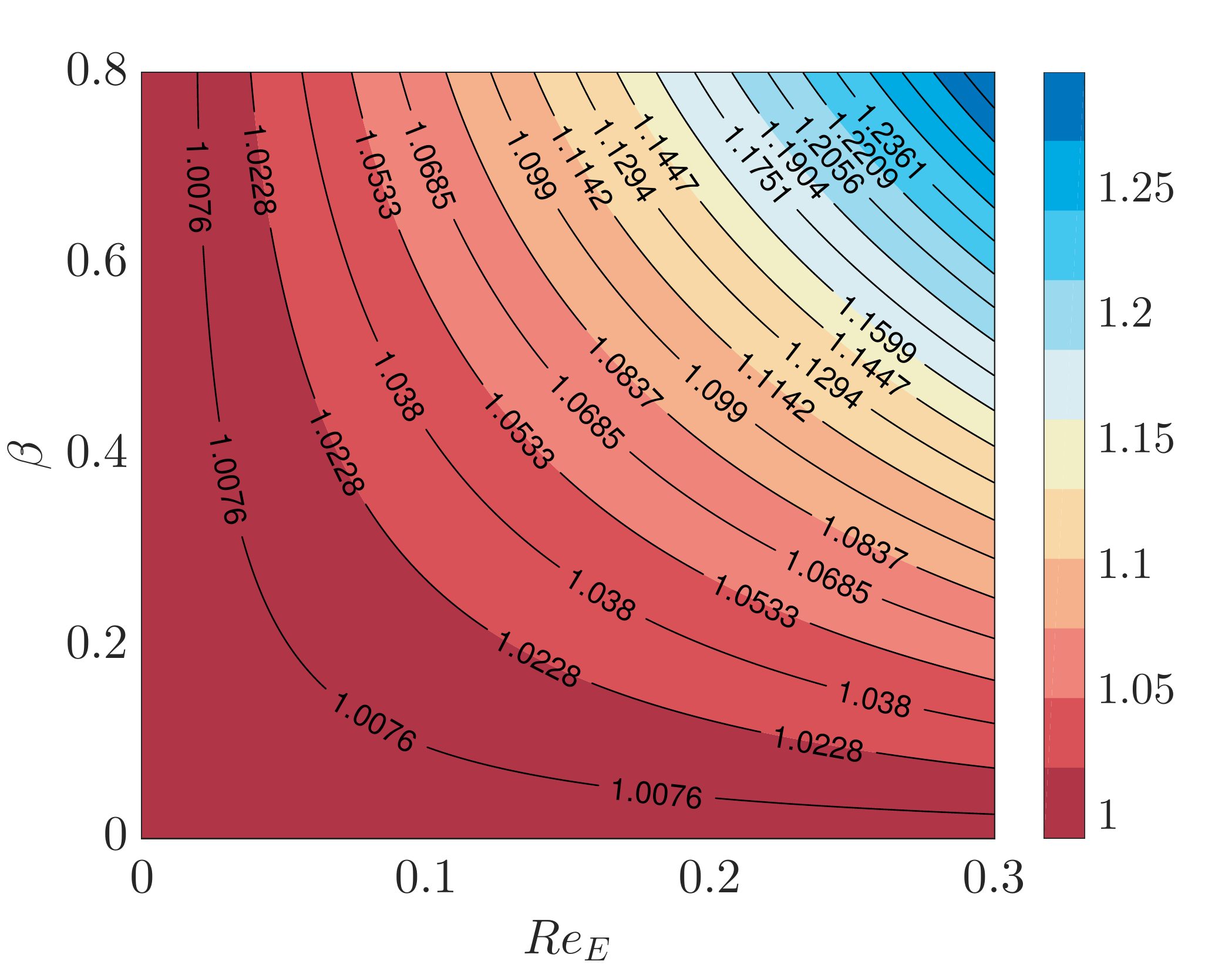}
		\vspace{12ex}
		\caption{}
		\label{fig:corr_Uz_upto_ReE_Beta_plane}
	\end{subfigure}
	\begin{subfigure}[!htb]{0.48\textwidth}
		\centering
		\includegraphics[width=1.02\textwidth]{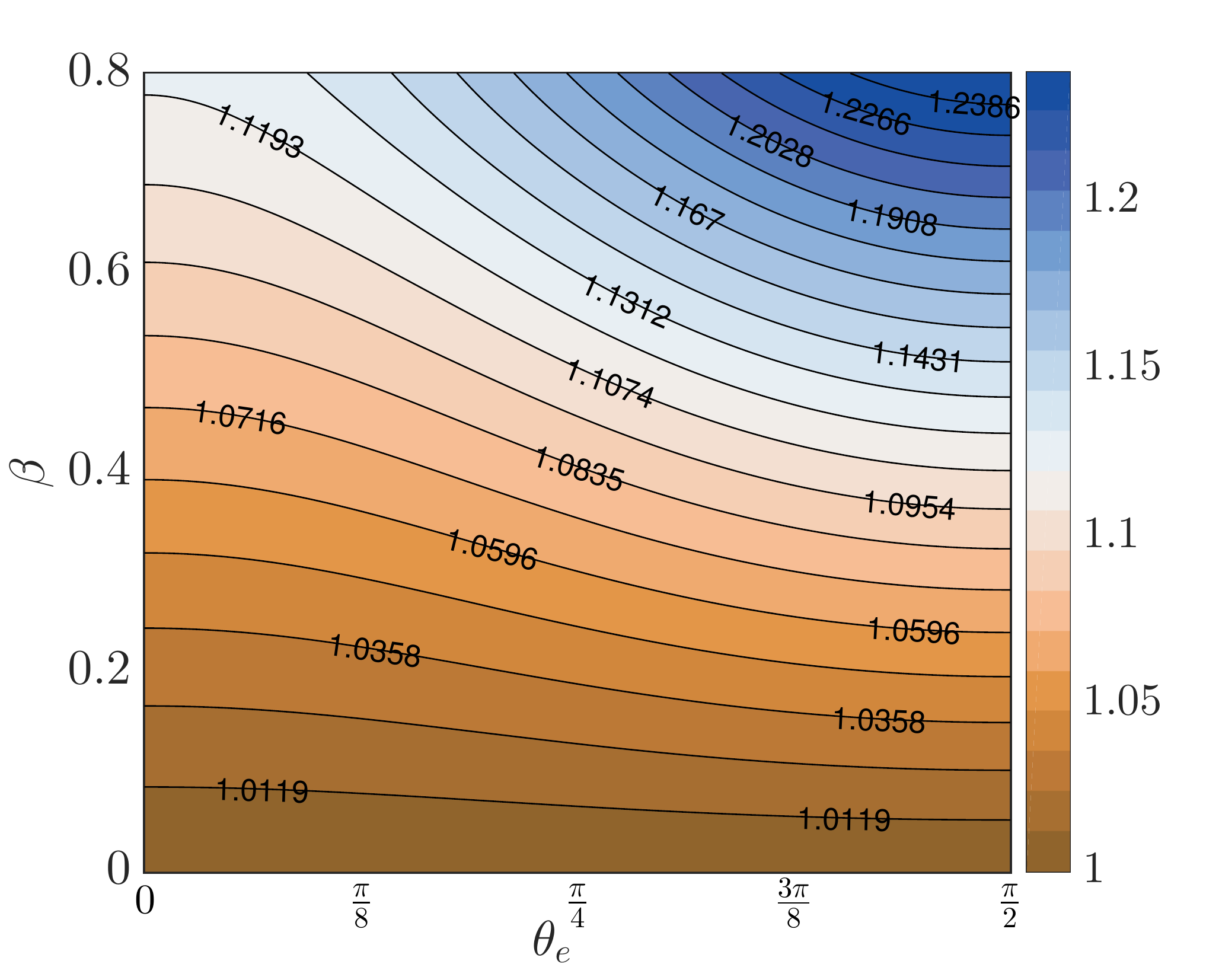}
		\vspace{12ex}
		\caption{}
		\label{fig:Corr_Uz_upto_ReE}
	\end{subfigure}
	\caption{Effect of surfactant parameters on longitudinal drop velocity (defined as $ U_{d,z}=U_{d,z}^{(0)}+Re_E\,U_{d,z}^{(Re_E)} $) for a non-deformable drop. Subplot (a) shows the normalized drop velocity vs. $ \beta $ for different values of $ k $. Here the drop velocity is normalized with respect to the magnitude of the imposed flow velocity at that location $ |V_\infty| $.  The corresponding variation in $U_{d,z}^{(Re_E)} $ is provided in the inset. Here the tilt angle of the applied electric field is $ \theta_e=\pi/4 $. Subplot (b) :  Correction factor $ \mathcal{F}^{CC}_z $ variation on $Re_E-\beta $ plane for $ k=1, \theta_e=\pi/4. $ Subplot (c) :  Correction factor $ \mathcal{F}^{CC}_z $ variation on $ \theta_e-\beta $ plane for $ k=1. $  Other parameters were taken as $ \lambda=0.1, M=2, Re_E=0.2,R=0.1,S=20, x_d=4 \text{ and } H=10.$}
\end{figure}

In figure \ref{fig:Udz_vs_beta_vary_k} it is observed that, the drop lags behind the flow. However an increase in the elasticity parameter $ \beta $ or the physicochemical constant $ k $, causes the axial migration speed to enhance and thereby making the drop follow the applied flow speed. Figure \ref{fig:qs_vs_theta_phi_vary_PHI_beta}
portrays the fact the even for a clean drop, in the presence of charge convection there is clear asymmetry in the charge distribution about the transverse plane. This also becomes more evident if the the azimuthal distribution of $ q_S $ is compared  for $ \theta=\pi/4 $ and $ 3\pi/4$ in figure \ref{fig:qs_vs_phi_theta_pi_by_four_vary_beta}. This being the case for a surfactnat-free drop, a non-uniform distribution of surfactant molecules greatly affects the extent of the asymmetry in charge distribution  about the transverse plane, thus modifying the asymmetric distribution of tangential electric force itself (shown in figure \ref{fig:FEt_vs_theta_vary_beta}). 
In turn the drop force on the drop in the longitudinal direction also gets modified nature of which is decided a complex interplay among the electric, hydrodynamic and Marangoni stresses.  

In contrast to the cross-stream velocity, the drop exhibits longitudinal movement at any direction of the applied electric field $ 0 \le \theta_e \le \pi/2 $. Now from \eqref{eq:Udz-ReE-low}, the $ O(Re_E) $ axial velocity is dependent on the electric field components, apart from being a function of $ \beta, k $ and $ \lambda $. 
while the parametric dependence of $ \mathcal{F}^{CC}_{z} $ is presented in  \ref{fig:Corr_Uz_upto_ReE}  which shows that the longitudinal drop velocity gets enhanced  with increasing $ \beta $ for all values of $0 \le \theta_e \le \pi/2 $. 
Changing the direction of the applied electric field severely affects the surfactant distribution (as shown in figure \ref{fig:sph_surf_conc_M_tilt}), which in turn modifies the way the electrohydrodynamically induced flow interacts with the Marangoni flow. Thus the resultant interface stress gets altered, which again modifies the drag force on the drop dictating the drop velocity.  Now from the inset of figure \ref{fig:Udz_vs_beta_vary_k} we find that $ U_{d,z}^{(Re_E)} $ remains $ < 0 $ for the choice of the electrical parameter combination $ (R=0.1, S=20) $, but the magnitude of the same $(U_{d,z}^{(Re_E)})$ decreases. Also in figure \ref{fig:correction-fact-ReE-z} we find that it gets further reduced if the tilt angle is increased. Although the $O(Re_E)$ longitudinal velocity is associated with the phenomena of charge convection, the practical effect of the same can only be realized when it is added to the leading order term to give an overall physical velocity in the $ z $ direction as $  U_{d,z}= U^{(0)}_{d,z}+Re_E\, U^{(Re_E)}_{d,z}. $  Now the leading order velocity $ U^{(0)}_{d,z} $ is $ >0 $ for the said choice of parameters in the figure, with or without surfactant coating. As a resultant effect the reduction in $ U_{d,z} $ due to charge convection gets suppressed by Marangoni flow. In order to have a clear picture about the opposing effects of $ Re_E $ and $ \beta $ in modifying $ U_{d,z} $ we provide a map of the parameter $ \mathcal{F}_z^{CC} $ on the $ Re_E-\beta $ plane in figure \ref{fig:corr_Uz_upto_ReE_Beta_plane}. It shows that as the charge convection effect increases (quantified by $ Re_E $), the Marangoni flow (quantified by $ \beta $), in effect, causes more and more augmentation in the longitudinal drop velocity $(U_{d,z})$.     This behavior is increasingly dominant as the tilt angle of the applied electric field increases before reaching the maximum in the case of transverse electric field (at $ \theta_e=\pi/2 $) (see figure	\ref{fig:Corr_Uz_upto_ReE}).
\subsection{Effects on a deformable drop}
\label{ss:deformation-charac}
\subsubsection{Drop velocity alteration in the cross-stream direction}
\label{ssc:Ca-vel-x}
\begin{figure}\label{fig:Udx_CA}
	\centering
	\begin{subfigure}[!htb]{0.45\textwidth}
		\centering
		\includegraphics[width=1.08\textwidth]{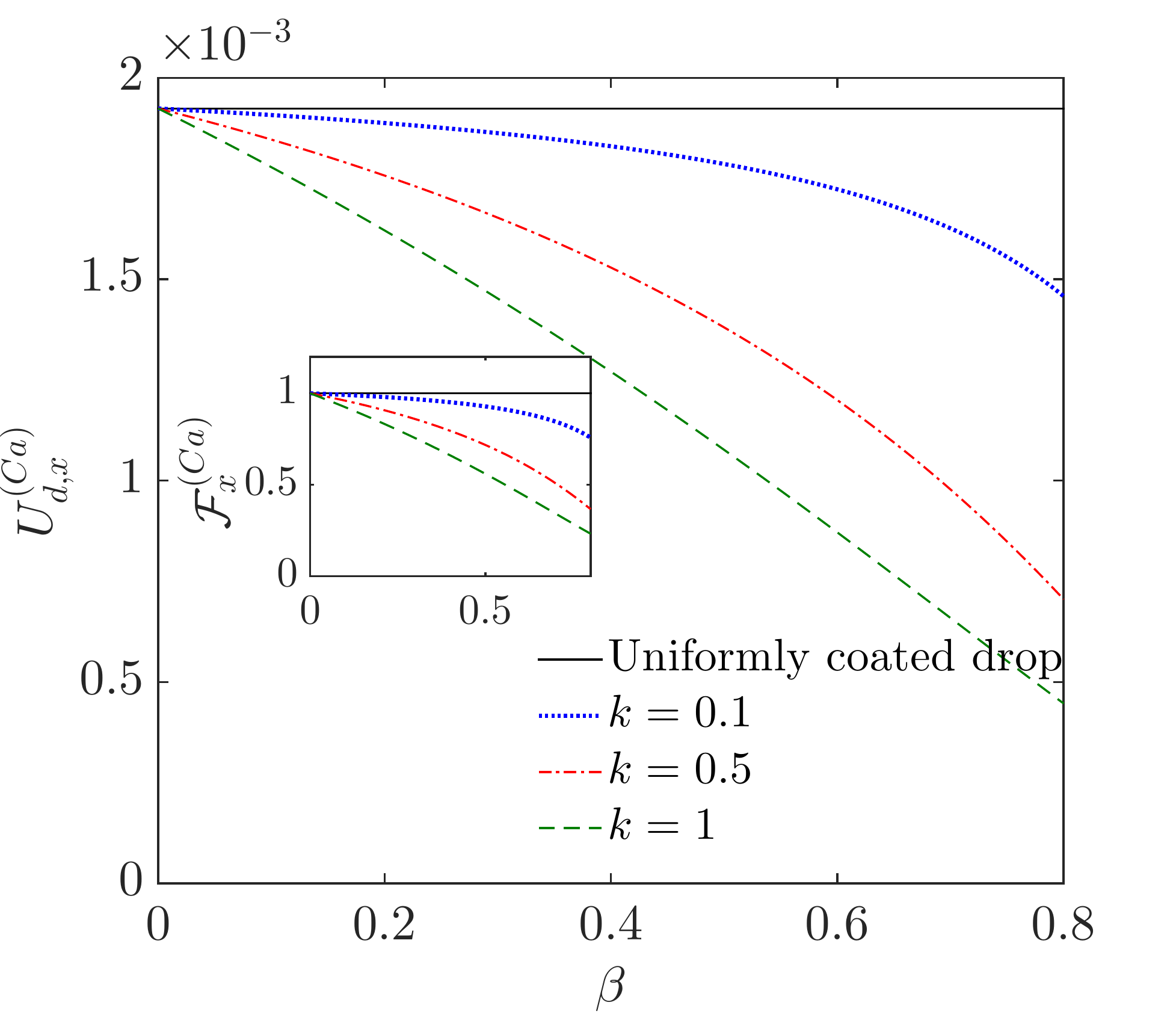}		
		\vspace{12ex}
		\caption{}
		\label{fig:UDx_CA_lambda_0p1_R_5_S_0p5_M_0p_one}
	\end{subfigure}
	\begin{subfigure}[!htb]{0.45\textwidth}
		\centering
		\includegraphics[width=1.08\textwidth]{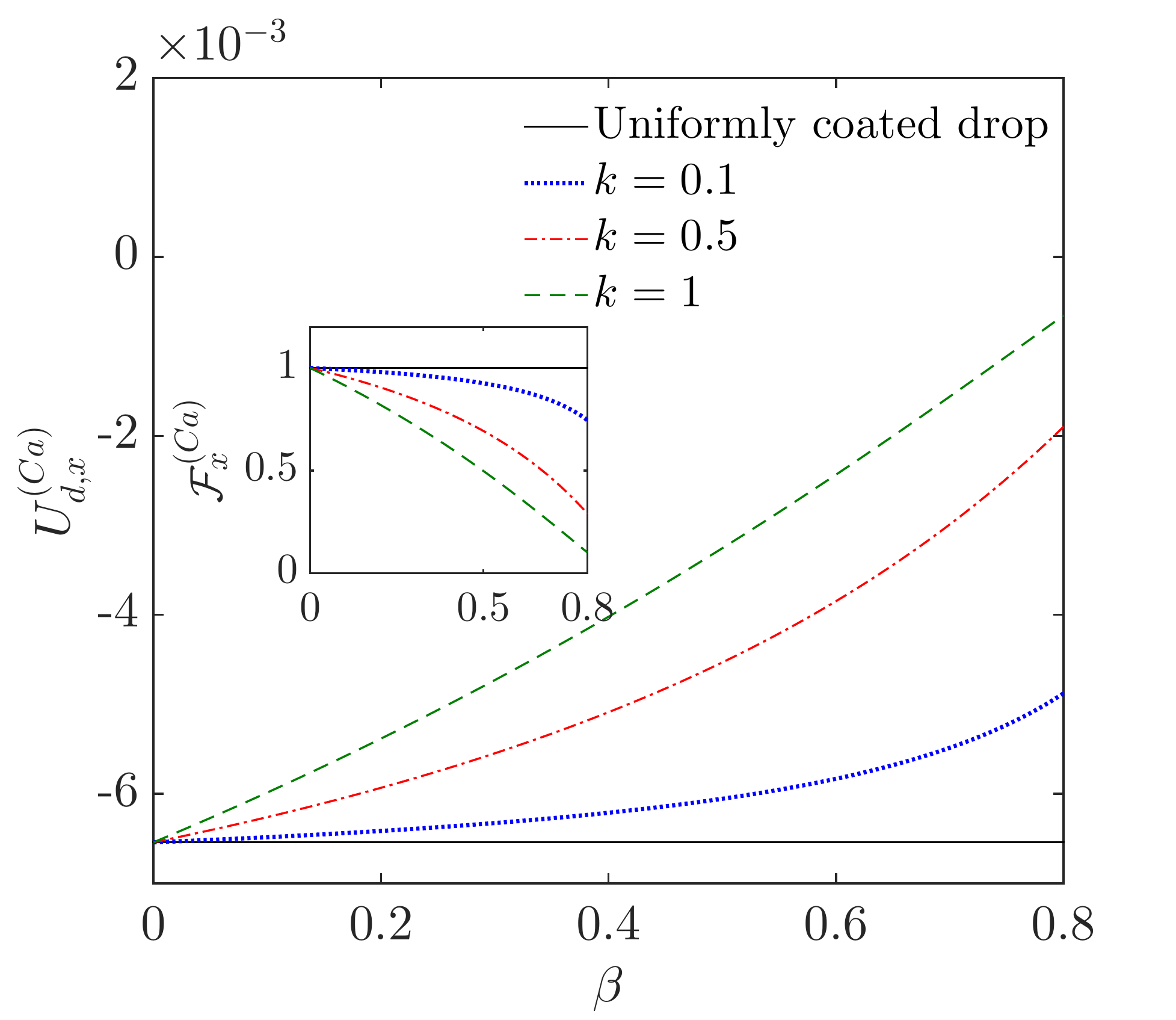}
		\vspace{12ex}
		\caption{}
		\label{fig:UDx_CA_lambda_0p1_R_0p1_S_10_M_0p_one}
	\end{subfigure}
	\\
	\begin{subfigure}[!htb]{0.55\textwidth}
		\centering
		\includegraphics[width=\textwidth]{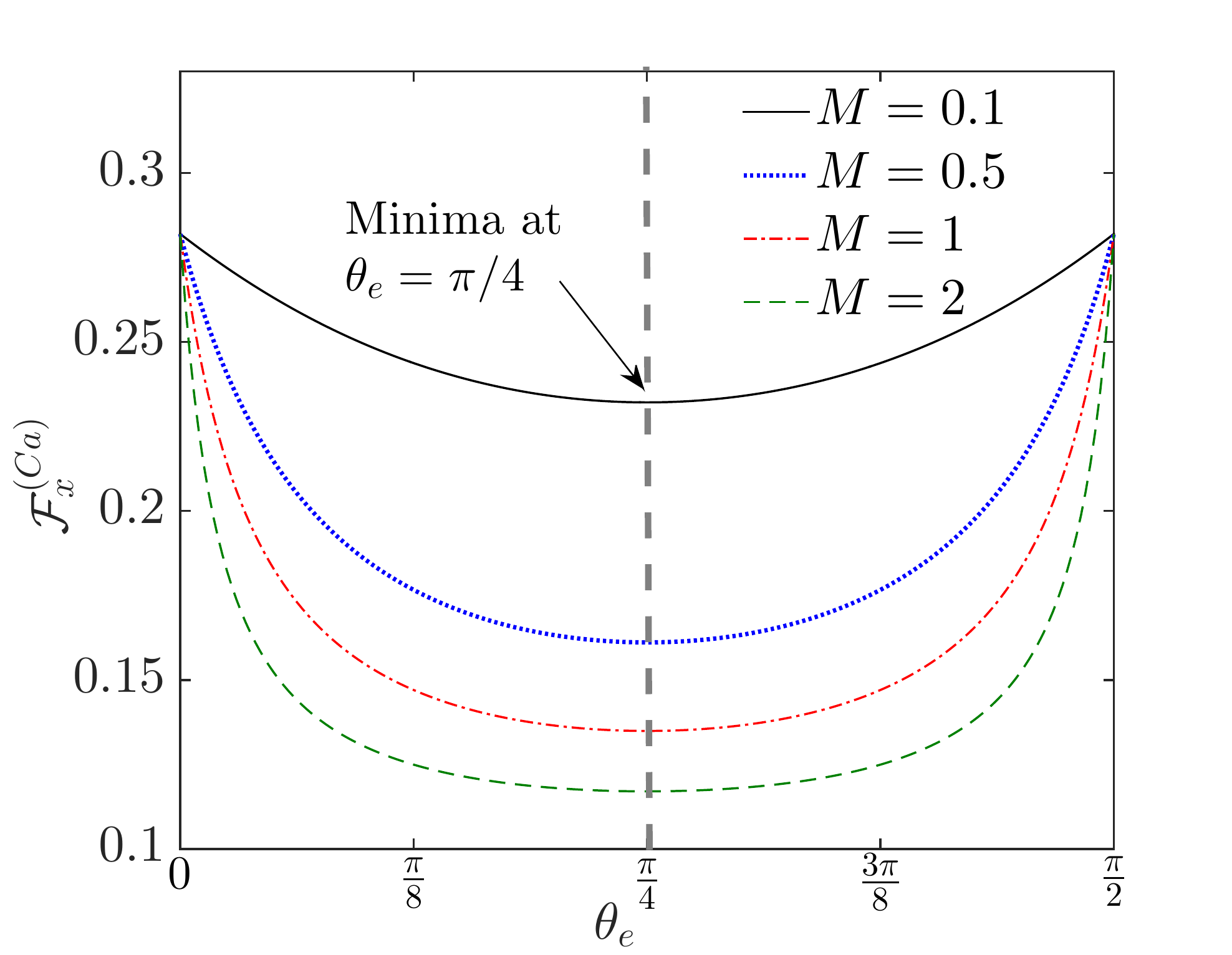}		
		\vspace{12ex}
		\caption{ }
		\label{fig:Udx_Ca_corr_vs_TILT_vary_M}
	\end{subfigure}
	\caption{Effect on surfactant parameters on the $ O(Ca) $ cross-stream velocity $ (U^{(Ca)}_x) $. Subplot (a): $ (U^{(Ca)}_x) $ vs. $ \beta $ for different values of $ k $.  In this case $ R=5,S=0.5 $ and $ \theta_e=\pi/4. $ Subplot (b): similar to (a) but with  $ R=0.1, S=10 $. For both (a) and (b) $ M=0.1 $ was taken. Subplot (c): Surfactant correction factor $ \mathcal{F}_x^{(Ca)} $ with tilt angle $ \theta_e $ for various values of the Mason number $ (M) $ and $ R=5,S=0.5 $. The other parameters are $ \lambda=0.1, x_d=4 $ and $ H=10. $} 
\end{figure}

Towards explaining the surfactant related effect we first look into the possible change in the electrohydrodynamically induced drop shape deformation that surfactants can bring in. In figures \ref{fig:A_YZ_vary_beta_K_R_5_S_0p5_M_0p_five} and \ref{fig:A_YZ_vary_beta_K_R_0p1_S_10_M_0p_five} we have shown the variations in the drop cross section perpendicular to the lateral migration direction $(A_{yz})$. Previously it was observed in figure	\ref{fig:SC_ST_var_beta_k} of section \ref{sssec:beta-k-physiscs} that the local surface tension $ (\gamma )$ of the drop reduces at some localized positions  where the surfactant concentration becomes high. As a consequence, the drop surface in those locations have to deform more in order maintain a balance the normal stress (please refer to \eqref{eq:n_stress_bal}). This mechanism is known as the `tip stretching' phenomena \citep{Pawar1996}. Now figure \ref{fig:A_YZ_vary_beta_K_R_5_S_0p5_M_0p_five} shows that an increase in  either  of $ \beta $ or $ k $, the drop cross section gets reduced for $ R=5 $ and $ S =0.5$ and thus $ U^{(Ca)}_{d,x} $ is expected to increase due to reduced flow resistance. This is in apparent contradiction to the increasing trend of $ |U^{(Ca)}_{d,x}| $ in figure 	\ref{fig:UDx_CA_lambda_0p1_R_0p1_S_10_M_0p_one}. This can be resolved by realizing firstly that the amount of shape change due to surfactants is very low in this case thus rendering the `tip stretching' to play a very minor role in this case. Secondly,  even in the case of a clean drop surface when an electric field is applied it causes a drop deformation and subsequent modifications in charge distribution on the drop surface results. This charge distribution gets altered significantly when Marangoni stress comes into play. As a consequence of such phenomena the electrohydynamic flow pattern is redistributed in a way that has the potential to affect the drag force on the drop and also the resulting drop velocity magnitude. As an example of such effect we appeal to figure  \ref{fig:us0_components_vs_theta_vary_beta_phi_pi_by_four} where the significant alterations in the surface flow due to surfactant was captured in terms of changes in the $ u_S^{(0)} $ components. In addition we present figure \ref{fig:explain-tangential} which shows that the $ O(Ca) $ correction to the electrohydrodynamic traction force distribution $ [[\mathbf{T}^{E,(Ca)}]] $ also gets modified when the surface elasticity $ (\beta) $ varies. In the case of $ R=5 $ and $ S =0.5$, the small amount of surfactant induced deformation effect is surpassed by the above effect. 

In contrary to above, for a choice of parameters $ R=0.1,S=10 $ the drop cross-section gets increased with surfactant effects (refer to figure \ref{fig:A_YZ_vary_beta_K_R_0p1_S_10_M_0p_five}) although again the amount of modulations is very low. This effect,  competes with the electrohydrodynamic flow modification effect and finally due to a resultant consequence a  reduction in $ |U^{(Ca)}_x| $ is observed.   

Figure \ref{fig:Udx_Ca_corr_vs_TILT_vary_M}  depicts an increasing trend of surfactant effect with increase in Mason number. It can be explained as follows. As the Mason number increases, charge convection becomes more and  more prominent and also it causes tremendous changes in the surfactant distribution. This leads to a greater gradients in surface tension and the Marangoni stress has a greater scope to influence the drop velocity. It (figure \ref{fig:Udx_Ca_corr_vs_TILT_vary_M})  also suggests, when the applied electric field is either axial $ (\theta_e=0 )$ or transverse $ (\theta_e=\pi/2) $, changing the Mason number $ (M)$ has no effect on the surfactant contribution. It is due to the fact that in those cases the deformation-triggered cross-stream velocity is not at all affected by electrical effects and is generated solely due to hydrodynamic effects \citep{Mandal2016c}. When the electric field is applied at an angle $ \theta_e > 0$, the parameter $ \mathcal{F}^{(Ca)}_x $ decreases continuously before reaching at a minima at $ \theta_e=\pi/4 $. In a symmetric manner, a further increase in tilt angle causes $ \mathcal{F}^{(Ca)}_x $ to increases, physically depicting a fall in the surfactant-induced retardation effect. 

\subsubsection{Drop velocity alteration in the longitudinal direction}
\label{ssc:Ca-vel-z}  
\begin{figure}[!htb]
	\centering
	\begin{subfigure}[!htb]{0.45\textwidth}
		\centering
		\includegraphics[width=1.12\textwidth]{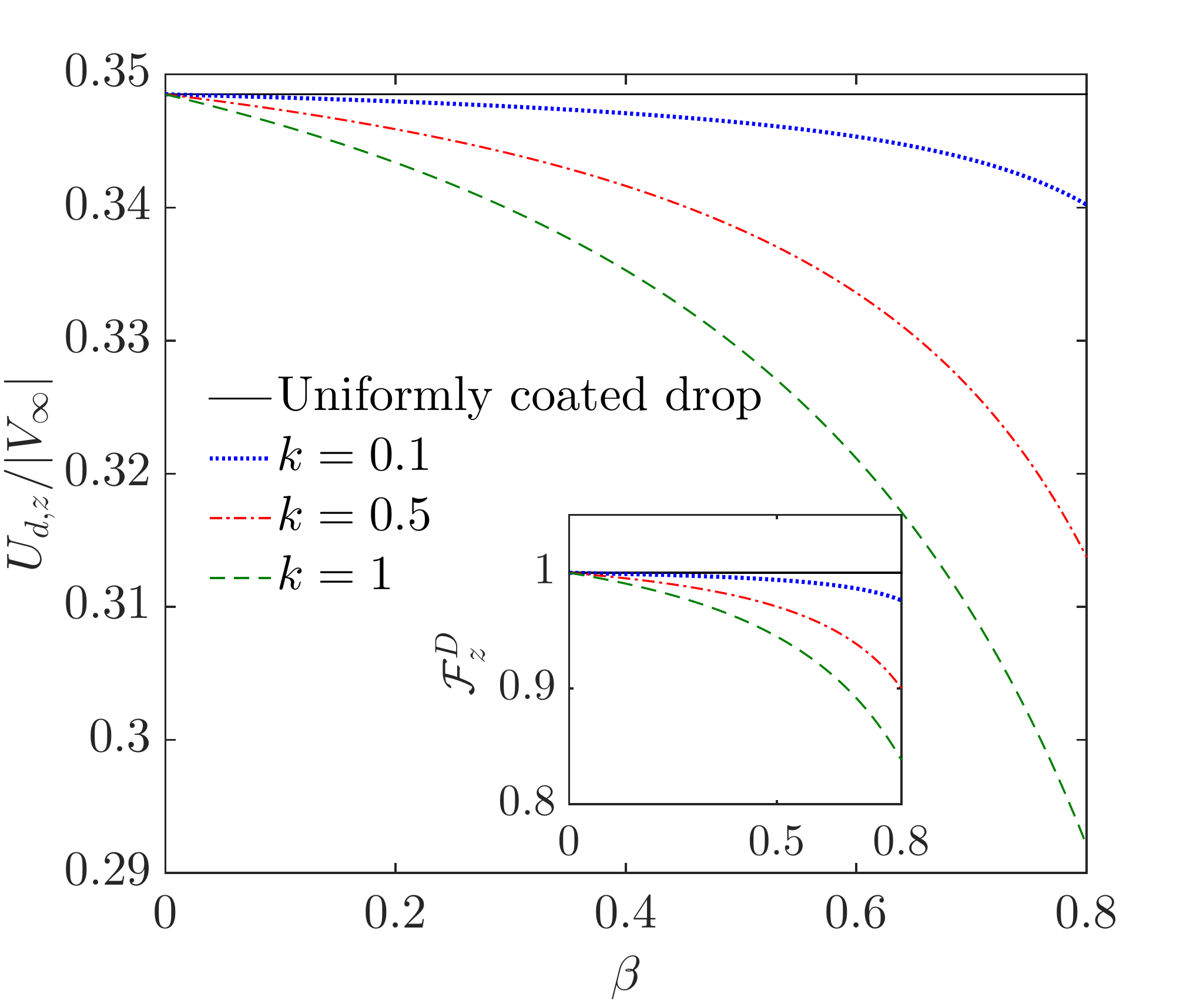}
		\vspace{8ex}
		\caption{}
		\label{fig:UDZ_upto_CA_lambda_0p1_R_0p1_S_20_M_one_Ca_0p_two}
	\end{subfigure}
	\quad 
	\begin{subfigure}[!htb]{0.45\textwidth}
		\centering
		\vspace{2.7ex}
		\includegraphics[width=1.19\textwidth]{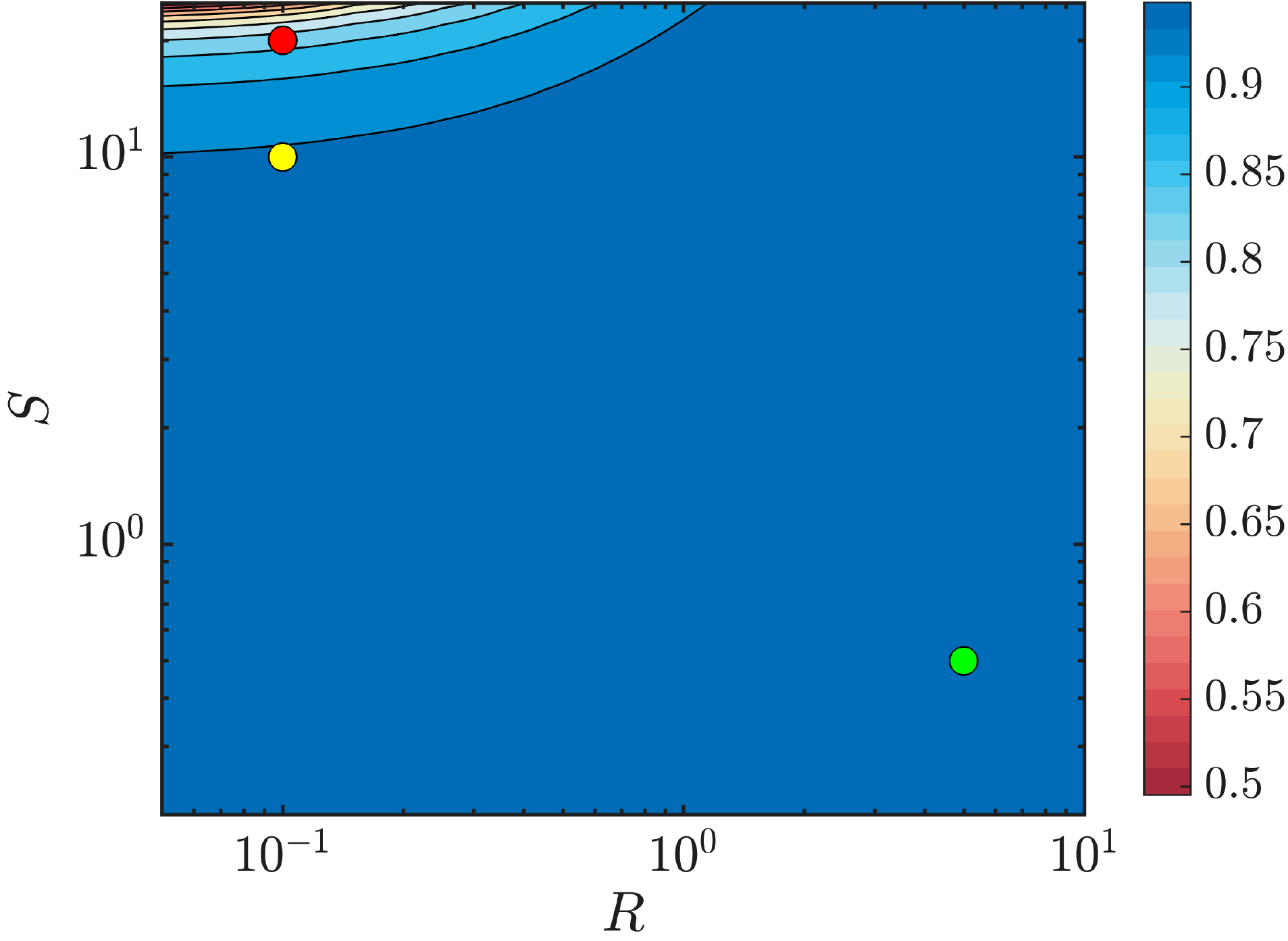}
		\vspace{8ex}
		\caption{}
		\label{fig:Corr_Uz_upto_CA_new}
	\end{subfigure}
	\caption{(a) Variation of longitudinal drop velocity defined as $ U_{d,z} =U^{(0)}_{d,z}+Ca\,U^{(Ca)}_{d,z}$, with $ \beta $ for various $ k $. Again the drop velocity is normalized with respect to the magnitude of the imposed flow velocity at that location $ |V_\infty|$. Here $ R=0.1$ and $S=20 $. (b) Variation of the surfactant correction factor $ \mathcal{F}_z^{D} $ on $ R-S $ plane. The red, yellow and green circles represent the three cases $ R=0.1,S=20; R=0.1,S=10 $ and $ R=5,S=0.5$, respectively. The other parameters were chosen as  $ \lambda=0.1, M=1, Ca=0.2, x_d=4, H=10 $ and $ \theta_e=\pi/4. $ }
	\label{fig:UDZ-CA}
\end{figure}

It is evident from figure \ref{fig:UDZ_upto_CA_lambda_0p1_R_0p1_S_20_M_one_Ca_0p_two} that in the absence of surfactant induced Marangoni effects, the drop lags behind the flow (i.e. $U_{d,z}/|V_\infty|<1$) for the choice of electrohydrodynamic parameters. An  increase in either $ \beta $ or $ k $ indicates a corresponding enhancement in the Marangoni stresses. Now the variation of projected area perpendicular to the drop motion direction shows that $A_{xy}$ increases with increasing values of surfactant parameters (shown in figure 		\ref{fig:A_XY_vary_beta_K_R_0p1_S_20}). Similar to the discussion in the preceding section (\ref{ssc:drop-vel-x})), a combination of the mechanisms of tip stretching and redistribution of electrohydrodynamic stresses on the deformed surface, causes an increase in the drag force on the drop in a direction opposite to its longitudinal motion. This creates further retardation in the longitudinal motion of the drop.

Similar to the case of $ O(Re_E) $ (section \ref{ssc:drop-vel-z}), in order to quantify the surfactant induced modifications in the longitudinal drop velocity where the shape deformation is the dominant mechanism, we define another overall surfactant correction parameter as  
\begin{equation}\label{eq:corr_upto_CA_z}
\mathcal{F}^{D}_{z}=\frac{U_{d,z}^{(0)}+Ca \, U_{d,z}^{(Ca)}}{U_{d,z}^{(0)}\big{|}_\text{Clean}+Ca \, U_{d,z}^{(Ca)}\big{|}_\text{Clean}}.
\end{equation} and investigate the effects of the electrical property ratios $  R $ and $ S $ in figure \ref{fig:Corr_Uz_upto_CA_new}. It is observed that for the choice of parameters in figure \ref{fig:UDZ_upto_CA_lambda_0p1_R_0p1_S_20_M_one_Ca_0p_two} $(R=0.1, S=20)$, a significant effect of surfactant results. However, for some other choices similar to those used in the case of $ O(Ca) $ cross-stream velocity ($ (R,S)=(5,0.5) $ and $ (R,S)=(0.1,10)$) in section \ref{ssc:Ca-vel-x}, the surfactants play a very minimal role in altering the longitudinal drop velocity ($ U_{d,z}=U_{d,z}^{(0)}+Ca\,U_{d,z}^{(Ca)} $). This indicates that the capability of surface active agents in modifying the drop motion in each direction can be selectively tuned with a suitable choice of electrical conductivity and permittivity ratios of the drop-surrounding fluid pair.

It is worth to mention that in general the charge convection ($O(Re_E)$) and shape deformation ($O(Ca)$) may be significant for a system simultaneously. Having a deep insight of the Marangoni effects on each of the mechanisms, as discussed in the preceding sections, will be advantageous in grasping the overall picture. 
 
\subsection{Drop trajectory}
\label{sc:trajectory}
\begin{figure}
	\centering
	\begin{subfigure}[!htb]{0.47\textwidth}
		\centering
		\includegraphics[width=1.075\textwidth]{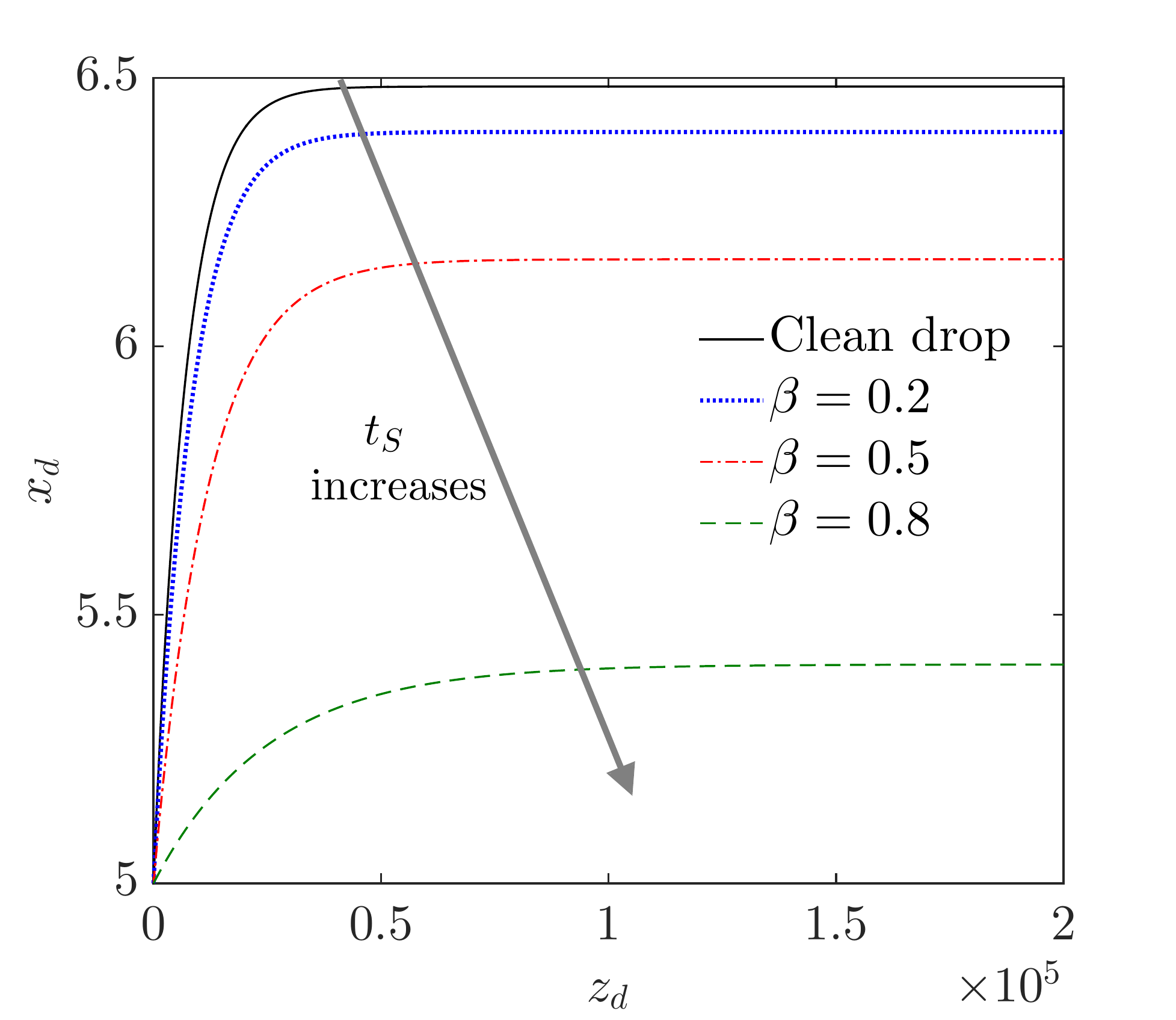}		
		\vspace{12ex}
		\caption{}
		\label{fig:Trajectory_R_5_S_0p_five_M_p_five}
	\end{subfigure}
	\begin{subfigure}[!htb]{0.47\textwidth}
		\centering
		\includegraphics[width=1.075\textwidth]{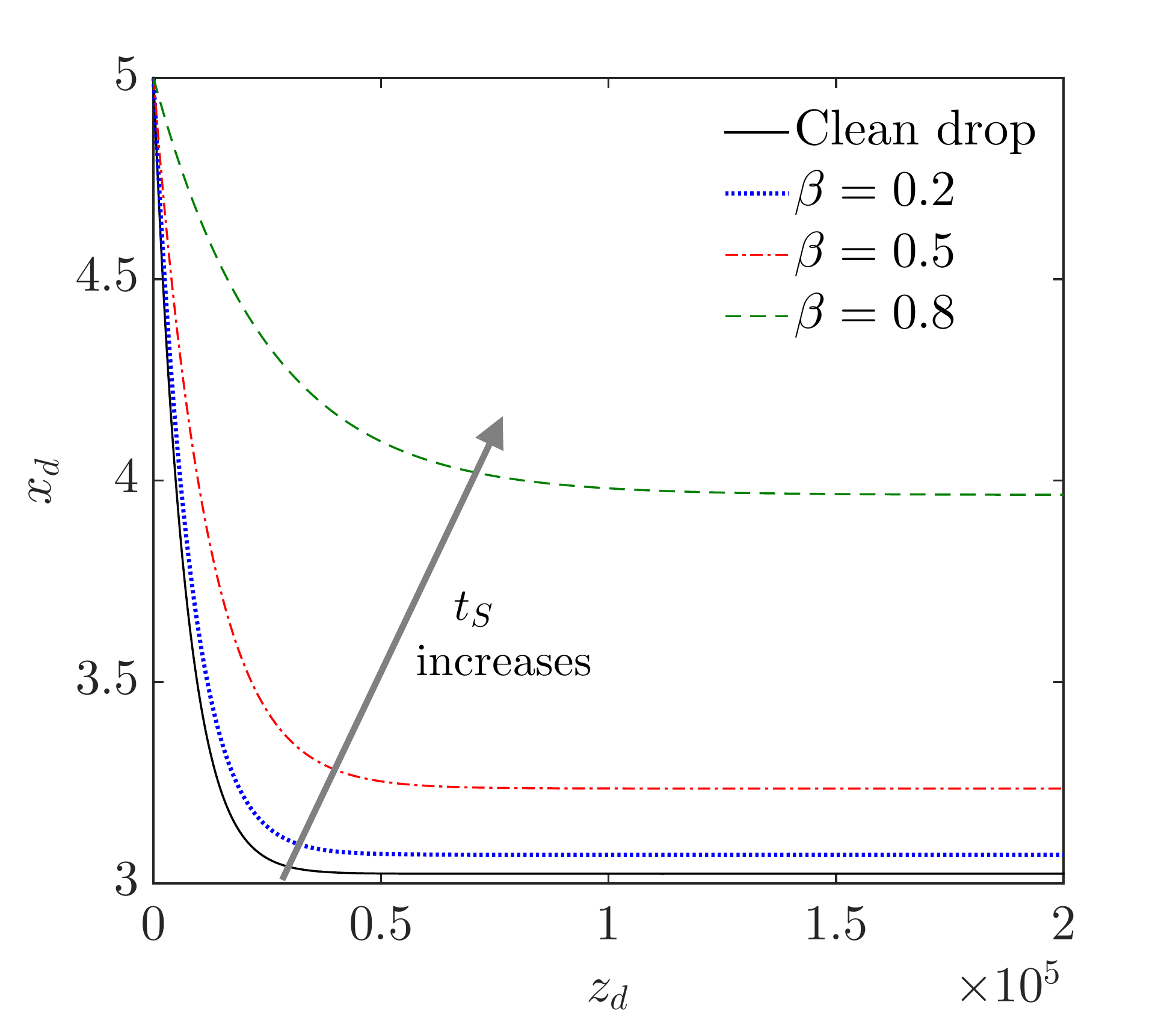}
		\vspace{12ex}
		\caption{}
		\label{fig:Trajectory_R_0p_5_S_5_M_p_one}
	\end{subfigure}
	\\
	\begin{subfigure}[!htb]{0.47\textwidth}
		\centering
		\includegraphics[width=1.075\textwidth]{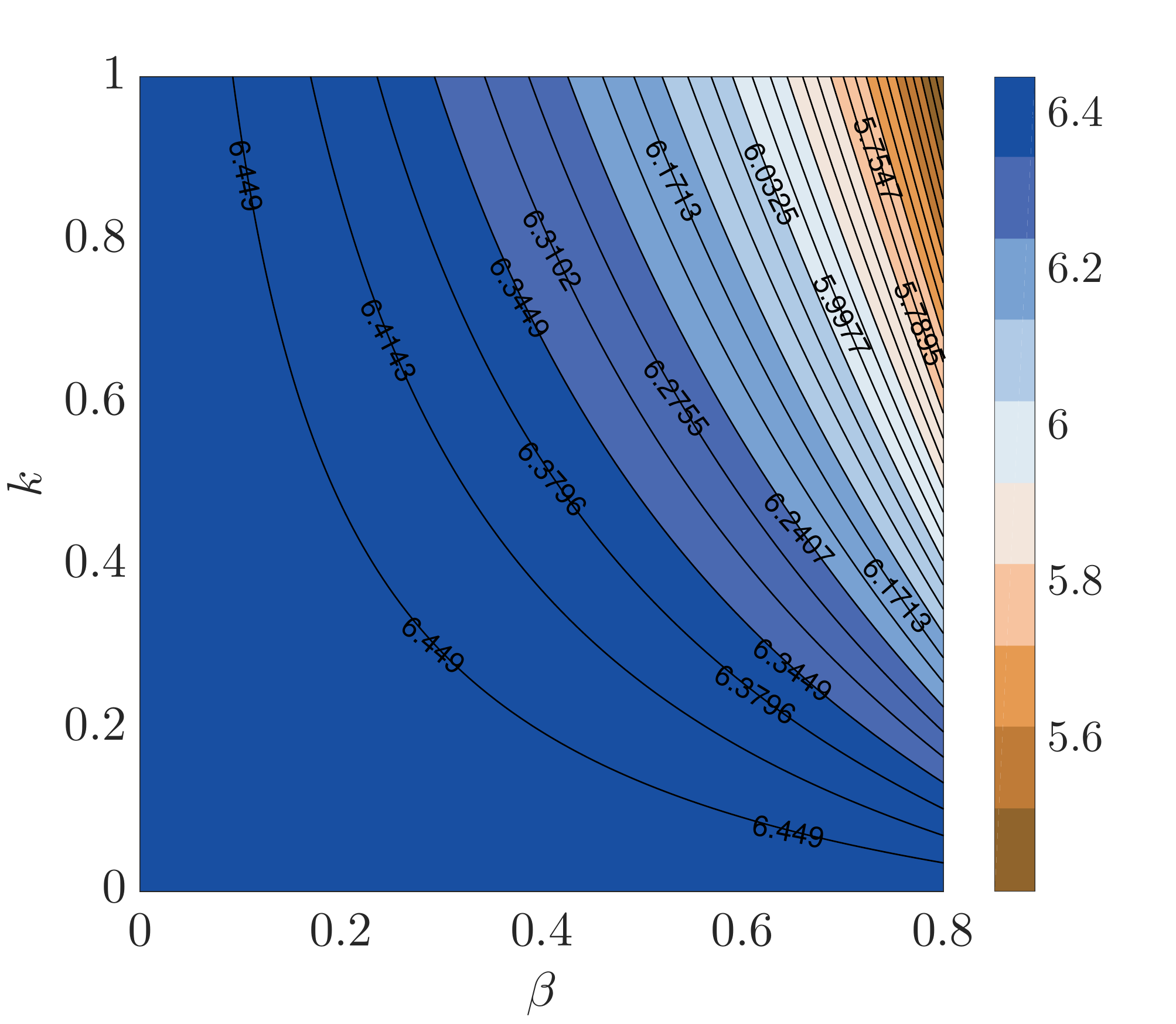}		
		\vspace{12ex}
		\caption{}
		\label{fig:Xd_inf_R_5_S_0p_five_M_p_five}
	\end{subfigure}
	\begin{subfigure}[!htb]{0.47\textwidth}
		\centering
		\includegraphics[width=1.075\textwidth]{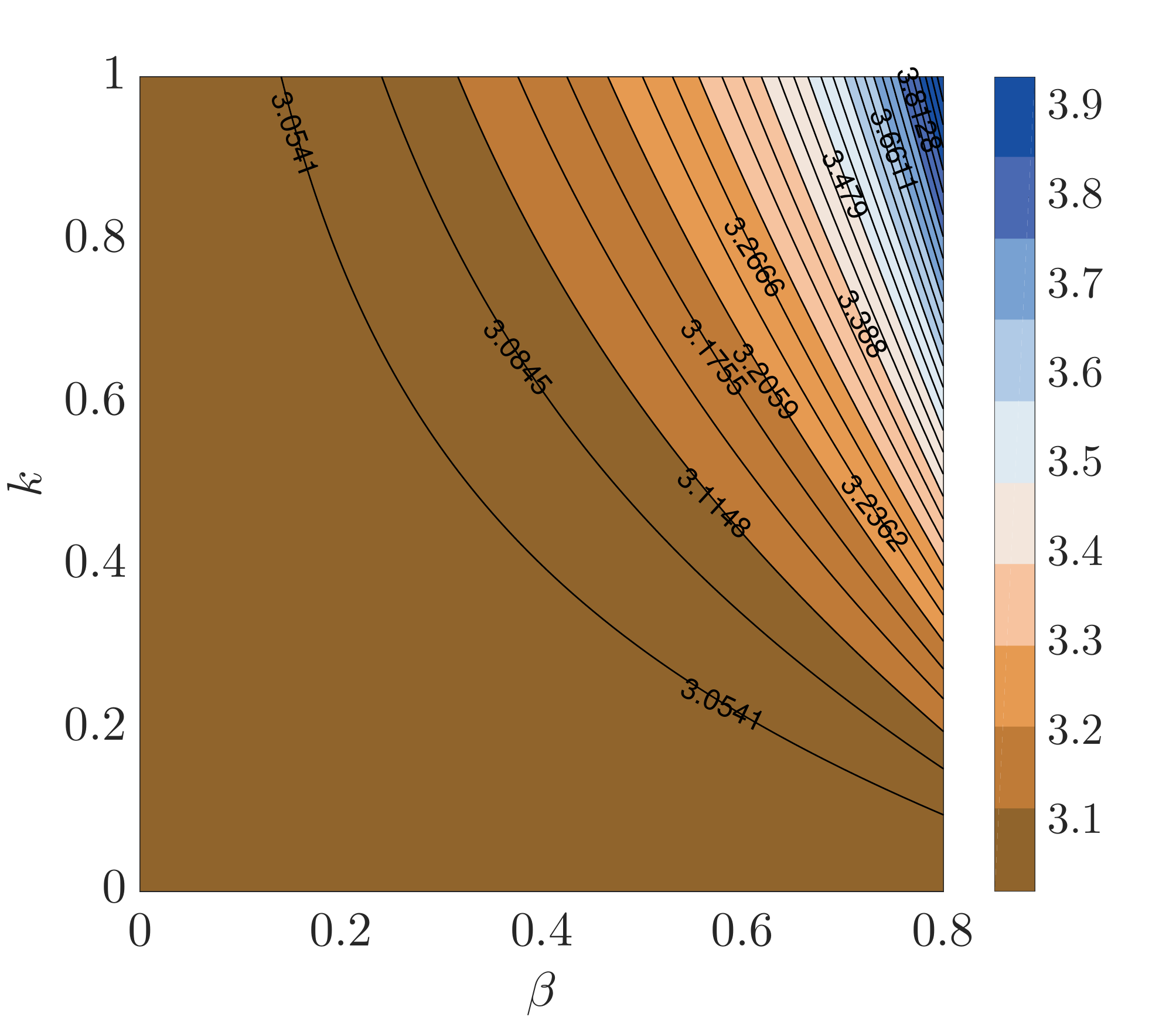}
		\vspace{12ex}
		\caption{}
		\label{fig:Xd_inf_R_0p_5_S_5_M_p_one}
	\end{subfigure}
	\caption{Effect of surfactant on the drop trajectory when both the charge convection and shape deformation are considered. Subplot (a),(b): Drop trajectory for various elasticity parameter $ (\beta) $ and $ k=1 $. Subplots (c),(d): Variation of final steady state transverse position of the drop $ x_{d,\infty} $ on the $ \beta-k $ plane. For (a),(c) $ R=5, S=0.5, M=0.5 $ and for (b)(d) $ R=0.5, S=5, M=0.1 $.  The other parameters are given as  $\lambda=0.1, \theta_e=\pi/4, Ca=0.1 $ and   $Re_E=0.1$.} 
\end{figure}
\begin{figure}
	\centering
	\begin{subfigure}[!htb]{0.47\textwidth}
		\centering
		\includegraphics[width=1.1\textwidth]{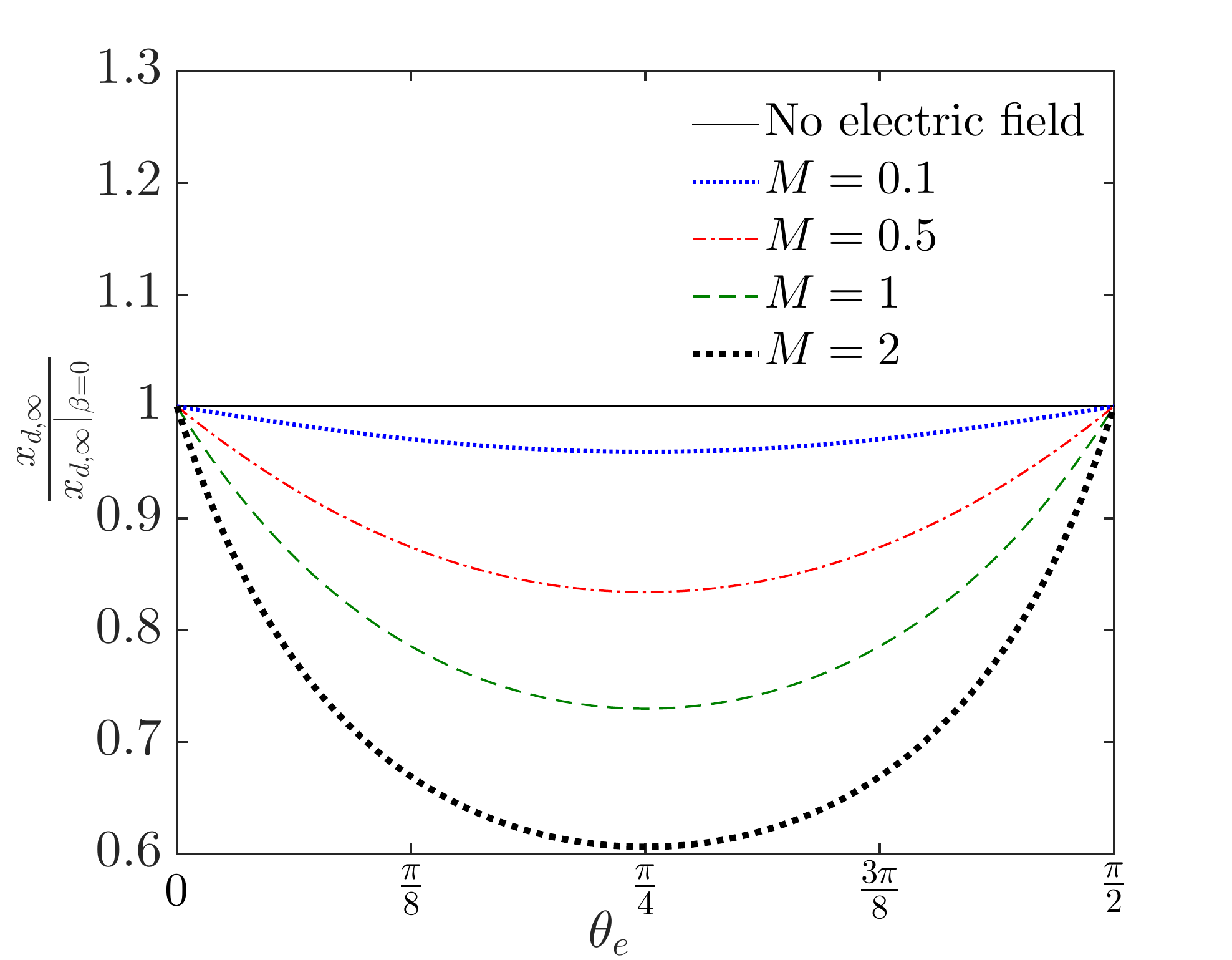}		
		\vspace{12ex}
		\caption{}
		\label{fig:xd_inf_correction}
	\end{subfigure}
	\begin{subfigure}[!htb]{0.47\textwidth}
		\centering
		\includegraphics[width=1.02\textwidth]{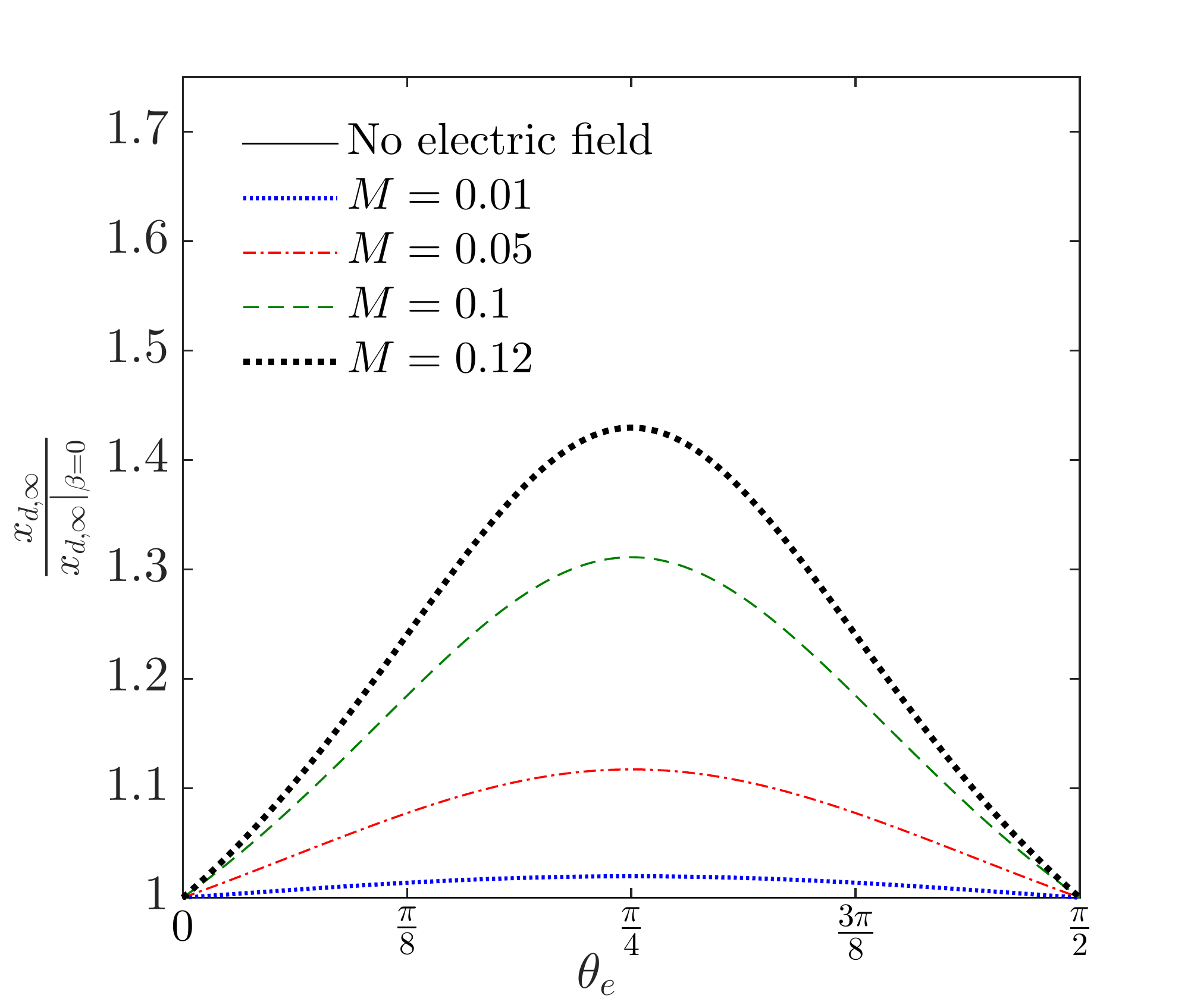}
		\vspace{12ex}
		\caption{}
		\label{fig:xd_inf_correction_R_0p5_S_five}
	\end{subfigure}
	\caption{Variation in the final steady state transverse position of the drop with respect to the surfactant free case $ ({x_{d,\infty}}/{x_{d,\infty,\text{Clean}}} )$
		for varying direction of the  applied electric field and various values of Mason number $ (M) $. For subplot (a): $ R=5, S=0.5 $ and  for (b): $ R=0.5, S=5 $. The other parameters are $ \lambda=0.1,
		\beta=0.8,k=1; Re_E=0.1 \text{ and } Ca=0.1 $.} 
\end{figure}
{In all the demonstrative examples so far we have considered a fixed value of the drop location  $ (x_d,z_d) $ at that specific instant of time. However the drop tries to adjust itself with the continuously varying surrounding flow conditions during its motion. The surfactant distribution, surface charge and the interfacial flow velocity are altered in such way that the drop begins to move in the longitudinal as well as in the cross-stream direction.} Under the quasi-steady state assumption the drop trajectory was obtained in section \ref{sss:combination-surf-E}.
Figure  \ref{fig:Trajectory_R_5_S_0p_five_M_p_five} shows that an increase in $ \beta $ causes the drop to settle down to a steday state transverse location much lower than a clean drop.  This can be explained again by following the figure \ref{fig:UDX_total_5_0p_five} which shows that the magnitude of the cross-stream migration velocity  $ (|U_{d,x}|) $ gets reduced for an increase in the elasticity parameter $ (\beta) $. In addition, the time required to reach the final steady state position $ t_S $ is increased for the same conditions. With an other parameter choice ($R=0.5, S=5, M=0.1$), figure \ref{fig:Trajectory_R_0p_5_S_5_M_p_one} demonstrates that  the downward drop motion away from the centerline is slowed down and finally it reaches a lateral position $ (x_{d,\infty}) $ which is much higher than a clean drop.
Comparing figures 
\ref{fig:xd_inf_correction}  and \ref{fig:xd_inf_correction_R_0p5_S_five}, it is found that for similar change in $ x_{d,\infty} $ for different values of $ R,S $, the Mason number range is significantly lower. This suggests that the surfactant induced change in steady state  drop-position in lateral direction is dictated by a compley interplay among electrical properties. As the electrical effects increase compared to the hydrodynamic effect, presence of the surfactant molecules can alter the final drop location in a more dominant fashion.   

The combined action of charge convection and the shape deformation leads to a resultant cross stream velocity defined as  $ U_{d,x}=Re_E\,U_{d,x}^{(Re_E)}+Ca\,U_{d,x}^{(Ca)} $. For a choice of $R=5, S=0.5, M=0.5$ the $ U_{d,x} $ is $ >0 $ (shown in figure \ref{fig:UDX_total_5_0p_five}) and hence the drop trajectory shows a motion where the drop migrates downwards away from the centerline. For the other choice of parameters, as shown in \ref{fig:UDX_total_0p_five_five}, the sign of $ U_{d,x} $ is just the reverse $ (U_{d,x}<0) $. Accordingly the drop motion shows an upward trend in figure \ref{fig:Trajectory_R_0p_5_S_5_M_p_one}. 

\section{Conclusions}
In this paper we consider a physical microfluidic system of drop transport through a parallel plate channel. The motion of the drop is actuated with the simultaneous presence of  a pressure driven flow $ (\mathbf{V}_\infty) $ and an uniform electric field $ (\mathbf{E}_\infty) $  whose direction can be controlled. Additionally the drop surface is contaminated with the presence of a non-ionic surfactant coating. The nonlinearities associated with the charge convection, surfactant transport and shape deformation make the system of govering equations and boundary conditions highly coupled. Thus the hydrodynamic equations in the creeping flow limit are first solved concurrently with the electrical potential and the convection-diffusion equation of surfactant transport. In each order of perturbation ($ O(1),O(Re_E) \text{ and } O(Ca)$), we employ the force free condition to obtain the drop velocity components.  The obtained solution of the velocity and pressure field in both inside the drop and outer fluid are then utilized to solve the normal stress balance equation. Consequently a description of the deformed shape is obtained within the purview of the small deformation theory. Finally a  description of the drop trajectory $ (x_d(t),z_d(t)) $ was provided in the `quasi-steady-state' condition. Under various limiting situations, our resuting expressions of various key parameters match against the existing works in the literature thus indicating the validity of the present mathematical calculations. During the discussion of the results, the effect of various electrical $ (R,S,M,\theta_e,Re_E) $ and surfactant related parameters $ (\beta, k, \lambda) $ were demonstrated and possible physical mechanisms were critically investigated. The important observations can be summarized as follows:

\begin{enumerate}[label=(\roman*),leftmargin=0pt,itemindent=3em]
	\item 
	The surfactant distribution is found to be tremendously affected by the electrical parameters such as the $ Re_E, M $ and $ \theta_e $. Due to a complex coupled nature of the surfactant transport and charge convection mechanisms, the surface tension gradients gets enhanced with an increase in the charge convection effect. Similarly with increasing dominance of the electrical stresses relative to the hydrodynamic stresses (quantified by $ M $), the surface tension gradient $ (|\gamma_{\text{max}}-\gamma_{\text{min}}|) $ also increases. Also the nature of surfactant distribution closely follows the direction of the applied electric field $ (\theta_e) $. 
	 
	 \item 
	 Owing to a noticeable influence of the surfactant distribution on the electrohydrodynamic flow pattern, the accumulation of surface charges is also found to be altered a lot. With an increased sensitivity of the surface tension on the surfactant concentration, the drop surface gets depleted of charges in some places, while more accumulation is caused in some other locations. 
	 
	\item 

In the leading order the longitudinal component is the only existing component of the drop velocity $ (U^{(0)}_{d,z}) $.  The correction term due to a presence of non-uniform surfactant distribution is free from any electrical effects and becomes a linear superposition to the corresponding clean drop-drop case. Also surfacants will try to retard the axial migration of the drop by reducing the magnitude of $  (U^{(0)}_{d,z}) $. 
	
	\item
	
	In the $ O(Re_E) $, both the cross stream and longitudinal components of the drop velocity $ (U_{d,x}^{(Re_E)},U_{d,z}^{(Re_E)}) $ are found to be altered in the presence of surfactants. It is found that the magnitude of $ U_{d,x}^{(Re_E)} $ always gets reduced due to Marangoni effects. The surfactant correction term in this context $ (\mathcal{F}_x^{(Re_E)})$ is again not affected by the electrical parameters. When the drop internal viscosity is too high as compared to the surrounding fluid $ (\lambda \ll 1) $, the Marangoni effect is much stronger while it diminishes as $\lambda \to \infty$. 
	
	\item
	
	The surfactant induced modification in the $ O(Re_E) $ longitudinal drop velocity $ (U_{d,z}^{(Re_E)}) $, on the other hand is dependent on the applied  electric field direction $ \theta_e $ as well. The amount of reduction in  $ (U_{d,z}^{(Re_E)}) $ is maximum if an axial electric field is applied, while is continuously less influential as the tilt angle reaches towards the lateral configuration $ \theta_e=\pi/2 $. However, since in the present case the charge convection effects itself reduces the drop velocity, surfactants act to cause an augmentation in the overall axial migration sped of a spherical drop ($U_{d,z}=U_{d,z}^{(0)}+Re_E\,U_{d,z}^{(Re_E)}$). 
	
	\item
When only the drop surface is prone to deformation,  both the cross-stream $ (U_{d,x}^{(Ca)}) $ and longitudinal $(U_{d,z}= U_{d,z}^{(0)}+Ca\,U_{d,z}^{(Ca)})$ components of the drop velocity experience a decrease in their magnitudes with directions remaining unaltered. The ratio of electrical conductivities $ (R) $ and the permittivities $ (S) $ of the drop-surrounding pair  have a key role to play in it.
		
		\item 
		
The surfactant effect on the shape-deformation triggered cross-stream migration velocity, is of maximum intensity when the tilt angle $ \theta_e $ of the applied electric field is $\pi/4$. The said effect have negligible influence as the external electric field approaches either a perfect axial $ (\theta_e=0) $ or lateral direction $ (\theta_e=\pi/2)$.
 A further enhancement in such effect results with increasing Mason number $ (M) $.		
		\item
		
The quasi-steady state drop trajectory under the simultaneous action of a background plane Poiseuielle flow and a tilted electric field, is also greatly influenced by the non-uniform distribution of surface-active molecules along the surface. The surfactant effects, quantified by an increase  in $\beta$ or $k$, can either cause the drop to move towards or away from the channel centerline depending on the particular choice of parameters $ R $ and $ S $.	Similar to some other important quantities, the surfactant-induced relative increase or decrease in the steady state final transverse position of the drop, is increasingly pronounced at higher values of the Mason number $(M)$ and near $ \theta_e=\pi/4 $.  The time required for the final steady state transverse position in the channel $ (t_S) $ is reduced in the presence of surfactant-induced Marangoni stress.

\qquad Our study reveals that adding non-ionic surfactant on the surface of a electric field actuated drop in a plane Poiseuielli flow, provides a means for selectively controlling the velocity components. Most interestingly the electrical property ratios $ (R,S) $, the Mason number $ (M) $ and the tilt angle $ (\theta_e) $ of the externally applied uniform electric field bring remarkable changes in the surfactant effects. In practice, such physical consequences can be exploited in achieving an additional degree of control in optimizing drop motion through common microfluidic devices.   
	\\ 
\end{enumerate}

\begin{appendices}
	\renewcommand{\thesection}{\Alph{section}}
	\renewcommand{\thesubsection}{\thesection.\arabic{subsection}}
	\renewcommand\thefigure{A-\arabic{figure}}    
	\setcounter{figure}{0} 
	\renewcommand{\theequation}{A-\arabic{equation}}
	\setcounter{equation}{0}  

	\section{ Harmonic constants $ (\Gamma^{(Ca)}_{n,m}) $ in \eqref{eq:leading-surf-distr}}
	\label{sec:GCa_nm}
	\begin{equation}
	\label{eq:leading-surf-harmonics}
	\begin{split}
	& \Gamma^{(Ca)}_{1,0}=-{\frac {{\it k}\, \left( -1+\beta \right) c_{{2}}}{{\it k}\,\beta-3\,\beta\,\lambda-2\,\beta+3\,\lambda+2}};\qquad \Gamma^{(Ca)}_{1,1}=0;\\
	& \Gamma^{(Ca)}_{2,0}={\frac {3M \left( {E_{{x}}}^{2}-2\,{E_{{z}}}^{2} \right)  \left( -
			1+\beta \right)  \left( R-S \right) {\it k}}{ 2\left(  \left( {\it k}-5\,\lambda-5 \right) \beta+5\,\lambda+5 \right)  \left( R+2\right) ^{2}}};\\
	& \Gamma^{(Ca)}_{2,1}=-{\frac { \left( -1+\beta \right)  \left( 18\,RME_{{x}}E_{{z}}-18\,MSE_{{x}}E_{{z}}+5\,{R}^{2}c_{{1}}+20\,Rc_{{1}}+20\,c_{{1}}\right) {\it k}}{6 \left( {\it k}\,\beta-5\,\beta\,\lambda-5\,\beta+5\,\lambda+5 \right)  \left( R+2 \right) ^{2}}};\\
	& \Gamma^{(Ca)}_{2,2}=-{\frac {3M{E_{{x}}}^{2} \left( -1+\beta \right)  \left( R-S\right) {\it k}}{4 \left(  \left( {\it k}-5\,\lambda-5 \right) \beta+5\,\lambda+5 \right)  \left( R+2 \right) ^{2}}};\\
	&\Gamma^{(Ca)}_{3,0}={\frac {7{\it k}\, \left( -1+\beta \right) c_{{2}}}{ \left( 12\,{\it k}-84\,\lambda-84 \right) \beta+84\,\lambda+84}};\qquad \Gamma^{(Ca)}_{3,1}=0;\\
	&\Gamma^{(Ca)}_{3,2}=-{\frac {7{\it k}\, \left( -1+\beta \right) c_{{2}}}{ \left( 72\,{\it k}-504\,\lambda-504 \right) \beta+504\,\lambda+504}} \quad \text{and} \quad \Gamma^{(Ca)}_{3,3}=0.
	\end{split}
	\end{equation}

	\section{Harmonic constants $( L^{(Ca)}_{n,m}) $  in \eqref{eq:leading-shape}}
	\label{sec:LCa_nm}
	\begin{equation}
	\label{eq:sol-leading-shape}
	\begin{split}
	& L^{(Ca)}_{2,0}=L^{(Ca)}_{2,0}\big{|}_\text{Clean}+{\frac {3\,M \left( {2E_{{z}}}^{2}-\,{E_{{x}}}^{2} \right) {\it k}\,
			\beta\, \left( \lambda+4 \right)  \left( R-S \right) }{ 40\left(\,
			\lambda+1 \right)  \left(  \left( -5\,\beta+5 \right) \lambda+5+
			\left( {\it k}-5 \right) \beta \right)  \left( R+2 \right) ^{2}}};\\
	& L^{(Ca)}_{2,1}=L^{(Ca)}_{2,1}\big{|}_\text{Clean}+ {\frac {3\,{\it k}\,\beta\, \left( \lambda+4 \right) \left( {\dfrac {5\, \left( R+2 \right) ^{2}c_{{1}}}{18}}+ME_{{x}}E_{{z
			}} \left( R-S \right)  \right)}{ \left( 20\,
			\lambda+20 \right)  \left(  \left( -5\,\beta+5 \right) \lambda+5+
			\left( {\it k}-5 \right) \beta \right)  \left( R+2 \right) ^{2}}
	};\\
	& L^{(Ca)}_{2,2}=L^{(Ca)}_{2,2}\big{|}_\text{Clean}+{\frac {3\,M{E_{{x}}}^{2}{\it k}\,\beta\, \left( \lambda+4 \right) 
			\left( R-S \right) }{80\, \left( R+2 \right) ^{2} \left( \lambda+1
			\right)  \left(  \left( -5\,\beta+5 \right) \lambda+5+ \left( {\it 
				k}-5 \right) \beta \right) }};\\
	&L^{(Ca)}_{3,0}=L^{(Ca)}_{3,0}\big{|}_\text{Clean}-{\frac {{\it k}\,\beta\,c_{{2}} \left( 2\,\lambda+5 \right) }{\left( 120\,\lambda+120 \right)  \left(  \left( -7\,\beta+7 \right) 
			\lambda+7+ \left( {\it k}-7 \right) \beta \right) }}
	;\\
	& \text{and} \quad L^{(Ca)}_{3,2}=L^{(Ca)}_{3,2}\big{|}_\text{Clean}+{\frac {{\it k}\,\beta\,c_{{2}} \left( 2\,\lambda+5 \right) }{
			\left( 720\,\lambda+720 \right)  \left(  \left( -7\,\beta+7 \right) 
			\lambda+7+ \left( {\it k}-7 \right) \beta \right) }},
	\end{split}
	\end{equation}
	where in each case the first term describes the same harmonic for a surfactant-free drop $ (\beta =0) $ and the second term stands for the correction due to surfactant.
		\section{Justification of the quasi-steady state assumption}
	\label{sec:quasi}
	The various important time scales as mentioned in section \ref{sc:trajectory} can be written as follows - \\
	 (i) viscous time scale or the time scale for the velocity field to establish: $ t_v=\dfrac{\rho_e a^2}{\mu_e} $;
	 (ii) charge relaxation time scale or the time required for the charge distribution to establish: $ t_e=\dfrac{\epsilon_e}{\sigma_e}$; 
	 (iii) time scale for drop shape deformation: $t_{D}=\dfrac{\mu_e a}{\gamma_{eq}}$; 
	 (iv) time scale for the diffusion of the surfactant molecules: $ t_{sd}=\dfrac{a^2}{D_S} $; the time scales for (v)  the drop motion in the longitudinal direction: $ t_{z}=\dfrac{a}{\widetilde{V}_0\,U_{d,z}} $ and (vi) in the cross-stream direction: $t_{x}=\dfrac{a}{\widetilde{V}_0\,U_{d,x}}$.

	Now considering the relevant mechanisms for drop migration in the cross-stream direction we get, $ U_{d,x} \sim Re_E M \text{ or } Ca M$ and $U_{d,z} \sim 1$. Thus $ t_v $ can be compared to $ t_z $  or $ t_x $ as $ t_v \ll t_z \text{ or } Re \ll 1 $ and $ t_v \ll t_x \text{ or } (Re Re_E M, Re Ca M) \ll 1 $. Among these the condition $ Re Re_E M, Re Ca M \ll 1 $ is more restrictive in the sense that the cross- stream motion of the drop is the main source of the unsteady characteristics. Along similar lines the $ t_e $ can be compared to $ t_z,t_x $ as $ t_e \ll t_z \text{ or } Re_E \ll 1 $ and $ t_e \ll t_x \text{ or } (Re_E^2 M, Re_E Ca M) \ll 1 $ where the criteria $ (Re_E^2 M, Re_E Ca M) \ll 1 $ is more stringent. Also for the drop deformation and surfactant distribution to take place instantaneously with respect to the drop motion time scale we obtain the most stringent criterion as:
    $ t_{D} \ll t_x \text{ or } (Ca^2\,M, Ca\,Re_E\,M \ll 1)  $ and $ t_{sd} \ll t_x \text{ or } (Pe\,Ca\,M, Pe\,Re_E\,M \ll 1) $. Considering the domain of the present analysis ($Re \ll 1, Ca \ll 1, Re_E \ll 1, Pe \ll 1 $ and $ M \sim 1 $) it is observed that all the above criterion are fully satisfied thus making the quasi-steady state assumption applicable in the present scenario.
		\section{Model validation}
	\label{sec:Validation}
	
	In the present work we have considered a two dimensional Poiseuielle profile as the imposed flow.  Under a similar condition \citet{Chan1979} found the existence of cross stream velocity of a drop initially placed at an off-centerline position in the absence of both electric field and surfactant coating. Their expressions of drop velocity components for the case of a Newtonian fluid, in the leading order as well as $ O(Ca) $ and the $ O(Ca) $ shape function ($ f^{(Ca)} $), can be reached if we substitute $ M=0 $ and $ \beta=0 $ in our results. On the other hand the expressions of all the important physical quantities in \citet{Mandal2016}, in all the three orders of perturbation, can be restored if only a $ \beta =0 $ substitution is made. Such a  substitution, in effect,  leads to the characteristics of a clean drop suspended in a plane Poiseuielle flow and acted upon by a tilted electric field. 
	
	In the simplified case of clean deformable drop in the absence of electric field, the present calculations can be easily extended to a more complicated case of a 3-D Poiseuielle flow \citep{Chan1979} given as
	\begin{equation}\label{3-D Poiseuielle flow}
	\widetilde{\boldsymbol{V}}_\infty=\widetilde{V}_0\left (c'_0+c'_1\,x +c'_2\left( x^2 + y^2 \right)  \right )\mathbf{i}_z
	\end{equation}
	or, equivalently a circular Poiseuielle flow  in the form  
	\begin{equation}
	\frac{\widetilde{V}_\infty}{\widetilde{U}_b}=\left [1-\left (\frac{r}{R_0}  \right )^{2}\sin^2\theta-\left (\frac{b}{R_0}  \right )^{2}-\frac{2rb}{R^2_0}\sin\theta \cos\phi  \right ]\mathbf{i}_z.
	\end{equation}
	In the above equations $ c'_0=1-\dfrac{b^2}{R^2_0}\; $, $ c'_1=-\dfrac{2 a b}{R^2_0}\; $, $ c'_2=-\dfrac{a^2}{R^2_0}\; $, $ \widetilde {U}_b $ is the characteristic flow speed at a dimensionless distance $ b $ from drop centroid and $ R_0 $ is the dimensionless radial distance from channel centerline where the imposed velocity becomes zero. After making such modifications in the imposed flow, we obtain the deformation induced cross-stream migration velocity as 
	\begin{equation}\label{3d-Udx}
	U^{(Ca)}_{d,x} = -{\frac {c'_1\,c'_2}{ \left( 1+\lambda \right) ^{2} \left( 2+3\lambda \right) } \left({\frac { \left( 16+19\lambda \right) \left( -24{\lambda}^{3}-73{\lambda}^{2}-36\lambda+13 \right) }{42\left( 2+3\lambda \right)  \left( 4+\lambda \right) }}+{\frac {\left( 10+11\lambda \right)  \left( 3{\lambda}^{2}-\lambda+8\right) }{105}} \right) },
	\end{equation} which matches exactly with the calculation of \citet{Chan1979}. 
	In the next step when the surfactant effects are considered, our resulting expressions again match with those of \citet{Das2017a}. 
	
	\section{Behaviour of interface velocity and electrical traction vector components}
	\label{sec:us_FE}
	\begin{figure}
		\centering
		\begin{subfigure}[!htb]{0.45\textwidth}
			\centering
			\includegraphics[width=1.1\textwidth]{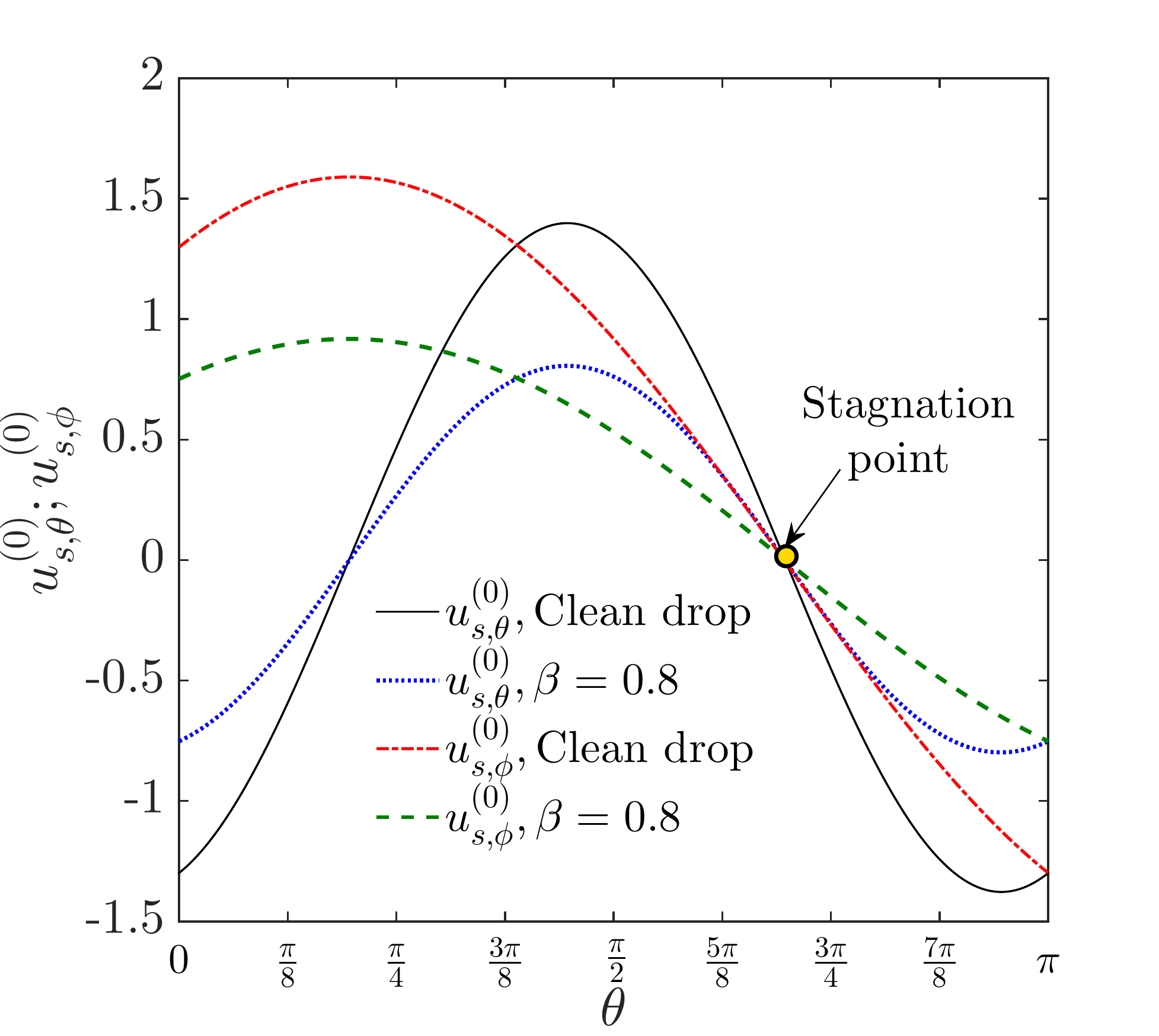}		
			\vspace{8ex}
			\caption{}
			\label{fig:us0_components_vs_theta_vary_beta_phi_pi_by_four}
		\end{subfigure}
		\quad 
		\begin{subfigure}[!htb]{0.45\textwidth}
			\centering
			\includegraphics[width=1.1\textwidth]{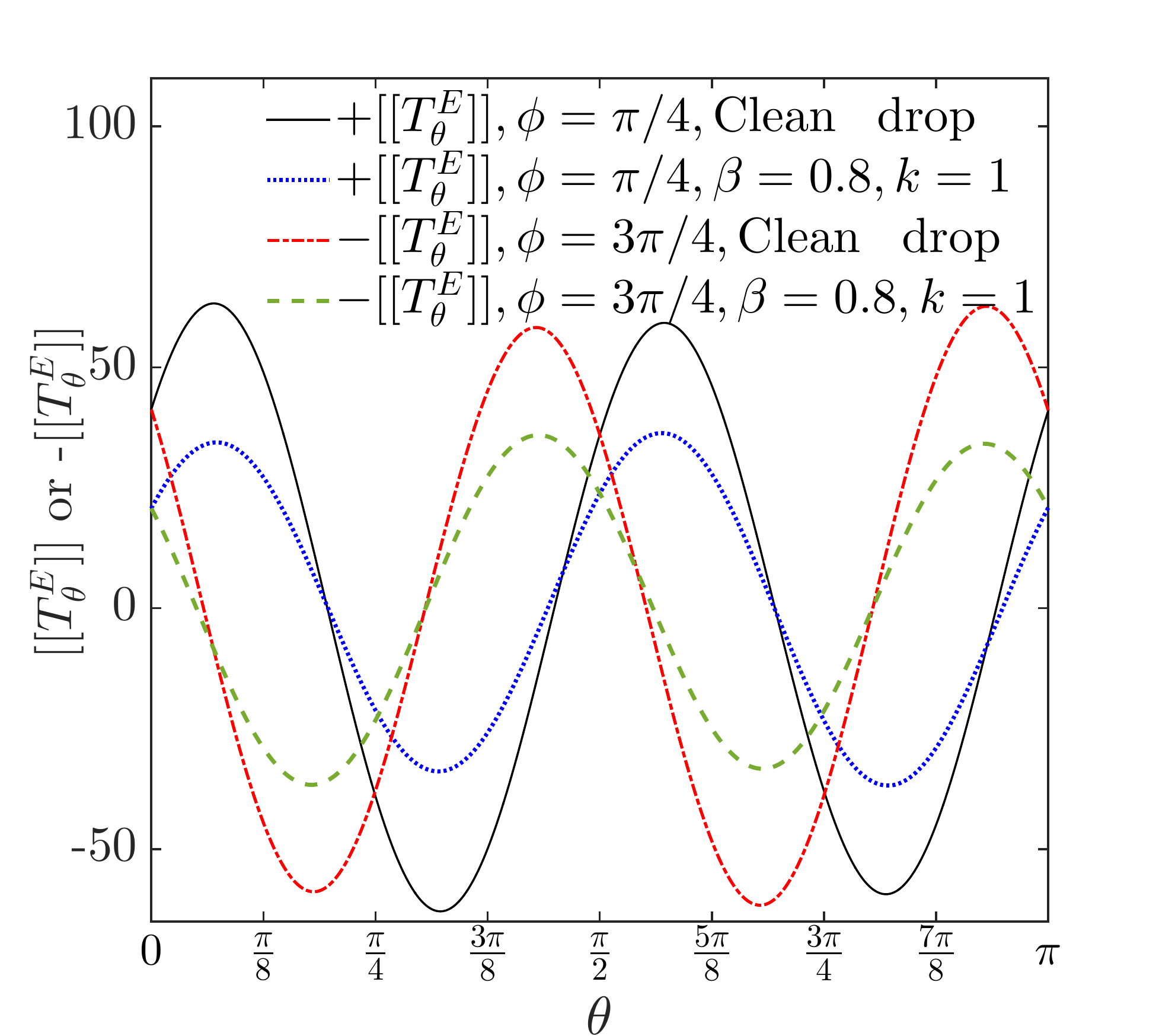}
			\vspace{8ex}
			\caption{}
			\label{fig:FEt_vs_theta_vary_beta}
		\end{subfigure}
		\caption{Subplot (a): Effect of surfactant on the variation of leading order surface velocity components $ (u^{(0)}_{S,\theta},u^{(0)}_{S,\phi}) $ in the polar direction $ (\theta) $. Subplot (b): Variation of the polar component of the electric traction vector ($ [[T^{E}_\theta]]=[[T^{E,0}_{\theta}]]+Re_E\,[[T^{E,Re_E}_{\theta}]] $) for different $ \beta $. In both the subplots $ \theta_e=\pi/4, \phi=\pi/4, k=1, Re_E=0.2, M=1, R=0.1, S=10, x_d=5 $ and $ H=10 $. In  subplot (b) the negative of $ [[T^{E}_\theta]] $ is shown for $ \phi=3\pi/4$. The extra negative sign is used to check for any antisymmetry about the axial plane.}
	\end{figure}
\begin{figure}[!htb]	
		\centering
		\includegraphics[width=1.1\textwidth]{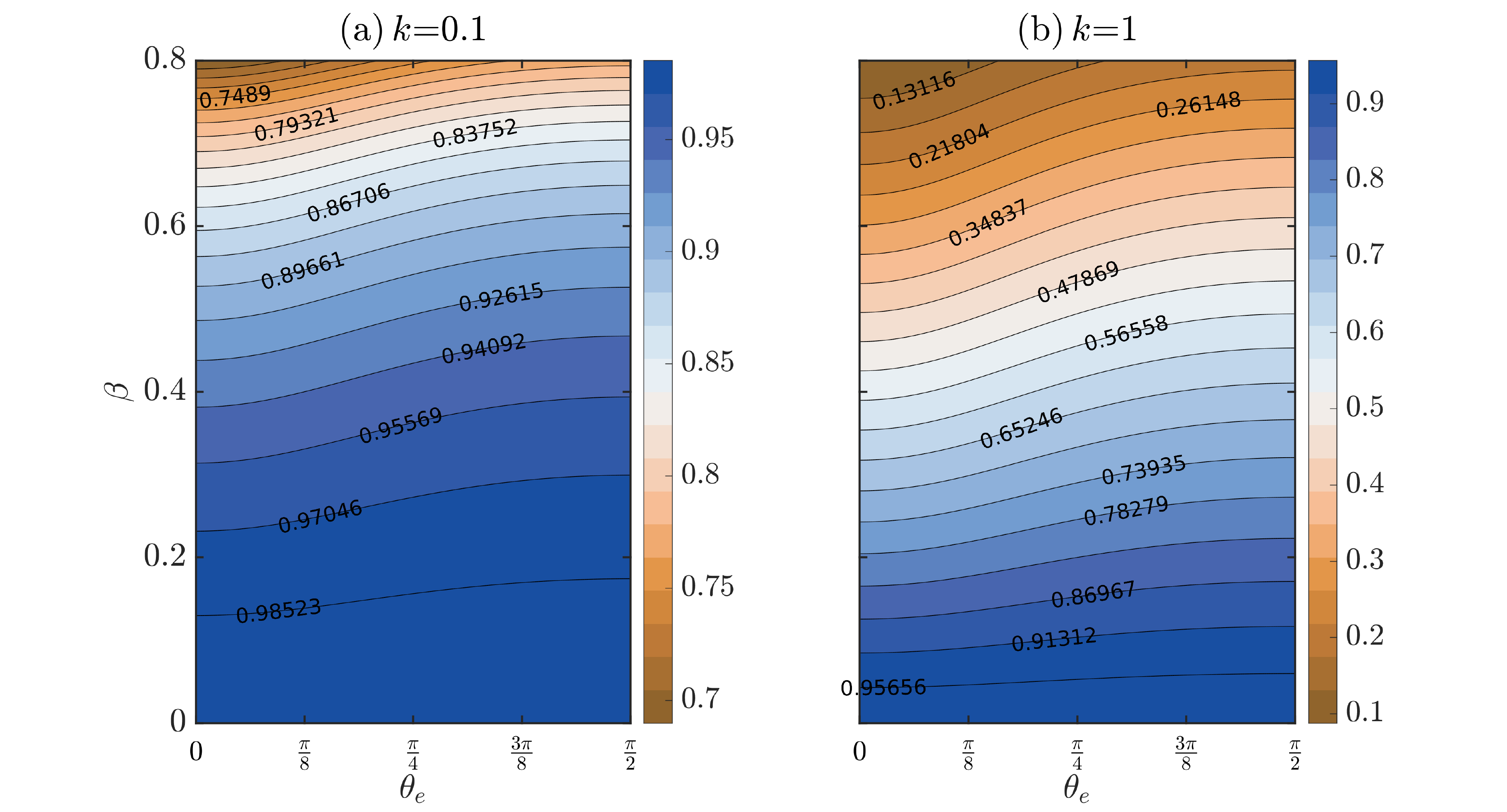}
		\caption{Variation of the correction parameter $ \mathcal{F}^{(Re_E)}_z $ on $ \theta_e - \beta $ plane for three different values of $ k $ and viscosity ratio, $ \lambda = 0.1 $.  }
		\label{fig:correction-fact-ReE-z}
\end{figure}
	\begin{figure} \label{fig:explain-deform}
		\centering
		\begin{subfigure}[!htb]{0.45\textwidth}
			\centering
			\includegraphics[width=1.09\textwidth]{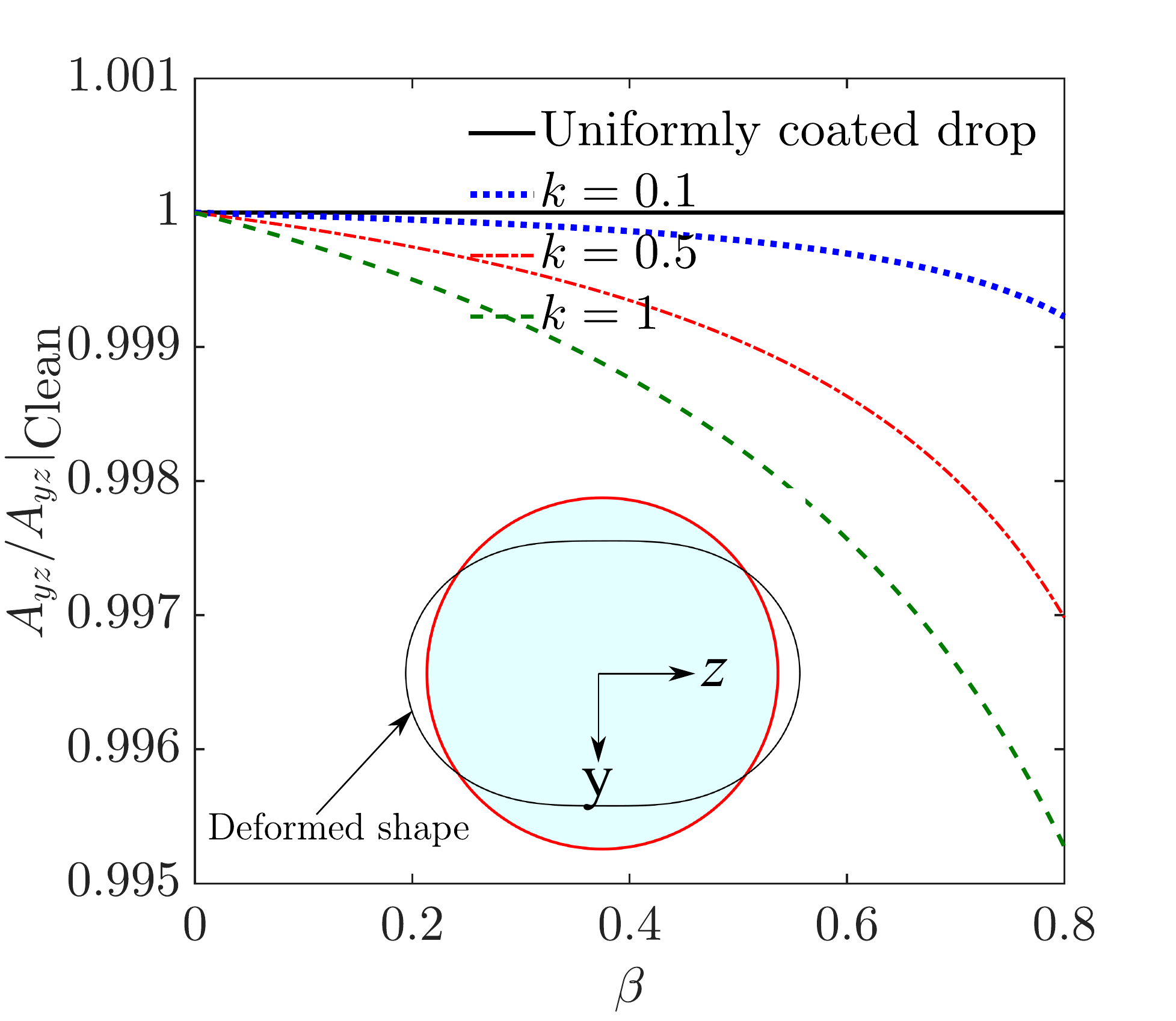}		
			\vspace{12ex}
			\caption{}
			\label{fig:A_YZ_vary_beta_K_R_5_S_0p5_M_0p_five}
		\end{subfigure}
		\begin{subfigure}[!htb]{0.45\textwidth}
			\centering
			\includegraphics[width=1.1\textwidth]{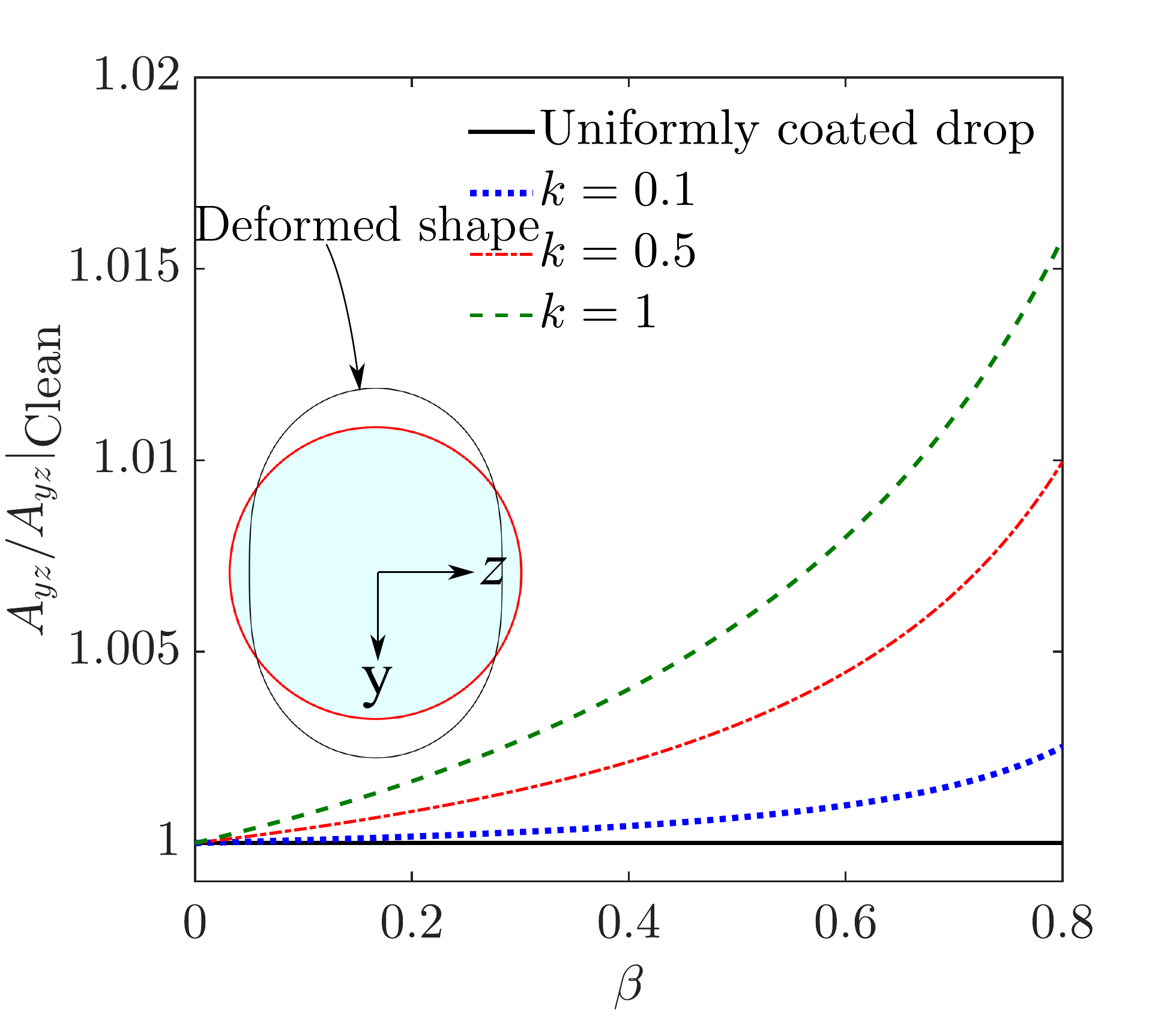}
			\vspace{12ex}
			\caption{}
			\label{fig:A_YZ_vary_beta_K_R_0p1_S_10_M_0p_five}
		\end{subfigure}
		\caption{In subplots (a) and (b) the variation of the drop cross sectional area perpendicular to the cross-stream motion $ (A_{yz}) $ is shown for different values of $ \beta $ and $ k $. Schematic representation of deformed drop shape for the corresponding cases are provided as insets in each subplot.  Subplots (c) and (d) depict the variations in the polar component of the electrical traction vector with $ \theta $ for $ k=1 $ and various $ \beta $.  In subplots (a),(c) the parameters are taken same as in figure \ref{fig:UDx_CA_lambda_0p1_R_5_S_0p5_M_0p_one} and in subplots (b),(d) the parameters are same as figure \ref{fig:UDx_CA_lambda_0p1_R_0p1_S_10_M_0p_one}.} 
	\end{figure}

\begin{figure} 
	\centering
	\begin{subfigure}[!htb]{0.45\textwidth}
		\centering
		\includegraphics[width=1.08\textwidth]{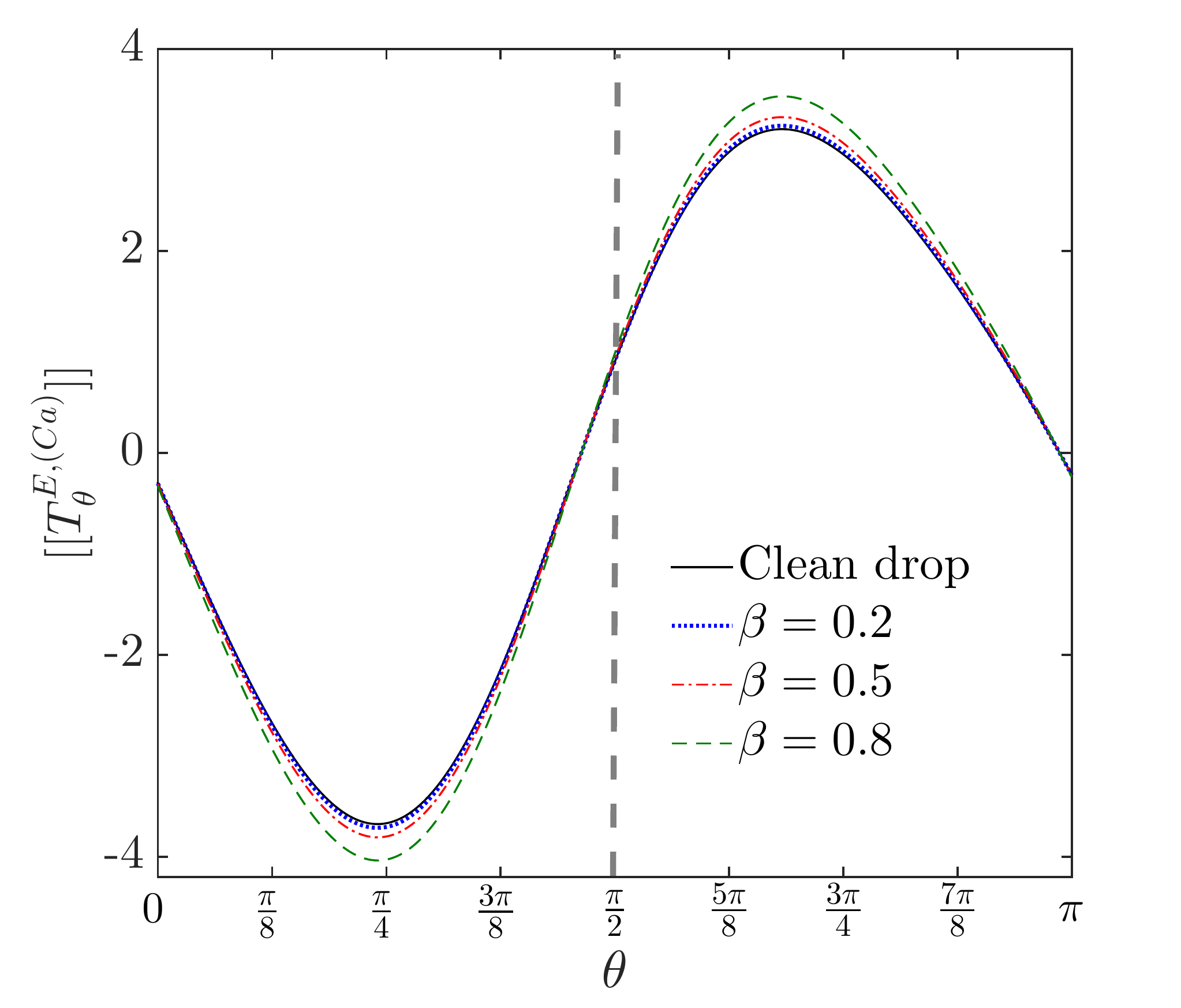}
		\vspace{11.5ex}
		\caption{}
		\label{fig:FeT_vs_theta_R_0p1_S_10_M_0p_one}
	\end{subfigure}
	\begin{subfigure}[!htb]{0.45\textwidth}
		\centering
		\vspace{-1.5ex}
		\includegraphics[width=1.05\textwidth]{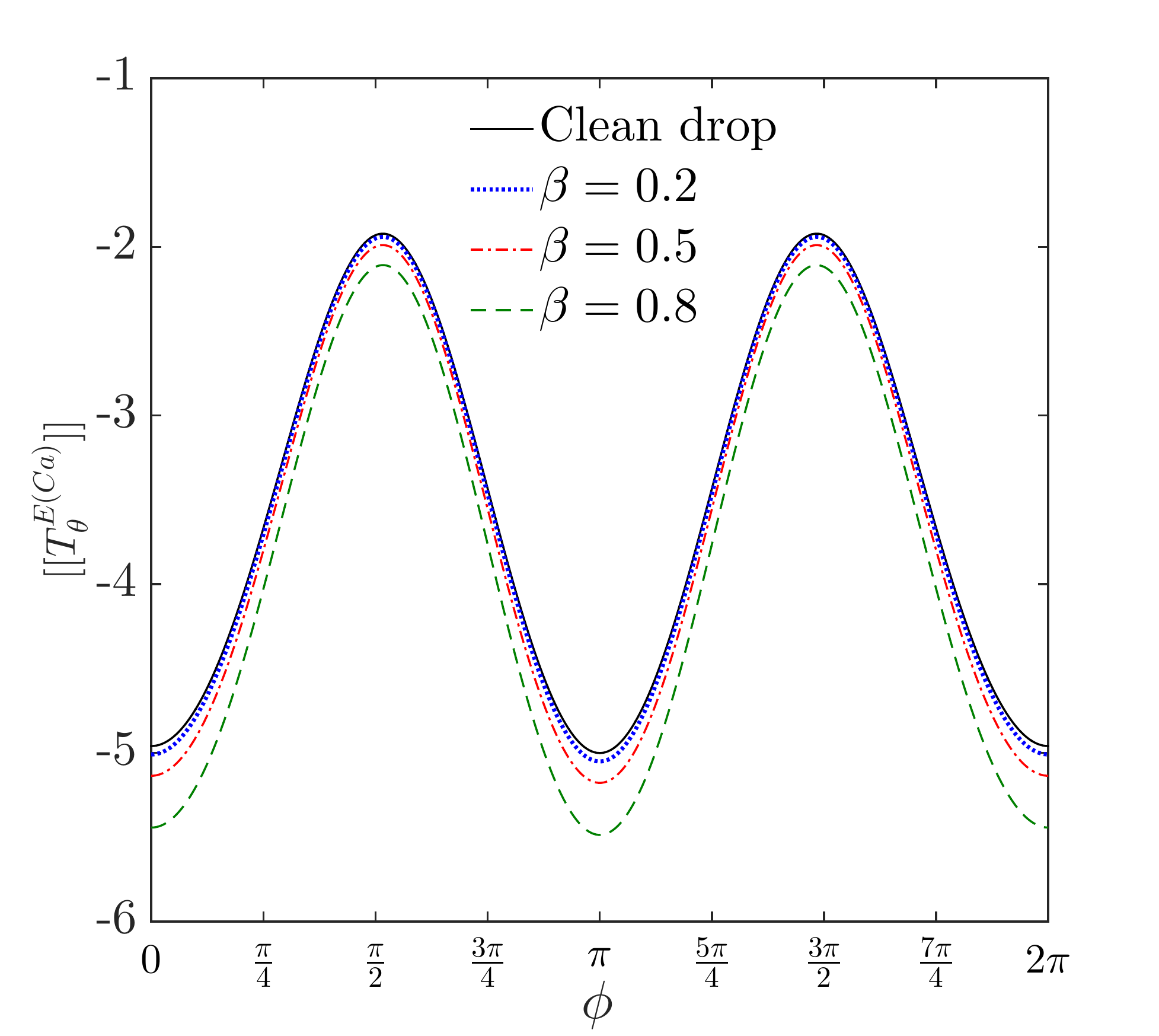}
		\vspace{12ex}
		\caption{}
		\label{fig:FeT_theta__vs_phi_R_0p1_S_10_M_0p_one}
	\end{subfigure}
	\\
	\begin{subfigure}[!htb]{0.45\textwidth}
		\centering
		\includegraphics[width=1.1\textwidth]{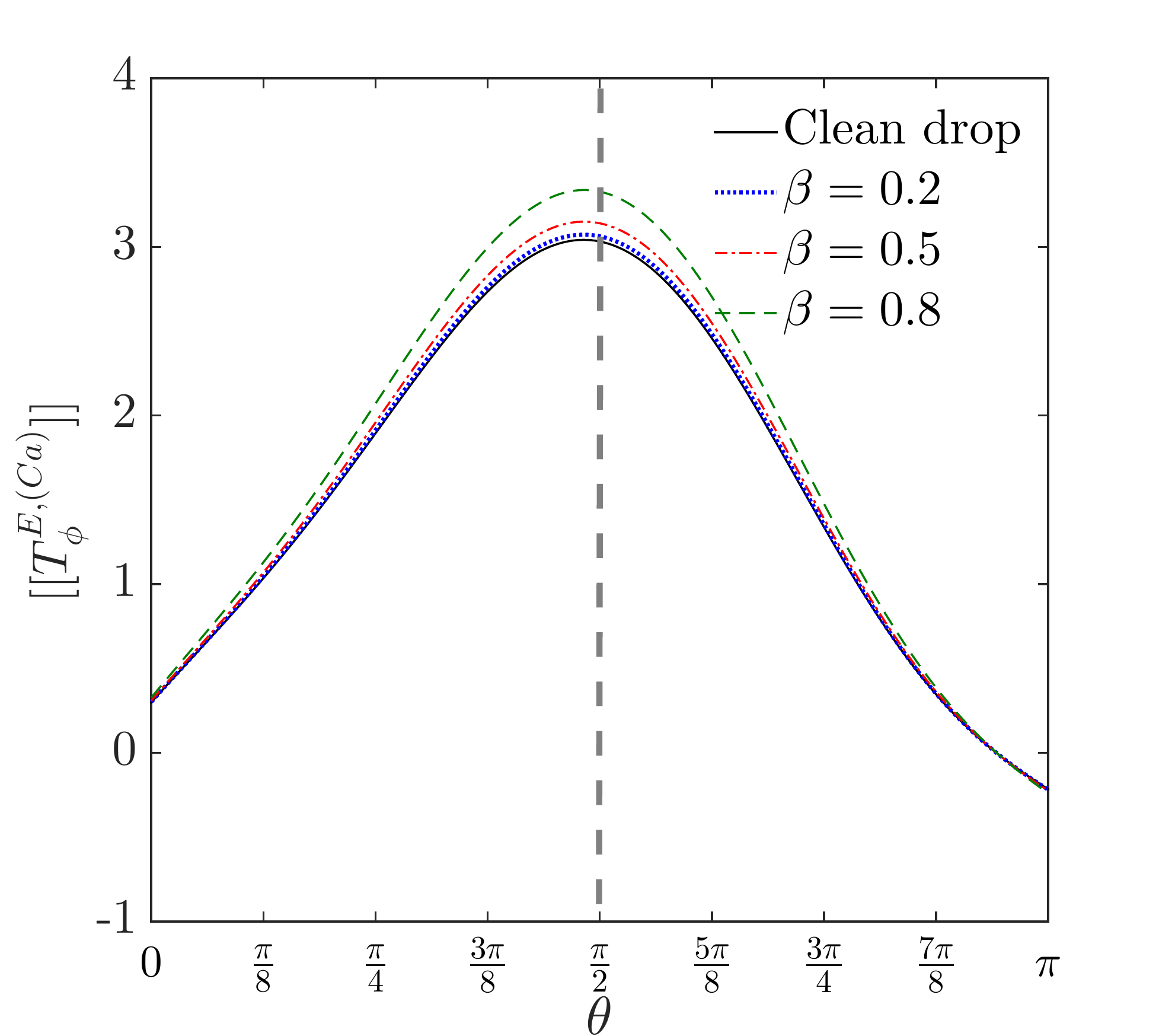}
		\vspace{11ex}
		\caption{}
		\label{fig:FeT_PHI__vs_theta_R_0p1_S_10_M_0p_one}
	\end{subfigure}
	\begin{subfigure}[!htb]{0.45\textwidth}
		\centering
		\includegraphics[width=1.08\textwidth]{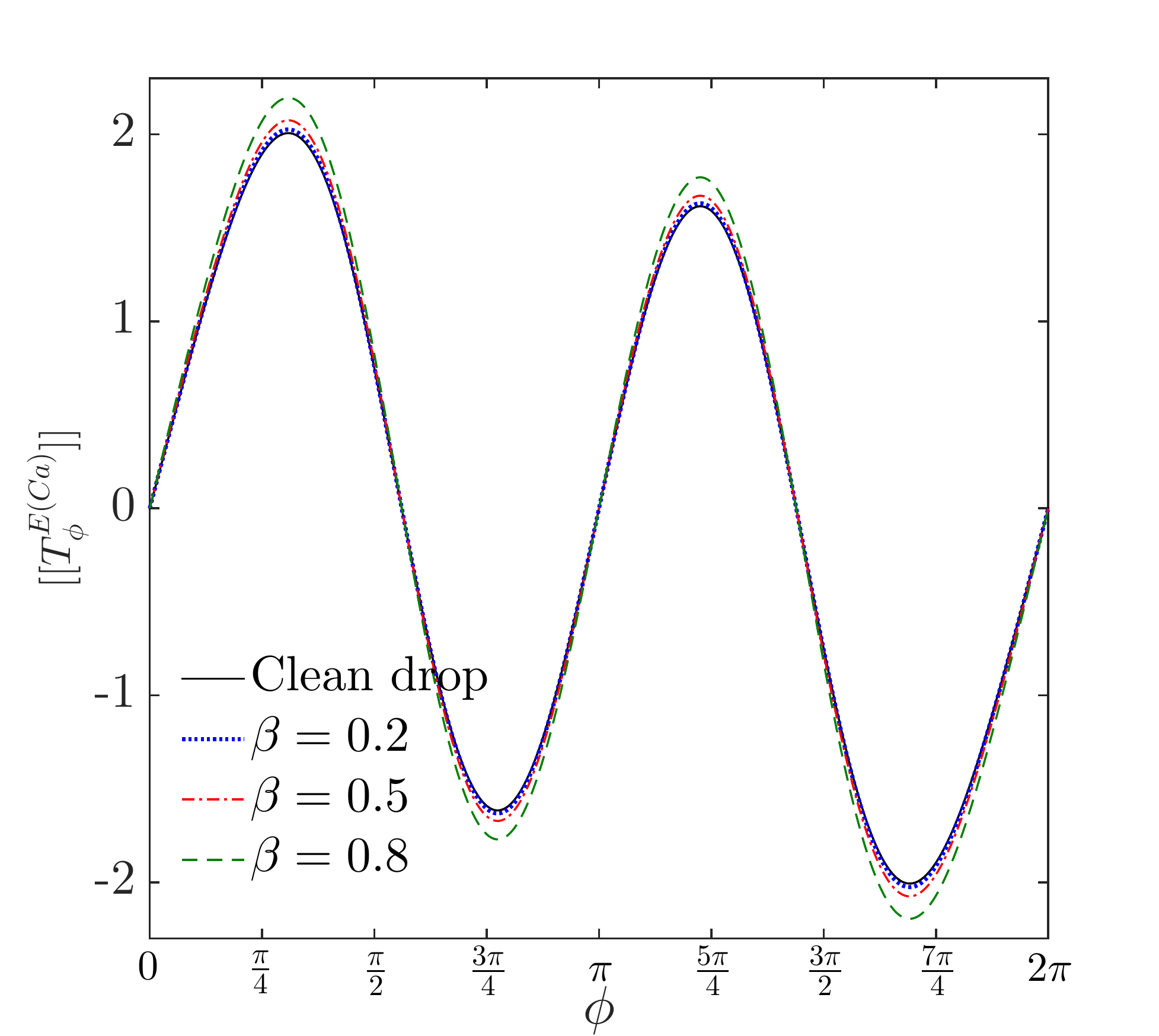}
		\vspace{12ex}
		\caption{}
		\label{fig:FeT_PHI__vs_phi_R_0p1_S_10_M_0p_one}
	\end{subfigure}
	\caption{Subplot (a): Polar component of the $ O(Ca) $ electrical traction force $ [[T^{E(Ca)}_\theta]] $ vs polar angle $ \theta $ at $ \phi=\pi/4 $;  (b) Polar component of the $ O(Ca) $ electrical traction force $ [[T^{E(Ca)}_\theta]] $ vs azimuthal angle $ \phi $ at $ \theta=\pi/4 $; (c) Azimuthal component of the $ O(Ca) $ electrical traction force $ [[T^{E(Ca)}_\phi]] $ vs polar angle $ \theta $ at $ \phi=\pi/4 $; (d) Azimuthal component of the $ O(Ca) $ electrical traction force $ [[T^{E(Ca)}_\phi]] $ vs azimuthal angle $ \phi $ at $ \theta=\pi/4 $. In all the subplots $ k=1 $ is used. Other parameters are taken same as in figure  \ref{fig:UDx_CA_lambda_0p1_R_0p1_S_10_M_0p_one}.} 
	\label{fig:explain-tangential}
\end{figure}
\begin{figure}[!htb]	
		\centering
		\includegraphics[width=0.55\textwidth]{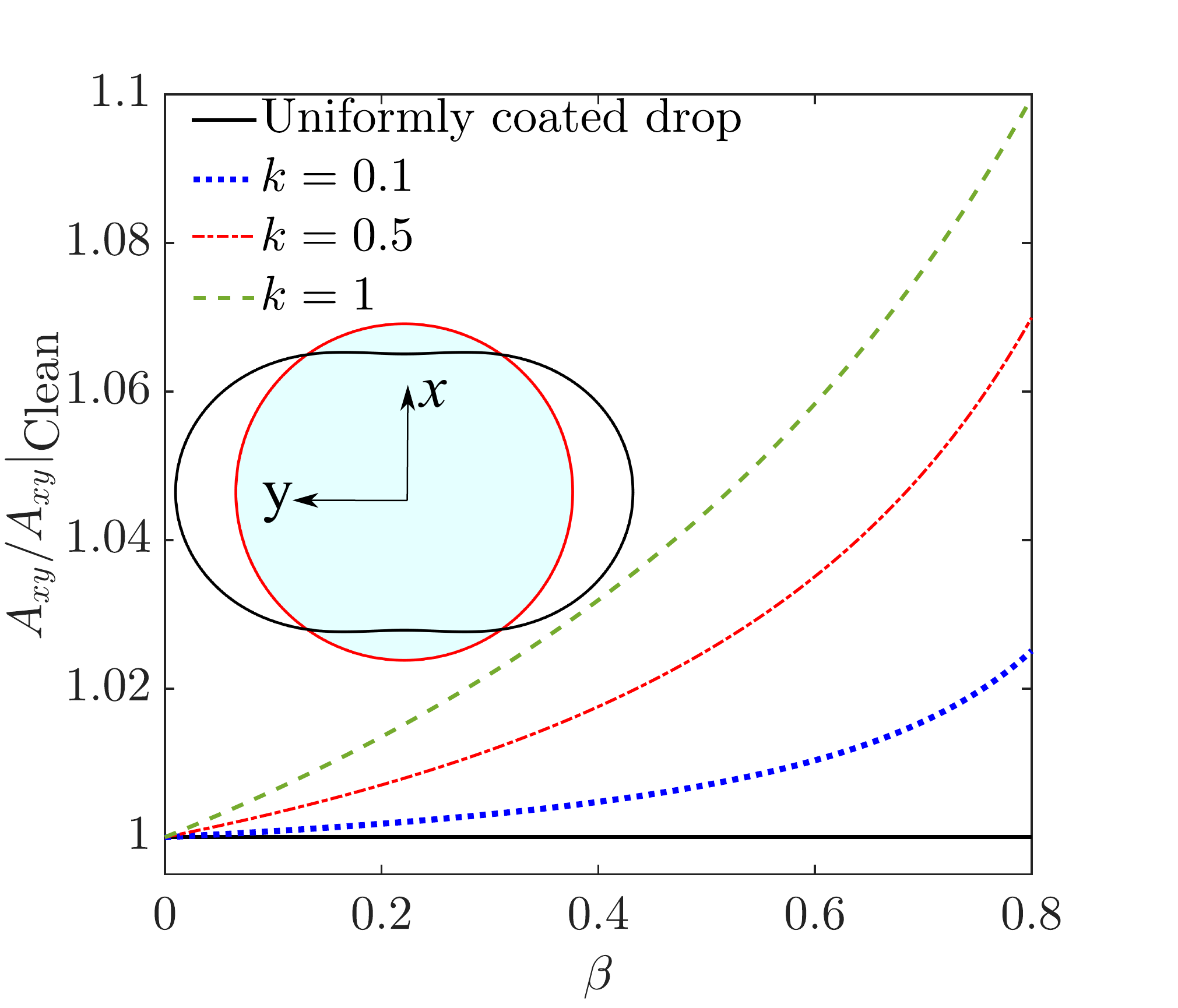}
		\caption{Variation in the relative projected area of the drop on the $ x-y $ plane. Parameters are same as in figure \ref{fig:UDZ-CA}. Schematic representation of the deformed drop shape for the choice of parameters are provided as an inset.}
		\label{fig:A_XY_vary_beta_K_R_0p1_S_20}
\end{figure}
\begin{figure}
		\centering
		\begin{subfigure}[!htb]{0.45\textwidth}
			\centering
			\includegraphics[width=1.1\textwidth]{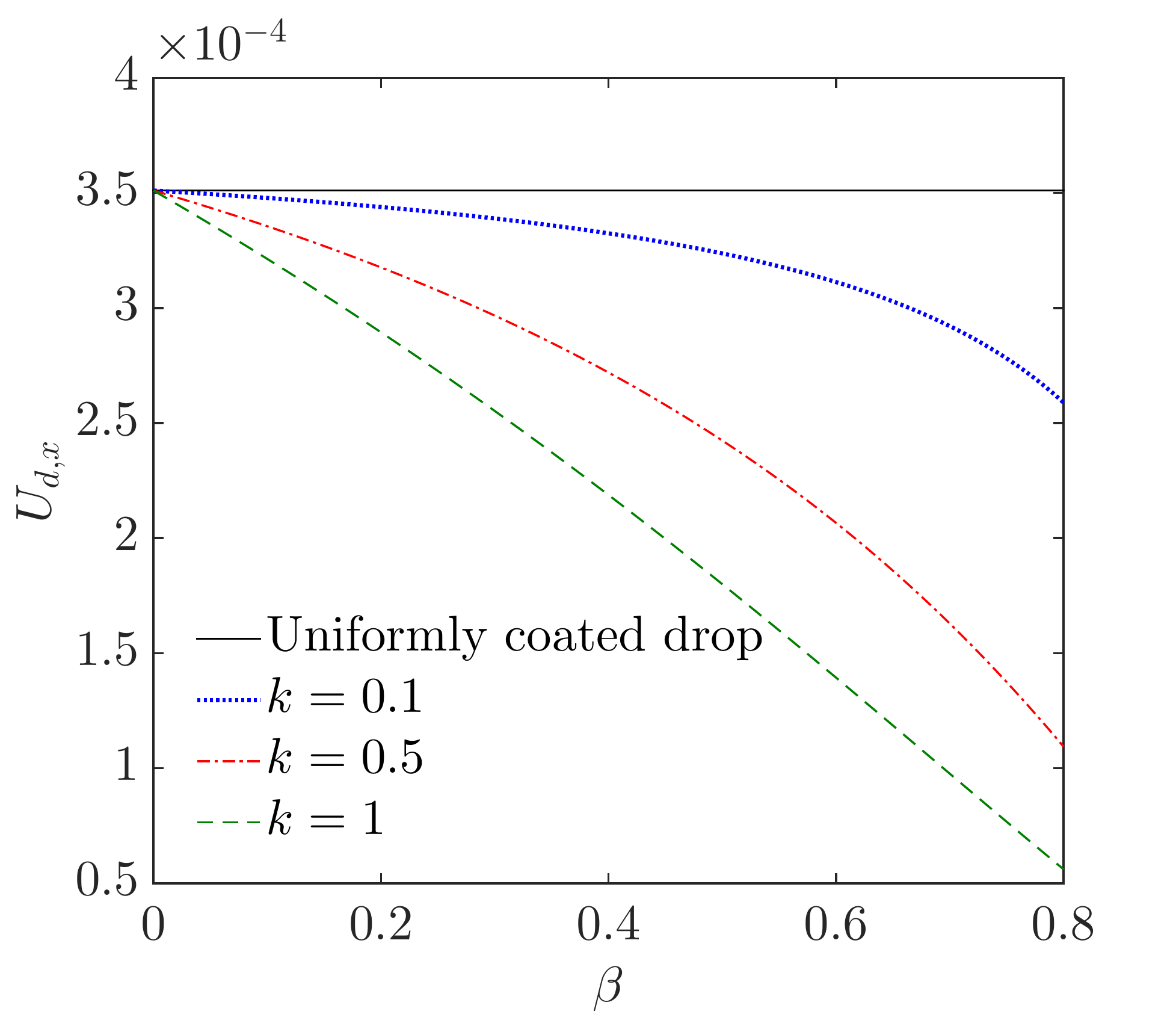}		
			\vspace{8ex}
			\caption{}
			\label{fig:UDX_total_5_0p_five}
		\end{subfigure}
		\quad 
		\begin{subfigure}[!htb]{0.45\textwidth}
			\centering
			\includegraphics[width=1.1\textwidth]{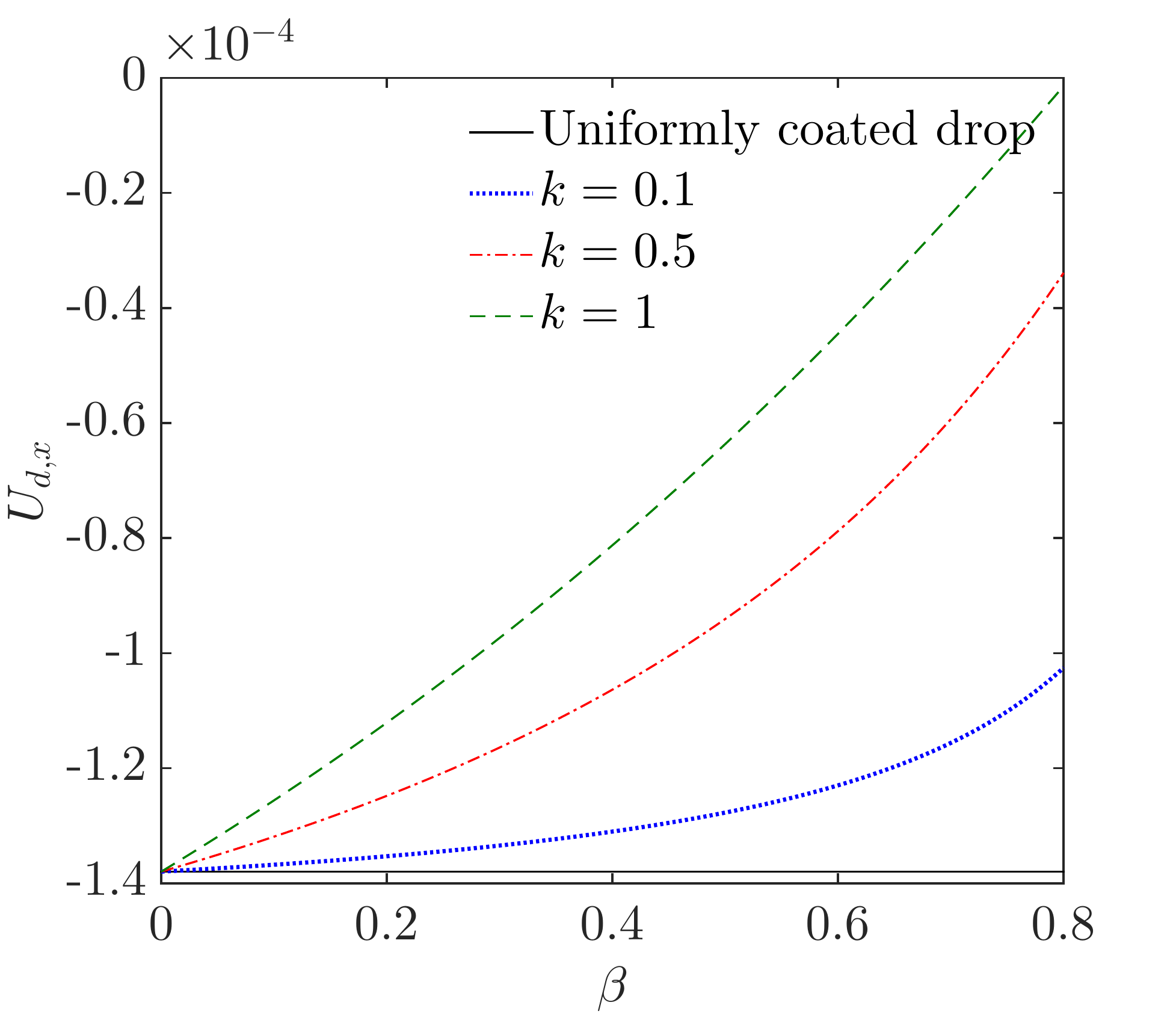}
			\vspace{8ex}
			\caption{}
			\label{fig:UDX_total_0p_five_five}
		\end{subfigure}
		\caption{Effect of surfactant on the total cross stream velocity defined as $ U_{d,x}=Re_E\,U_{d,x}^{(Re_E)}+Ca\,U_{d,x}^{(Ca)} $ variation of leading order surface velocity components with $ \beta$ and $ k $. The parameters for subplots (a) and (b) are taken similar to the figures \ref{fig:Trajectory_R_5_S_0p_five_M_p_five}
			and	\ref{fig:Trajectory_R_0p_5_S_5_M_p_one}, respectively.  } 
	\end{figure}
	\FloatBarrier
\end{appendices}


\bigskip
\newpage
\begin{thebibliography}{65}
	\expandafter\ifx\csname natexlab\endcsname\relax\def\natexlab#1{#1}\fi
	\def\au#1{#1} \def\ed#1{#1} \def\yr#1{#1}\def\at#1{#1}\def\jt#1{\textit{#1}}
	\def\bt#1{#1}\def\bvol#1{\textbf{#1}} \def\vol#1{#1} \def\pg#1{#1}
	\def\publ#1{#1}\def\arxiv#1{#1}\def\org#1{#1}\def\st#1{\textit{#1}}
	
	\bibitem[Ajayi(1978)]{Ajayi1978}
	{\sc \au{Ajayi, OO}} \yr{1978} A note on taylor's electrohydrodynamic theory.
	\bt{In {\em Proceedings of the Royal Society of London A: Mathematical,
			Physical and Engineering Sciences\/}}, ,  \vol{vol. 364},  \pg{pp. 499--507}.
	The Royal Society.
	
	\bibitem[Anna(2016)]{Anna2016}
	{\sc \au{Anna, Shelley~Lynn}} \yr{2016}  \at{{Droplets and Bubbles in
			Microfluidic Devices}}.  \jt{Annual Review of Fluid Mechanics}
	\bvol{48}~(1),  \pg{285--309}.
	
	\bibitem[Bandopadhyay {\em et~al.\/}(2016{\natexlab{{\em a\/}}})Bandopadhyay,
	Mandal, Kishore \& Chakraborty]{Bandopadhyay2016}
	{\sc \au{Bandopadhyay, Aditya}, \au{Mandal, Shubhadeep}, \au{Kishore, N.K.~K.}
		\& \au{Chakraborty, Suman}} \yr{2016{\natexlab{{\em a\/}}}}  \at{{Uniform
			electric-field-induced lateral migration of a sedimenting drop}}.
	\jt{Journal of Fluid Mechanics}  \bvol{792}~(2016),  \pg{553--589}.
	
	\bibitem[Bandopadhyay {\em et~al.\/}(2016{\natexlab{{\em b\/}}})Bandopadhyay,
	Mandal, Kishore \& Chakraborty]{Bandopadhyay2016a}
	{\sc \au{Bandopadhyay, Aditya}, \au{Mandal, Shubhadeep}, \au{Kishore, N.~K.} \&
		\au{Chakraborty, Suman}} \yr{2016{\natexlab{{\em b\/}}}}  \at{Uniform
		electric-field-induced lateral migration of a sedimenting drop}.  \jt{Journal
		of Fluid Mechanics}  \bvol{792},  \pg{553–589}.
	
	\bibitem[Banerjee {\em et~al.\/}(2012)Banerjee, Kreit, Liu, Heikenfeld \&
	Papautsky]{Banerjee2012}
	{\sc \au{Banerjee, Ananda}, \au{Kreit, Eric}, \au{Liu, Yuguang},
		\au{Heikenfeld, Jason} \& \au{Papautsky, Ian}} \yr{2012}  \at{Reconfigurable
		virtual electrowetting channels}.  \jt{Lab on a Chip}  \bvol{12}~(4),
	\pg{758--764}.
	
	\bibitem[Baret(2011)]{Baret2011}
	{\sc \au{Baret, J~C}} \yr{2011}  \at{{Surfactants in droplet-based
			microfluidics.}}  \jt{Lab on a Chip} ~(422),  \pg{422--433}.
	
	\bibitem[Baroud {\em et~al.\/}(2010)Baroud, Gallaire \& Dangla]{Baroud2010}
	{\sc \au{Baroud, Charles~N.}, \au{Gallaire, Francois} \& \au{Dangla,
			R{\'{e}}mi}} \yr{2010}  \at{{Dynamics of microfluidic droplets}}.  \jt{Lab on
		a Chip}  \bvol{10}~(16),  \pg{2032}.
	
	\bibitem[Basaran(2002)]{Basaran2002}
	{\sc \au{Basaran, Osman~A}} \yr{2002}  \at{Small-scale free surface flows with
		breakup: Drop formation and emerging applications}.  \jt{AIChE Journal}
	\bvol{48}~(9),  \pg{1842--1848}.
	
	\bibitem[Chan \& Leal(1979)]{Chan1979}
	{\sc \au{Chan, P. C.-H.} \& \au{Leal, L.~G.}} \yr{1979}  \at{{The motion of a
			deformable drop in a second-order fluid}}.  \jt{Journal of Fluid Mechanics}
	\bvol{92}~(01),  \pg{131----170}.
	
	\bibitem[Chen {\em et~al.\/}(2014)Chen, He, Bertola \& Wang]{Chen2014}
	{\sc \au{Chen, Simeng}, \au{He, Xinting}, \au{Bertola, Volfango} \& \au{Wang,
			Moran}} \yr{2014}  \at{{Electro-osmosis of non-Newtonian fluids in porous
			media using lattice Poisson-Boltzmann method}}.  \jt{Journal of Colloid and
		Interface Science}  \bvol{436},  \pg{186--193}.
	
	\bibitem[Das {\em et~al.\/}(2017{\natexlab{{\em a\/}}})Das, Mandal \&
	Chakraborty]{Das2017a}
	{\sc \au{Das, Sayan}, \au{Mandal, Shubhadeep} \& \au{Chakraborty, Suman}}
	\yr{2017{\natexlab{{\em a\/}}}}  \at{Cross-stream migration of a
		surfactant-laden deformable droplet in a poiseuille flow}.  \jt{Physics of
		Fluids}  \bvol{29}~(8),  \pg{082004}.
	
	\bibitem[Das {\em et~al.\/}(2017{\natexlab{{\em b\/}}})Das, Mandal, Som \&
	Chakraborty]{Das2017b}
	{\sc \au{Das, Sayan}, \au{Mandal, Shubhadeep}, \au{Som, SK} \& \au{Chakraborty,
			Suman}} \yr{2017{\natexlab{{\em b\/}}}}  \at{Migration of a surfactant-laden
		droplet in non-isothermal poiseuille flow}.  \jt{Physics of Fluids}
	\bvol{29}~(1),  \pg{012002}.
	
	\bibitem[Eggleton {\em et~al.\/}(1999)Eggleton, Pawar \& Stebe]{Eggleton1999}
	{\sc \au{Eggleton, Charles~D.}, \au{Pawar, Yashodhara~P.} \& \au{Stebe,
			Kathleen~J.}} \yr{1999}  \at{{Insoluble surfactants on a drop in an
			extensional flow: a generalization of the stagnated surface limit to
			deforming interfaces}}.  \jt{Journal of Fluid Mechanics}  \bvol{385},
	\pg{79--99}.
	
	\bibitem[Ervik {\em et~al.\/}(2018)Ervik, Penne, Helles{\o}, Munkejord \&
	M{\"u}ller]{Ervik2018}
	{\sc \au{Ervik, {\AA}smund}, \au{Penne, Torstein~Eidsnes}, \au{Helles{\o},
			Svein~Magne}, \au{Munkejord, Svend~Tollak} \& \au{M{\"u}ller, Bernhard}}
	\yr{2018}  \at{Influence of surfactants on the electrohydrodynamic stretching
		of water drops in oil}.  \jt{International Journal of Multiphase Flow}
	\bvol{98},  \pg{96--109}.
	
	\bibitem[Feng(1999)]{Feng1999}
	{\sc \au{Feng, J.~Q.}} \yr{1999}  \at{{Electrohydrodynamic behaviour of a drop
			subjected to a steady uniform electric field at finite electric Reynolds
			number}}.  \jt{Proceedings of the Royal Society A: Mathematical, Physical and
		Engineering Sciences}  \bvol{455}~(1986),  \pg{2245--2269}.
	
	\bibitem[Fischer \& Erni(2007)]{Fischer2007}
	{\sc \au{Fischer, Peter} \& \au{Erni, Philipp}} \yr{2007}  \at{Emulsion drops
		in external flow fields?the role of liquid interfaces}.  \jt{Current Opinion
		in Colloid \& Interface Science}  \bvol{12}~(4-5),  \pg{196--205}.
	
	\bibitem[Griggs {\em et~al.\/}(2007)Griggs, Zinchenko \& Davis]{Griggs2007}
	{\sc \au{Griggs, Andrew~J}, \au{Zinchenko, Alexander~Z} \& \au{Davis,
			Robert~H}} \yr{2007}  \at{Low-reynolds-number motion of a deformable drop
		between two parallel plane walls}.  \jt{International journal of multiphase
		flow}  \bvol{33}~(2),  \pg{182--206}.
	
	\bibitem[Ha \& Yang(1995)]{Ha1995}
	{\sc \au{Ha, JW} \& \au{Yang, SM}} \yr{1995} {Effects of Surfactant on the
		Deformation and Stability of a Drop in a Viscous Fluid in an Electric Field}.
	
	\bibitem[Ha \& Yang(1998)]{Ha1998}
	{\sc \au{Ha, Jong-Wook} \& \au{Yang, Seung-Man}} \yr{1998}  \at{Effect of
		nonionic surfactant on the deformation and breakup of a drop in an electric
		field}.  \jt{Journal of colloid and interface science}  \bvol{206}~(1),
	\pg{195--204}.
	
	\bibitem[Haber \& Hetsroni(1971)]{Haber1971}
	{\sc \au{Haber, S.} \& \au{Hetsroni, G.}} \yr{1971}  \at{{The dynamics of a
			deformable drop suspended in an unbounded Stokes flow}}.  \jt{Journal of
		Fluid Mechanics}  \bvol{49}~(02),  \pg{257--277}.
	
	\bibitem[Haber \& Hetsroni(1972)]{Haber1972}
	{\sc \au{Haber, S.} \& \au{Hetsroni, G.}} \yr{1972}  \at{{Hydrodynamics of a
			drop submerged in an unbounded arbitrary velocity field in the presence of
			surfactants}}.  \jt{Applied Scientific Research}  \bvol{25}~(1),
	\pg{215--233}.
	
	\bibitem[Hanna \& Vlahovska(2010)]{Hanna2010}
	{\sc \au{Hanna, James~A.} \& \au{Vlahovska, Petia~M.}} \yr{2010}
	\at{{Surfactant-induced migration of a spherical drop in Stokes flow}}.
	\jt{Physics of Fluids}  \bvol{22}~(1),  \pg{1--7}.
	
	\bibitem[Happel \& Brenner(1983)]{Happel1983}
	{\sc \au{Happel, J} \& \au{Brenner, H}} \yr{1983} {\em {Low Reynolds number
			hydrodynamics, 1965}\/}.
	
	\bibitem[Hetsroni \& Haber(1970)]{Hetsroni1970}
	{\sc \au{Hetsroni, G.} \& \au{Haber, S.}} \yr{1970}  \at{{The flow in and
			around a droplet or bubble submerged in an unbound arbitrary velocity
			field}}.  \jt{Rheologica Acta}  \bvol{9}~(4),  \pg{488--496}.
	
	\bibitem[Hetsroni {\em et~al.\/}(1971)Hetsroni, Wacholder \&
	Haber]{Hetsroni1971}
	{\sc \au{Hetsroni, G.}, \au{Wacholder, E.} \& \au{Haber, S.}} \yr{1971}
	\at{{The Hydrodynamic Resistance of a Fluid Sphere Submerged in Stokes
			Flows}}.  \jt{ZAMM ??? Journal of Applied Mathematics and Mechanics /
		Zeitschrift f??r Angewandte Mathematik und Mechanik}  \bvol{51}~(1),
	\pg{45--50}.
	
	\bibitem[Hoburg \& Melcher(1977)]{Hoburg1977}
	{\sc \au{Hoburg, JF} \& \au{Melcher, JR}} \yr{1977}  \at{Electrohydrodynamic
		mixing and instability induced by co-linear fields and conductivity
		gradients}.  \jt{The Physics of Fluids}  \bvol{20}~(6),  \pg{903--911}.
	
	\bibitem[Lamb(1975)]{Lamb1975}
	{\sc \au{Lamb, Horace}} \yr{1975} {\em Hydrodynamics\/}, 6th edn.
	\publ{Cambridge University Press}.
	
	\bibitem[Lanauze {\em et~al.\/}(2015)Lanauze, Walker \& Khair]{Lanauze2015}
	{\sc \au{Lanauze, Javier~A}, \au{Walker, Lynn~M} \& \au{Khair, Aditya~S}}
	\yr{2015}  \at{{Nonlinear electrohydrodynamics of slightly deformed oblate
			drops}}.  \jt{Journal of Fluid Mechanics}  \bvol{774},  \pg{245--266}.
	
	\bibitem[Leal(2007)]{Leal2007}
	{\sc \au{Leal, L~Gary}} \yr{2007} {\em Advanced transport phenomena: fluid
		mechanics and convective transport processes\/}.  \publ{Cambridge University
		Press}.
	
	\bibitem[Lequeux(1998)]{Lequeux1998}
	{\sc \au{Lequeux, Fran{\c{c}}ois}} \yr{1998}  \at{Emulsion rheology}.
	\jt{Current opinion in colloid \& interface science}  \bvol{3}~(4),
	\pg{408--411}.
	
	\bibitem[Li \& Pozrikidis(1997)]{Li1997}
	{\sc \au{Li, Xiaofan} \& \au{Pozrikidis, C.}} \yr{1997}  \at{{The effect of
			surfactants on drop deformation and on the rheology of dilute emulsions in
			Stokes flow}}.  \jt{Journal of Fluid Mechanics}  \bvol{341}~(1997),
	\pg{165--194}.
	
	\bibitem[Magnaudet(2003)]{Magnaudet2003}
	{\sc \au{Magnaudet, Jacques}} \yr{2003}  \at{Small inertial effects on a
		spherical bubble, drop or particle moving near a wall in a time-dependent
		linear flow}.  \jt{Journal of Fluid Mechanics}  \bvol{485},  \pg{115--142}.
	
	\bibitem[Mandal {\em et~al.\/}(2016{\natexlab{{\em a\/}}})Mandal, Bandopadhyay
	\& Chakraborty]{Mandal2016b}
	{\sc \au{Mandal, Shubhadeep}, \au{Bandopadhyay, Aditya} \& \au{Chakraborty,
			Suman}} \yr{2016{\natexlab{{\em a\/}}}}  \at{{Dielectrophoresis of a
			surfactant-laden viscous drop}}.  \jt{Physics of Fluids}  \bvol{28}~(6).
	
	\bibitem[Mandal {\em et~al.\/}(2016{\natexlab{{\em b\/}}})Mandal, Bandopadhyay
	\& Chakraborty]{Mandal2016}
	{\sc \au{Mandal, Shubhadeep}, \au{Bandopadhyay, Aditya} \& \au{Chakraborty,
			Suman}} \yr{2016{\natexlab{{\em b\/}}}}  \at{{The effect of uniform electric
			field on the cross-stream migration of a drop in plane Poiseuille flow}}.
	\jt{Journal of Fluid Mechanics}  \bvol{809}~(2016),  \pg{726--774}.
	
	\bibitem[Mandal {\em et~al.\/}(2016{\natexlab{{\em c\/}}})Mandal, Bandopadhyay
	\& Chakraborty]{Mandal2016c}
	{\sc \au{Mandal, Shubhadeep}, \au{Bandopadhyay, Aditya} \& \au{Chakraborty,
			Suman}} \yr{2016{\natexlab{{\em c\/}}}}  \at{{The effect of uniform electric
			field on the cross-stream migration of a drop in plane Poiseuille flow}}.
	\jt{Journal of Fluid Mechanics}  \bvol{809},  \pg{726--774}.
	
	\bibitem[Mandal {\em et~al.\/}(2016{\natexlab{{\em d\/}}})Mandal, Bandopadhyay
	\& Chakraborty]{Mandal2016a}
	{\sc \au{Mandal, Shubhadeep}, \au{Bandopadhyay, Aditya} \& \au{Chakraborty,
			Suman}} \yr{2016{\natexlab{{\em d\/}}}}  \at{{The effect of uniform electric
			field on the cross-stream migration of a drop in plane Poiseuille flow}}.
	\jt{Journal of Fluid Mechanics}  \bvol{809}~(Leal),  \pg{726--774}.
	
	\bibitem[Melcher \& Taylor(1969)]{Melcher1969}
	{\sc \au{Melcher, JR} \& \au{Taylor, GI}} \yr{1969}  \at{Electrohydrodynamics:
		a review of the role of interfacial shear stresses}.  \jt{Annual review of
		fluid mechanics}  \bvol{1}~(1),  \pg{111--146}.
	
	\bibitem[Mhatre \& Thaokar(2013)]{Mhatre2013}
	{\sc \au{Mhatre, Sameer} \& \au{Thaokar, Rochish~M.}} \yr{2013}  \at{{Drop
			motion, deformation, and cyclic motion in a non-uniform electric field in the
			viscous limit}}.  \jt{Physics of Fluids}  \bvol{25}~(7).
	
	\bibitem[Mortazavi \& Tryggvason(2000)]{Mortazavi2000}
	{\sc \au{Mortazavi, Saeed} \& \au{Tryggvason, Gr{\'{e}}tar}} \yr{2000}  \at{{A
			numerical study of the motion of drops in Poiseuille flow. Part 1. Lateral
			migration of one drop}}.  \jt{Journal of Fluid Mechanics}  \bvol{411},
	\pg{325--350}.
	
	\bibitem[Mukherjee \& Sarkar(2013)]{Mukherjee2013}
	{\sc \au{Mukherjee, Swarnajay} \& \au{Sarkar, Kausik}} \yr{2013}  \at{Effects
		of matrix viscoelasticity on the lateral migration of a deformable drop in a
		wall-bounded shear}.  \jt{Journal of Fluid Mechanics}  \bvol{727},
	\pg{318--345}.
	
	\bibitem[Nganguia {\em et~al.\/}(2013)Nganguia, Young, Vlahovska,
	B{\l}awzdziewcz, Zhang \& Lin]{Nganguia2013}
	{\sc \au{Nganguia, Herve}, \au{Young, Y.~N.}, \au{Vlahovska, Petia~M.},
		\au{B{\l}awzdziewcz, Jerzy}, \au{Zhang, J.} \& \au{Lin, H.}} \yr{2013}
	\at{{Equilibrium electro-deformation of a surfactant-laden viscous drop}}.
	\jt{Physics of Fluids}  \bvol{25}~(9).
	
	\bibitem[Pak {\em et~al.\/}(2014{\natexlab{{\em a\/}}})Pak, Feng \&
	Stone]{Pak2014}
	{\sc \au{Pak, On~Shun}, \au{Feng, Jie} \& \au{Stone, Howard~A}}
	\yr{2014{\natexlab{{\em a\/}}}}  \at{Viscous marangoni migration of a drop in
		a poiseuille flow at low surface p{\'e}clet numbers}.  \jt{Journal of Fluid
		Mechanics}  \bvol{753},  \pg{535--552}.
	
	\bibitem[Pak {\em et~al.\/}(2014{\natexlab{{\em b\/}}})Pak, Feng \&
	Stone]{Pak2014a}
	{\sc \au{Pak, On~Shun}, \au{Feng, Jie} \& \au{Stone, Howard~a.}}
	\yr{2014{\natexlab{{\em b\/}}}}  \at{{Viscous Marangoni migration of a drop
			in a Poiseuille flow at low surface P{\'{e}}clet numbers}}.  \jt{Journal of
		Fluid Mechanics}  \bvol{753},  \pg{535--552}.
	
	\bibitem[Pawar \& Stebe(1996)]{Pawar1996}
	{\sc \au{Pawar, Yashodhara} \& \au{Stebe, Kathleen~J.}} \yr{1996}
	\at{{Marangoni effects on drop deformation in an extensional flow: The role
			of surfactant physical chemistry. I. Insoluble surfactants}}.  \jt{Physics of
		Fluids}  \bvol{8}~(November 1995),  \pg{1738--1751}.
	
	\bibitem[Poddar {\em et~al.\/}(2017)Poddar, Dhar \& Chakraborty]{Poddar2017}
	{\sc \au{Poddar, Antarip}, \au{Dhar, Jayabrata} \& \au{Chakraborty, Suman}}
	\yr{2017}  \at{Electro-osmosis of nematic liquid crystals under weak
		anchoring and second-order surface effects}.  \jt{Physical Review E}
	\bvol{96}~(1),  \pg{013114}.
	
	\bibitem[Ptasinski \& Kerkhof(1992)]{Ptasinski1992}
	{\sc \au{Ptasinski, KJ} \& \au{Kerkhof, PJAM}} \yr{1992}  \at{Electric field
		driven separations: phenomena and applications}.  \jt{Separation science and
		technology}  \bvol{27}~(8-9),  \pg{995--1021}.
	
	\bibitem[Saville(1997)]{Saville1997}
	{\sc \au{Saville, DA}} \yr{1997}  \at{Electrohydrodynamics: the taylor-melcher
		leaky dielectric model}.  \jt{Annual review of fluid mechanics}
	\bvol{29}~(1),  \pg{27--64}.
	
	\bibitem[Schwalbe {\em et~al.\/}(2011)Schwalbe, Phelan~Jr, Vlahovska \&
	Hudson]{Schwalbe2011}
	{\sc \au{Schwalbe, Jonathan~T}, \au{Phelan~Jr, Frederick~R}, \au{Vlahovska,
			Petia~M} \& \au{Hudson, Steven~D}} \yr{2011}  \at{Interfacial effects on
		droplet dynamics in poiseuille flow}.  \jt{Soft Matter}  \bvol{7}~(17),
	\pg{7797--7804}.
	
	\bibitem[Sengupta {\em et~al.\/}(2017)Sengupta, Walker \& Khair]{Sengupta2017}
	{\sc \au{Sengupta, Rajarshi}, \au{Walker, Lynn~M.} \& \au{Khair, Aditya~S.}}
	\yr{2017}  \at{The role of surface charge convection in the
		electrohydrodynamics and breakup of prolate drops}.  \jt{Journal of Fluid
		Mechanics}  \bvol{833},  \pg{29?53}.
	
	\bibitem[Spertell \& Saville(1974)]{Spertell1974}
	{\sc \au{Spertell, RB} \& \au{Saville, DA}} \yr{1974}  \at{The roles of
		electrohydrodynamic phenomena in the motion of drops and bubbles} .
	
	\bibitem[Stan {\em et~al.\/}(2011)Stan, Guglielmini, Ellerbee, Caviezel, Stone
	\& Whitesides]{Stan2011}
	{\sc \au{Stan, Claudiu~A.}, \au{Guglielmini, Laura}, \au{Ellerbee, Audrey~K.},
		\au{Caviezel, Daniel}, \au{Stone, Howard~A.} \& \au{Whitesides, George~M.}}
	\yr{2011}  \at{{Sheathless hydrodynamic positioning of buoyant drops and
			bubbles inside microchannels}}.  \jt{Physical Review E - Statistical,
		Nonlinear, and Soft Matter Physics}  \bvol{84}~(3).
	
	\bibitem[Stone \& Leal(1990)]{Stone1990a}
	{\sc \au{Stone, H.~a.} \& \au{Leal, L.~G.}} \yr{1990}  \at{{The effects of
			surfactants on drop deformation and breakup}}.  \jt{Journal of Fluid
		Mechanics}  \bvol{220},  \pg{161}.
	
	\bibitem[Subramanian \& Balasubramaniam(2001)]{Subramanian2001}
	{\sc \au{Subramanian, R~Shankar} \& \au{Balasubramaniam, Ramaswamy}} \yr{2001}
	{\em The motion of bubbles and drops in reduced gravity\/}.  \publ{Cambridge
		University Press}.
	
	\bibitem[Subramanian \& Balasubramaniam(2005)]{Subramanian2005}
	{\sc \au{Subramanian, R.~Shankar} \& \au{Balasubramaniam, R.}} \yr{2005} {\em
		{The Motion of Bubbles and Drops in Reduced Gravity}\/}.  \publ{Cambridge:
		Cambridge University Press}.
	
	\bibitem[Taylor(1966)]{Taylor1966}
	{\sc \au{Taylor, Geoffrey}} \yr{1966}  \at{Studies in electrohydrodynamics. i.
		the circulation produced in a drop by an electric field}.  \jt{Proceedings of
		the Royal Society of London A: Mathematical, Physical and Engineering
		Sciences}  \bvol{291}~(1425),  \pg{159--166}.
	
	\bibitem[Teigen \& Munkejord(2010)]{Teigen2010}
	{\sc \au{Teigen, Knut~Erik} \& \au{Munkejord, Svend~Tollak}} \yr{2010}
	\at{{Influence of surfactant on drop deformation in an electric field}}.
	\jt{Physics of Fluids}  \bvol{22}~(11).
	
	\bibitem[Tsouris {\em et~al.\/}(2003)Tsouris, Culbertson, DePaoli, Jacobson,
	De~Almeida \& Ramsey]{Tsouris2003}
	{\sc \au{Tsouris, C}, \au{Culbertson, CT}, \au{DePaoli, DW}, \au{Jacobson, SC},
		\au{De~Almeida, VF} \& \au{Ramsey, JM}} \yr{2003}  \at{Electrohydrodynamic
		mixing in microchannels}.  \jt{AIChE journal}  \bvol{49}~(8),
	\pg{2181--2186}.
	
	\bibitem[Tsukada {\em et~al.\/}(1993)Tsukada, Katayama, Ito \&
	Hozawa]{Tsukada1993}
	{\sc \au{Tsukada, Takao}, \au{Katayama, Tetsuya}, \au{Ito, Yukikazu} \&
		\au{Hozawa, Mitsunori}} \yr{1993}  \at{{Theoretical and Experimental Studies
			of Circulations Inside and Outside a Deformed Drop under a Uniform Electric
			Field.}}  \jt{JOURNAL OF CHEMICAL ENGINEERING OF JAPAN}  \bvol{26}~(6),
	\pg{698--703}.
	
	\bibitem[Tucker \& Moldenaers(2002)]{Tucker2002}
	{\sc \au{Tucker, Charles~L} \& \au{Moldenaers, Paula}} \yr{2002}
	\at{{Microstructural evolution in polymer blends}}.  \jt{Annu. Rev. Fluid
		Mech.}  \bvol{34},  \pg{177--210}.
	
	\bibitem[Uijttewaal \& Nijhof(1995)]{Uijttewaal1995}
	{\sc \au{Uijttewaal, W. S.~J} \& \au{Nijhof, E.~J.}} \yr{1995}  \at{The motion
		of a droplet subjected to linear shear flow including the presence of a plane
		wall}.  \jt{Journal of Fluid Mechanics}  \bvol{302},  \pg{45--63}.
	
	\bibitem[Wu \& Clark(2008)]{Wu2008}
	{\sc \au{Wu, Yiquan} \& \au{Clark, Robert~L}} \yr{2008}
	\at{Electrohydrodynamic atomization: a versatile process for preparing
		materials for biomedical applications}.  \jt{Journal of Biomaterials Science,
		Polymer Edition}  \bvol{19}~(5),  \pg{573--601}.
	
	\bibitem[Xu {\em et~al.\/}(2006)Xu, Li, Lowengrub \& Zhao]{Xu2006}
	{\sc \au{Xu, Jian~Jun}, \au{Li, Zhilin}, \au{Lowengrub, John} \& \au{Zhao,
			Hongkai}} \yr{2006}  \at{{A level-set method for interfacial flows with
			surfactant}}.  \jt{Journal of Computational Physics}  \bvol{212}~(2),
	\pg{590--616}.
	
	\bibitem[Xu \& Homsy(2006)]{Xu2006a}
	{\sc \au{Xu, Xiumei} \& \au{Homsy, G.~M.}} \yr{2006}  \at{{The settling
			velocity and shape distortion of drops in a uniform electric field}}.
	\jt{Journal of Fluid Mechanics}  \bvol{564}~(May),  \pg{395}.
	
	\bibitem[Yariv \& Almog(2016)]{Yariv2016}
	{\sc \au{Yariv, Ehud} \& \au{Almog, Yaniv}} \yr{2016}  \at{{The effect of
			surface-charge convection on the settling velocity of spherical drops in a
			uniform electric field}}.  \jt{Journal of Fluid Mechanics}  \bvol{797},
	\pg{536--548}.
	
	\bibitem[Zhang {\em et~al.\/}(2015)Zhang, He, Ghadiri \& Hassanpour]{Zhang2015}
	{\sc \au{Zhang, L.}, \au{He, L.}, \au{Ghadiri, M.} \& \au{Hassanpour, A.}}
	\yr{2015}  \at{{Effect of surfactants on the deformation and break-up of an
			aqueous drop in oils under high electric field strengths}}.  \jt{Journal of
		Petroleum Science and Engineering}  \bvol{125},  \pg{38--47}.
	\end{thebibliography}
\end{document}